\documentclass[aps,prf,a4paper,groupedaddress]{revtex4-2}
\usepackage[utf8]{inputenc} % encodage à privilégier pour la portabilité et +
\usepackage[lmargin=2cm,rmargin=2cm,tmargin=2cm,bmargin=2cm]{geometry}
\usepackage{natbib}
\usepackage{subfig}
\usepackage{graphicx} % inclusion des figures
\usepackage[svgnames,dvipsnames]{xcolor}  % jeux de couleurs avec des noms conviviaux
\usepackage{fancyhdr}
\usepackage{fancybox}
\usepackage[labelfont=bf]{caption} % référencer des parties d’une figure
\usepackage[T1]{fontenc} % encodage européen des caractères (Cork)
\usepackage{comment}
\usepackage{tikz}
\usetikzlibrary{shapes.geometric,calc}
\usetikzlibrary{shapes}
\usepackage{pgfplots} % pour les graphiques couplé avec Tikz

\usepackage{amsmath,amsfonts,amssymb}
\usepackage{float} % fixer les figures

\usepackage{colortbl} % colorier un tableau
\usepackage[colorlinks=false,linkcolor=blue ,filecolor=red ]{hyperref}

 \fancypagestyle{plain}

\newcommand\cchi{\raisebox{1.5pt}{$\chi$}}

\begin{document}
\title{Flow structure and loads over inclined cylindrical rodlike particles and fibers}

\author{Mohammed Kharrouba$^{1,2}$}
\author{Jean-Lou Pierson$^1$\footnote{{Email address for correspondence: jean-lou.pierson@ifpen.fr}}}
\author{Jacques Magnaudet$^2$\footnote{{Email address for correspondence: magnau@imft.fr}}\vspace{2mm}\\}
\affiliation{$^1$IFP Energies Nouvelles, 69360 Solaize, France}
\affiliation{$^2$Institut de M\'ecanique des Fluides de Toulouse (IMFT), Universit\'e de Toulouse, CNRS, Toulouse, France}
%\author{\vspace{3mm} Mohammed Kharrouba$^{1,2,\dagger}$, Jean-Lou Pierson$^1$, Jacques Magnaudet$^2$\\
%\normalsize $^1$ IFP Energies nouvelles, 69360 Solaize, France\\ 
%\normalsize $^2$ Institut de M\'ecanique des Fluides de Toulouse (IMFT), Universit\'e de Toulouse, CNRS, Toulouse, France\\
%\normalsize $^\dagger$ E-mail: mohammed.kharrouba@ifp.fr
%\tiny\\
%\date{}
\maketitle

\section*{Abstract} The flow past a fixed finite-length circular cylinder, the axis of which makes a nonzero angle with the incoming stream, is studied through fully-resolved simulations, from creeping-flow conditions to strongly inertial \color{black} regimes. The investigation focuses on the way the body aspect ratio $\cchi$ (defined as as the length-to-diameter ratio), the inclination angle $\theta$ with respect to the incoming flow and the Reynolds number $\text{Re}$ (based on the cylinder diameter) affect the flow structure past the body and therefore the hydrodynamic loads acting on it. \color{black} The configuration $\theta=0^\circ$ (where the cylinder is aligned with the flow) is first considered from creeping-flow conditions up to $\text{Re}=400$, with aspect ratios up to $20$ ($10$) for $\text{Re}\leq10$ ($\text{Re}\geq10$). In the low-to-moderate Reynolds number regime ($\text{Re}\lesssim5$), %the specific case where the cylinder is aligned with the upstream flow ($\theta=0^\circ$), 
influence or the aspect ratio, inclination (from $0^{\circ}$ to $30^{\circ}$), and inertial effects is examined by comparing numerical results for the axial and transverse force components and the spanwise torque with theoretical predictions based on the slender-body approximation, possibly incorporating finite-Reynolds-number corrections. Semiempirical models based on these predictions and incorporating finite-length and inertial corrections extracted from the numerical data are derived. \color{black} %When the cylinder is inclined with respect to the incoming flow and inertial effects are dominant %, three distinct types of vortex patterns, all preserving a symmetry plane, are observed in the wake 
%past cylinders with aspect ratios in the range $3-7$, depending on $\text{Re}$ and $\theta$ (which spans the range $0^{\circ}-30^{\circ}$). %Whatever $\theta\neq0^\circ$, the wake structure involves a steady pair of counter-rotating streamwise vortices up to $\text{Re}\approx330$. 
For large enough Reynolds numbers ($\text{Re}\gtrsim10^2$), separation takes place along the upstream part of the lateral surface of the cylinder, deeply influencing the surface stress distribution. %Beyond $\text{Re}\approx330$, increasing the inclination angle first leads to a regime in which double-sided hairpin-shaped vortices are shedded downstream, while another regime characterized by the shedding of single-sided  hairpin vortices takes over for large enough inclinations. 
Numerical results are used to build empirical models for the force components and the torque, valid for moderately inclined cylinders ($|\theta|\lesssim30^\circ$) of arbitrary aspect ratio up to $\text{Re}\approx300$ \color{black} and matching those obtained at low-to-moderate Reynolds number. \color{black} %While the force perpendicular to the cylinder axis follows a remarkably simply law, the axial force and the spanwise torque exhibit complex dependencies with respect to the three control parameters. In particular, finite-length effects are observed to have a strong influence.% which may even reverse the sign of the torque for short cylinders ($\cchi\approx3$) and large enough Reynolds numbers. 
%\begin{equation}
%\frac{\partial\boldsymbol{\omega}_a}{\partial t}+(\mathbf{u}\cdot\nabla)\boldsymbol{\omega}_a=(\boldsymbol{\omega}_a\cdot\nabla)\mathbf{u}+\nu\nabla^2\boldsymbol{\omega}_a\quad\quad\quad\quad (2.1)
%\end{equation}

\section{Introduction}
\label{intro}
The flow past rodlike cylindrical particles or cylindrical fibers with circular cross section is involved in many industrial and natural processes such as bubbling fluidized beds, pulp and paper making or the sedimentation of ice crystals in clouds. Despite the large number of studies devoted to the flow past a circular cylinder held perpendicular to the incoming flow in the laminar and transitional regimes, much less is known when the body is arbitrarily inclined or even aligned with this incoming flow. Three dimensionless parameters then govern the problem when the upstream flow is steady and uniform: the aspect ratio $\cchi=L/D$ where $L$ is the length of the cylinder and $D$ its diameter, the inclination angle $\theta$  (pitch or yaw angle) which is the angle between the cylinder axis and the incoming velocity, and the Reynolds number $\text{Re}=\frac{\rho UD}{\mu}$, where $U$ is the norm of the upstream velocity, and $\rho$ and $\mu$ are the fluid density and dynamic viscosity, respectively.\\
\indent Up to now, the flow past long rigid cylinders and the loads acting upon them have been investigated in two markedly different contexts. The first of them is the dynamics of dilute suspensions of slender particles, the theoretical study of which was pioneered by Batchelor \cite{batchelor1970} and Cox \cite{cox1970} in the creeping-flow limit. The corresponding results for the force and torque acting on an isolated fiber (see also \cite{tillett1970}), based on the slender-body theory, were extended to low-but-finite Reynolds numbers by Khayat and Cox \cite{khayat1989}. Since then, these various predictions have been extensively used in numerical simulations, be it to study the influence of hydrodynamic interactions and concentration on the sedimentation of a fiber suspension \citep{mackaplow1998,butler2002} or to reveal the dispersion properties of such suspensions in isotropic \citep{shin2005} or wall-bounded \citep{marchioli2010} turbulence; see \cite{voth2017} for a review.\\
\indent The second stream of investigations, often motivated by vortex-induced vibrations and applications to fluid-structure interactions, has focused on larger Reynolds numbers. % corresponding to transitional regimes in the wake. 
\citet{ramberg1983} studied experimentally the flow past long inclined cylinders with various end shapes for Reynolds numbers in the range $1.5\times10^2-10^3$. His experiments provide a qualitative map of the wake topology and shedding process as a function of the inclination angle. %Under such conditions, the wake is unsteady in the classical configuration in which the cylinder axis is perpendicular to the incoming flow. 
In particular, they show that, unlike the classical vortex patterns observed when the cylinder is held perpendicular to the flow \citep{williamson1996}, the wake is dominated by a pair of counter-rotating vortices emanating from the ends when the inclination angle is small enough.  %\cite{auguste2010bif} studied a flow around a $\cchi=1/3$ cylinder aligned with the flow, and noticed that the flow remains steady and axisymmetric at low enough values of the Reynolds number. Increasing the Reynolds number, a steady bifurcation occurs and leads to a steady-state mode characterized by a reflectional symmetry plane. A second bifurcation takes place, where the flow within the wake and the forces exerted by the cylinder become unsteady and periodic. It was observed that the wake retains the symmetry plane selected by the previous regime. 
Numerical studies of the flow past an inclined cylinder in inertia-dominated regimes have also been reported \cite{vakil2009,pierson2019}. Based on the numerical data, these investigations proposed empirical expressions for the drag and lift forces valid throughout the considered range of $\cchi$ and $\text{Re}$ for arbitrary inclinations. Both studies examined the applicability and limitations of the so-called Independence Principle \citep{sears1948, ramberg1983,zdravkovich1997b}. This `principle', which states that the perpendicular force on a long cylinder depends solely on the normal velocity component of the incoming flow, was shown to apply only to large inclinations, $\theta\gtrsim45^\circ$. \citet{pierson2019} also considered the possibility to extend trigonometric relations valid under Stokes flow conditions to obtain the drag and lift forces at an arbitrary $\theta$ through simple linear combinations of the drag forces corresponding to the two extreme cases $\theta=0^\circ$ and $\theta=90^\circ$, an approach that has proved successful for prolate spheroids over a wide range of Reynolds numbers \citep{sanjeevi2017}.
%derived in the Stokes regime which fit well with their numerical results.
\vspace{2mm}\\ 
\indent In this paper, we expand on available studies by considering the flow past finite-length cylinders with flat ends, from creeping flow conditions up to \color{black} $\text{Re}\gtrsim 300$ \color{black} for moderate inclinations, namely $0^\circ\leq\theta\leq30^\circ$. The cylinder aspect ratio is varied from $2$ to $20$ in the reference case $\theta=0^\circ$, and from $3$ to $7$ in inclined configurations. By considering this wide range of Reynolds number, we aim at bridging the gap between conditions typical of submillimeter-diameter fibers relevant for instance to papermaking (for which $D$ stands typically in the range $15-30\,\mu$m) and millimeter-diameter rodlike particules relevant to fluidized beds (which have aspect ratios typically in the range $2-10$). The reason why we concentrate on low-to-moderate inclinations is two-fold. First, as already mentioned, the current knowledge of the flow structure and drag variations with $\text{Re}$ over a slender circular cylinder aligned with the flow is still far from complete. To the best of our knowledge, no study has considered this configuration from creeping-flow conditions up to Reynolds of some hundreds which are easily reached in some of the applications mentioned above. Similar to the case of a finite-length cylinder held perpendicular to the flow, a more complete description of this reference configuration is mandatory to improve the predictions of the trajectories of sedimenting rodlike particles and fibers spanning all possible orientations with respect to their path. Then, with the same final objective in mind, an important question in inertia-dominated regimes is to understand how  the flow structure and the loads on the body are affected by the loss of axial symmetry encountered as soon as the inclination angle becomes nonzero. In particular, given the three-dimensional nature of the flow past an inclined cylinder, it is not clear how far the trigonometric approach mentioned above can be used to predict realistically the loads acting on it, based only on results for the axisymmetric geometry corresponding to $\theta=0^\circ$ and the nearly two-dimensional geometry (for large $\chi$) corresponding to $\theta=90^\circ$. These are the main objectives of the present investigation.\\
% We use significant computational resources required to properly resolve the flow around the fiber. 
\indent The paper is organized as follows. We first present the numerical approach in Sec. \ref{num} and, in appendix \ref{app:num}, provide a validation of this approach by comparing some of the results obtained at moderate Reynolds number with those of \cite{pierson2019}. In Sec. \ref{aligned}, we specifically examine the case of a finite-length cylinder aligned with the incoming flow. We first consider (Sec. \ref{creep}) low-to-moderate Reynolds numbers, for which we use numerical results to improve over available drag predictions provided by the slender-body approximation (in appendix \ref{app:slender}, we extend the available theoretical prediction in the creeping-flow limit by computing explicitly the next-order finite-aspect-ratio correction). Then (Sec. \ref{moder}), we consider Reynolds numbers in the range $20-400$ and use numerical predictions for the pressure and viscous contributions to the drag to obtain an empirical drag law valid whatever $\cchi$ throughout this range of $\text{Re}$. The inclined configuration is examined in \color{black} Sec. \ref{incl1} in the low-to-moderate Reynolds number range. %\color{red} In Sec. \ref{incl1} we analyze the various flow patterns observed in the steady and unsteady regimes as the inclination increases, and the dynamics of the loads (force and torque) on the body in the unsteady regimes.
 \color{black} We compare numerical findings with available theoretical predictions and make use of results established in Sec. \ref{creep} to provide semiempirical laws for the drag, lift and torque as a function of the control parameters (since the law for the transverse force involves the drag on a cylinder held perpendicular to the flow, we discuss slender-body predictions in this specific configuration and extend them empirically in appendix \ref{perps}). In Sec. \ref{incl2}, we proceed with the flow past an inclined cylinder in the moderate-to-large Reynolds number range $10\lesssim\text{Re}\lesssim300$. We analyze the structure of the steady non-axisymmetric flow and its connection with the observed, sometimes nonintuitive, variations of the loads with $\text{Re}$, $\theta$ and $\cchi$. We finally provide empirical fits capable of reproducing these complex variations throughout the explored range of parameters. \color{black} A summary of the main results and a discussion of some open issues are provided in Sec. \ref{Summ}.%Some conclusions are provided in $Section 5$    
\section{Numerical methodology}
\label{num}
\begin{figure}[H]
	\centering
\begin{tikzpicture}[scale=1.0]
%%%Domaine
%\draw (0,0,0)--(8,0,0)--(8,4,0)--(0,4,0)--cycle; % face arrière
%\draw (0,0,4)--(8,0,4)--(8,4,4)--(0,4,4)--cycle; % face avant
%\node [cylinder, black, rotate=0, draw, minimum height=10cm, minimum width=5cm]  at (3, 1 ,0) {};
\draw (1.5,1) -- (3.5,1) -- (3.5,1.5) -- (1.5,1.5) -- (1.5,1) ;
\draw[red, <->] (1.5,1.25)   -- (3.5,1.25) node [at end, above left, red]   {$L$};
\draw[red, <->] (2.5,1)node [below, red] {$\frac{D}{2}$} -- (2.5,1.5) ;
\draw (-2,1) rectangle (7.5,3.5);
%arêteshorizontales,del’arrièreversl’avant
%\draw (0,0,0) -- (0,0,4); % bas gauche
%\draw (8,0,0) -- (8,0,4); % bas droit
%\draw (8,4,0) -- (8,4,4); % haut droit
%\draw (0,4,0) -- (0,4,4); % haut gauche
%%%Cube rafine
%%%Cylindre
%\node [cylinder, black, rotate=0, draw, minimum height=3cm, minimum width=0.5cm]  at (3, 2 ,2) (c) {}; %ne sera pas a lechelle mais pas grave
%\draw[dashed,blue] (1.5,1.75)--(1.5,2.5);
%\draw[dashed,blue] (1.5,2.5)--(4.5,2.5);
%\draw[dashed,blue] (4.5,2.5)--(4.5,1);
\draw[dashed] (1.25,1)--(1.25,1.75);
\draw[dashed] (1.25,1.75)--(3.75,1.75);
\draw[dashed] (3.75,1.75)--(3.75,1);
%\draw[<->, thick] (3.75,1.5)--(4.5,1.5) node[at start, above right]{$L_{x}$};
%\draw[<->, thick] (2,1.75)--(2,2.5) node[at start, above right]{$L_{y}$};
\draw[<->, thick] (-2,4)-- node[above]{$\mathcal{L}$} (7.5,4) ;
\draw[<->, thick] (9,1)-- node[right]{$\mathcal{R}$} (9,3.5) ;
\draw[<->, thick] (-2.0,1.25)--(1.5,1.25) node[at start, above right]{$L_{up}$};
\draw[<->, thick] (3.5,1.25)--(7.5,1.25) node[at end, above left]{$L_{down}$};
\draw[->, thick] (5,0)--(5.5,0) node[right]{$\bf{e_x}$};
\draw[->, thick] (5,0)--(5,0.5) node[right]{$\bf{e_y}$};
\draw[->, thick] (5,0)--(4.65,-0.28) node[left]{$\bf{e_z}$};
\draw[dashed] (-0.5,1,2)--(0.5,1,2);
\draw[->,black] (0.5,1,2)  arc (0:23:1.1)node[right]{$\theta$}; %draws an arc of radius 3 starting from (0,0)
\draw[->, thick] (-0.5,1,2)-- node[above]{$\mathbf{U}$} (0.5,1.5,2) ;
\draw (-2.5,2.25) node[]{Inlet};
\draw (3.5,3.7) node[]{Outlet};
\draw (8.3,2.25) node[]{Outlet};
\end{tikzpicture}
\caption{Scheme of the computational domain in an azimuthal plane (not to scale).}
\label{fig:domain2D}
\end{figure}
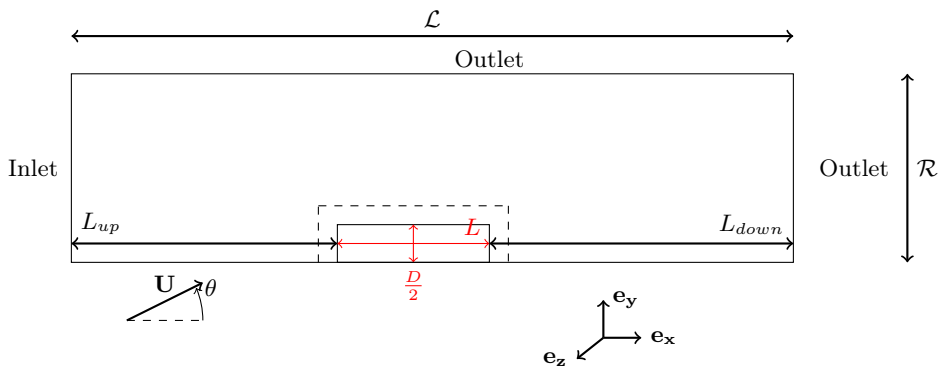

We consider the uniform incompressible steady flow of a Newtonian fluid past a finite-length circular cylinder. Computations are carried out with the JADIM code developed at IMFT. This code was used in the past to investigate various problems involved in the local dynamics of particle-laden and bubbly flows, among which the hydrodynamic forces acting on spheres in uniform or accelerated flows \citep{magnaudet1995}, the transition in the wake of spheres, disks and short cylinders \citep{fabre2008, auguste2010bif} and the path instabilities of freely-falling disks and light spheres \citep{auguste2013, auguste2018}. The code solves the three-dimensional unsteady Navier-Stokes equations using a finite volume discretization on a staggered grid. Centered schemes are used to discretize spatial derivatives in the momentum equation. Time advancement is achieved by combining a third-order Runge-Kutta algorithm for advective terms with a Crank-Nicolson scheme for viscous terms. The divergence-free condition is satisfied to machine accuracy at the end of each time step using a projection method. More details on the numerical methodology may be found in \cite{magnaudet1995} and \cite{calmet1997}.\\
\indent The present investigation makes use of a cylindrical computational domain with length $\mathcal{L}$ and radius $\mathcal{R}$ (Fig. \ref{fig:domain2D}). The length $\mathcal{L}$ may be decomposed as $\mathcal{L}=L_{up}+L+L_{down}$, where $L_{up}$ (resp. $L_{down}$) is the distance between the domain inlet (outlet) and the upstream (downstream) end of the cylinder. In what follows, for moderate-to large Reynolds numbers (say $\text{Re}>5$), we select $L_{up}=12D\cchi^{1/3}$ and $L_{down}=20D\cchi^{1/3}$. The reason why $L_{up}$ and $L_{down}$ are defined based on $D\cchi^{1/3}$, a length scale proportional to the diameter of the equivalent sphere, \textit{i.e.} the sphere with the same volume as the cylinder, is that this choice makes the size of the domain vary with the body aspect ratio while keeping the computational cost reasonable \citep{pierson2019}. With the above choice, $L_{down}$ is larger than $15D$ even for $\cchi=\mathcal{O}(1)$, which guarantees that the near wake is properly resolved for short-length cylinders \citep{auguste2010}. The radius of the numerical domain is  chosen as $\mathcal{R}=0.5D + 20D\cchi^{1/3}(1+0.8\sin\theta)$ (Fig. \ref{fig:domain2D}). It increases with the inclination angle to make sure that the wake is correctly captured whatever the body inclination. \color{black} At low Reynolds number, care must be taken of artificial confinement effects inherent to the $1/\mathfrak{r}$-decay of the disturbance (with $\mathfrak{r}$ the distance to the body center). For this reason, in the low-to-moderate-$\text{Re}$ regime  $\text{Re}\leq5$, we increase $L_{up}$ and $L_{down}$ by a factor of at least $2$, and increase $\mathcal{R}$ by a factor of at least $3$ compared to the above values. \color{black} 
A uniform fluid velocity making an angle $\theta$ with the body axis is specified on the inlet plane (Fig. \ref{fig:domain2D}). A no-slip boundary condition is imposed on the body, while a non-reflecting boundary condition is imposed on both the outlet plane and the lateral boundary \citep{magnaudet1995}. We define a Cartesian coordinate system ($x,y,z$) centered at the body geometrical center, with $x$ parallel to the body symmetry axis and $z$ perpendicular to the plane containing the direction of the upstream flow and the body axis. In this coordinate system, the incoming velocity is $\mathbf{U}=(U\cos\theta,U\sin\theta,0))$.
%for an incompressible and homogeneous fluid. A boundary-fitted method with second-order accuracy in time and space is used. The advection term is discretized using a third-order Runge-Kutta scheme while the viscous term is solved using a semi-implicit Cranck-Nicolson scheme. 
Simulations are performed with an axisymmetric cylindrical grid made involving regions with uniform and non-uniform cell distributions. In the cross-sectional plane $z=0$, a refined uniform distribution is used near the corners of the body, to properly capture the local flow (see \cite{auguste2013} for details). A slightly non-uniform distribution is imposed in a rectangular region extending up to $0.5D$ outside of the body in each direction (dashed line in Fig. \ref{fig:domain2D}). In this flow region, the cell aspect ratio is maintained below $4$ everywhere. Non-uniform cell distributions are used around the symmetry axis $y=z=0$ and near the body symmetry plane $x=0$. \color{black} In the low-to-moderate Reynolds number regime $\text{Re}\leq5$, we select a grid with $20$ cells per body diameter (earlier computations were performed with only $12$ cells per body diameter and minimal changes were noticed between the two resolutions). For larger Reynolds numbers, \color{black} we assume that the boundary layer  thickness scales as $D\text{Re}^{-1/2}$ and make sure that at least $8$ cells stand within it. The characteristic grid size in this region is thus $\Delta \approx D/(8\text{Re}^{1/2})$. Whatever the Reynolds number, the grid is non-uniform in the outer region (\textit{i.e.} beyond the dashed rectangle in Fig. \ref{fig:domain2D}), with cell sizes following a geometric law. For $\text{Re}>5$, the growth of the cells along the domain axis is controlled to guarantee that the wake is adequately resolved. In the azimuthal direction, $32$ to $128$ planes are used, depending on the Reynolds number. \color{black} No discernible difference in the solutions returned by the $32$ and the $64$ azimuthal resolutions was noticed up to $\text{Re}=5$. \color{black} The highest azimuthal resolution ensures that the cells closest to the body have approximatively the same size in all three directions. Although the code has been extensively validated in the past, an additional validation involving a grid convergence analysis is reported in appendix \ref{app:num}. Present results are found to agree well with those of \cite{pierson2019} obtained with a distinct numerical methodology.

%comparison of the present results to the one of \cite{pierson2019} is also 

%We find a good agreement between our numerical results; the gap is less than $2\%$ comparing the mean drag coefficient at different Reynolds number.

%Simulations are performed using a mixed regular and irregular grid.
%In the blue rectangle, an non uniform grid is used where the biggest cell is about five times bigger than the small one. As shown in figure \ref{fig:domain2D}, this blue zone in the wake is depending on the yawed angle and the aspect ratio following the law : $L_x=$4\raisebox{2pt}{$\chi$}$^{1/3}\cos\theta$ and $L_y=4\raisebox{2pt}{$\chi$}^{1/3}\sin\theta$ . 

%\begin{figure}[H]
%		\centering
%	\begin{tabular}{>{\centering\arraybackslash} m{2cm} >{\centering\arraybackslash} m{4cm}}
%	   \input{planz_azimutaux} &
%	\end{tabular} 
%	\caption{ Schematic of the azimuth plans. Section in the plane perpendicular to the cylinder axis
%\label{fig:plans_azimutaux}}
%\end{figure}
\section{Cylinder aligned with the upstream flow}
\label{aligned}
In this section we investigate the flow past a finite-length cylinder aligned with the incoming velocity. Our main purpose is to provide drag laws for cylinders of arbitrary aspect ratio beyond $2$ over a wide range of Reynolds number ($0.05\leq \text{Re} \leq 400$). We consider aspect ratios up to $20$ and discuss the results in ascending order of Reynolds number. %We divide the presentation in two sections In both section we provide a drag laws 
%Since the flow is axisymmetric in all the cases investigated here, we use a pure
%Aspect ratio up to 20 are investigated in this section. %The following aspect ratio are investigated ger
%J’ai choisi d’étudier que chi=3 ;5 ;7 dans le cas incliné par souci de temps de calcul ; 5 rapports de forme ca faisait bcp et en plus en terme de sillage y aura pas de nouveautés ( j’avais lancé quelques cas chi=2 au tout début de l’étude et ca ne se distingue pas de cgi=3)
%En ce qui concerne la comparaison avec tes résultats  j’ai fait tourner le cas 75° avec la methode corps libre mais en gardant le meme maillage utilisé par la methode corps fixe ;
%Mais je pense que je vais l’enlever parce qu’on se limite à 30° dans l’étude et ca n’a pas de sens de comparer à 75°.
%\subsection{Hydrodynamic forces}
%\subsection{Flow at small Reynolds numbers ($\text{Re} \ll 1$)}
\subsection{From creeping-flow conditions to $\mathcal{O}(10)$-Reynolds numbers}
\label{creep}
No exact results for the stress distribution on a finite-length cylinder exists in the Stokes regime. \citet[]{clift1978} reported a detailed comparison of numerical and experimental results with empirical laws from the literature aimed at estimating the drag force on a cylinder aligned with the upstream flow. Approximating the cylinder geometry with a prolate spheroid of same volume yields a relative error up to $15 \%$ on the drag for small aspect ratios ($\cchi \approx 2$). %\cite{clift1978} provide an empirical formula for the drag, based on experimental results by \cite{heiss1952}. However this formula fails for large aspect ratios ($\cchi \gtrsim 9$).\\
 To provide a drag law valid for $\cchi \gtrsim 2$, we start from the slender-body theory. This theory provides a convenient framework to compute the forces on a slender body under creeping-flow conditions through an expansion with respect to the small parameter $ 1/\ln(2\cchi)$ for $\cchi \gg 1$ \citep{batchelor1970,cox1970}. The solution corresponding to a a straight circular fiber was obtained to third order by \citet{batchelor1970} and \citet{keller1976}. In appendix \ref{app:slender}, we improve over these predictions by computing explicitly the fourth-order term. In what follows, we compare this improved prediction with several sources, namely the numerical results of Youngren and Acrivos \cite{youngren1975}, present numerical results obtained for $\text{Re}=0.05$, and experimental results by \citet{heiss1952}. 
 %Indeed it is well known than the spatial decay of the disturbance induced by a body at very small Reynolds numbers is very slow. In the case of a slender-body, the induced velocity perturbation decreases as $1/r$ where $r$ is the distance from the cylinder axis \citet{batchelor1970}. 
%We use $14$ cells per cylinder diameter in this regime. %The range of aspect ratio studied in this regime is $\{3;5;7;10\}$.
\begin{figure}
	\centering
		\includegraphics[scale=0.5]{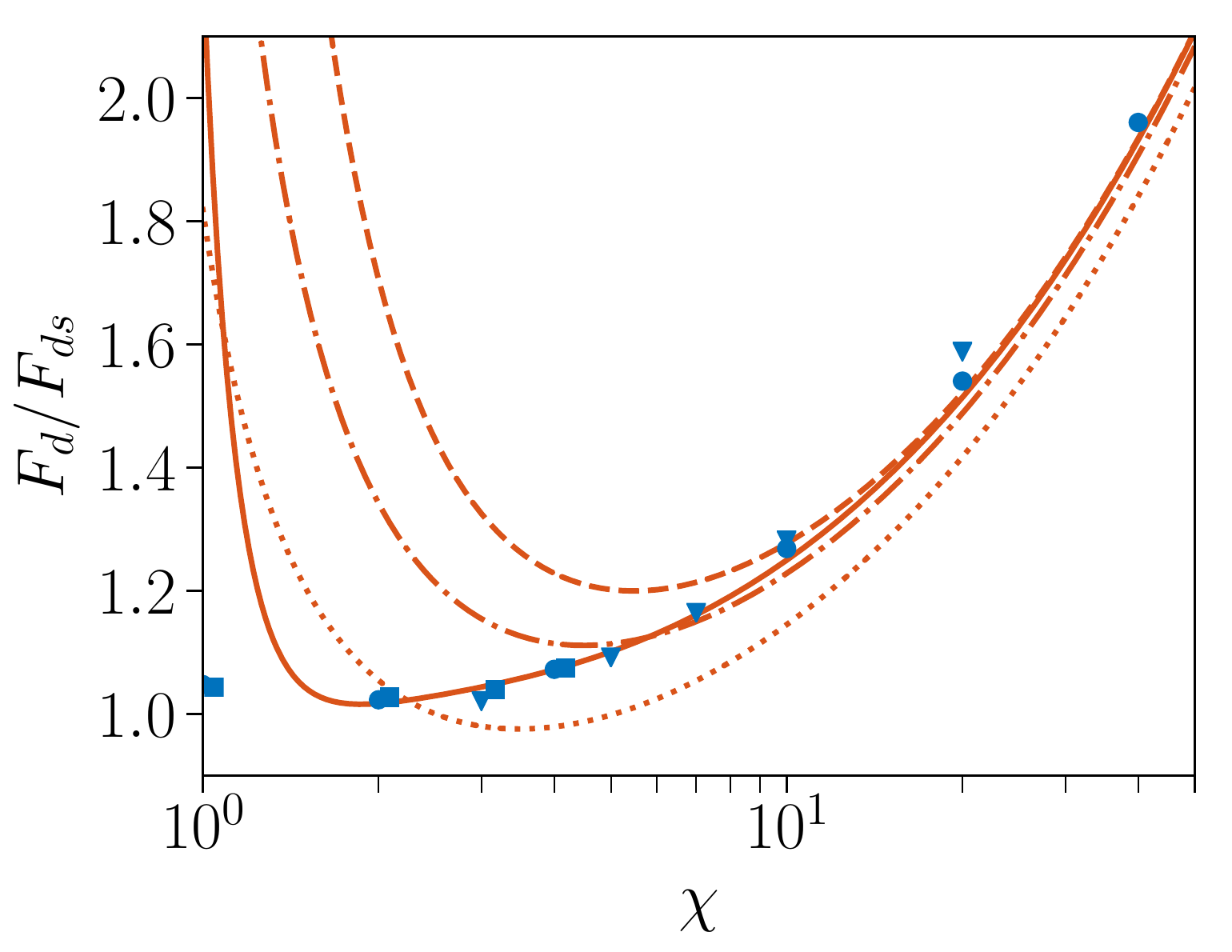}
%	\caption{ Drag force coefficient against $\chi$. - : Theoretical law \ref{eq:loi_keller} , $\star$ : numerical results of \cite{youngren1975},  $...$ :\cite{heiss1952}, $\bullet$ : Our numerical results.}
	\caption{Drag on a finite-length cylinder aligned with the flow direction, normalized by  the drag $F_{ds}$ of a sphere of same volume. Dotted, dash-dotted and dashed lines: predictions of $2^{nd}$-, $3^{rd}$- and $4^{th}$-order slender-body approximations, respectively; solid line: semiempirical formula (\ref{eq:slender_modified}); $\bullet$: numerical results of \cite{youngren1975}; $\blacksquare$: experimental results of \cite{heiss1952}; $\blacktriangledown$: present numerical results for $\text{Re}=0.05$.}
		\label{fig:forcere0}
\end{figure}
Figure \ref{fig:forcere0} displays the drag force $F_d\equiv \bf{F}\cdot\bf{e_x}$ on the body, normalized by the force $F_{ds}$ on a sphere of same volume (\textit{i.e.} with diameter $\mathfrak{D}/D=\left(\frac{3}{2}\cchi)^{1/3}\right)$), as a function of $\cchi$. Clearly, the second-order slender-body approximation (see appendix \ref{app:slender}) is quite inaccurate, even for large aspect ratios. The third-order approximation provides a better agreement, but significant deviations ($>5\%$) still exist for $\cchi=20$. %To further improve the agreement, the fourth-order term of the slender-body approximation is computed in appendix \ref{app:slender}, based on the iterative approach proposed by \citet{keller1976}. 
The fourth-order approximation computed in appendix \ref{app:slender} approaches the experimental and numerical results down to $\cchi \approx10$ significantly better. However, all slender-body approximations inherently diverge when $\cchi\rightarrow1/2$. Actually, Fig. \ref{fig:forcere0} indicates that they are all inaccurate for $\cchi\lesssim 5$; the higher the order of the expansion the larger the aspect ratio below which the slender-body approximation becomes inaccurate. To extend the domain of validity of the theoretical approximation towards short cylinders, we empirically correct the fourth-order approximation by introducing an \textit{ad hoc} additional term. %We sought the form of this term in such a way that it attenuates the divergence of the slender-body approximation for $\cchi\rightarrow1/2$ and becomes negligible for $\cchi=\mathcal{O}(10)$.
\color{black} This term must attenuate the divergence of the slender-body approximation for $\cchi\rightarrow1/2$ and become negligible for $\cchi=\mathcal{O}(10)$. Therefore we sought it in the form $\cchi^{1/3}(\cchi-1/2)^{-p}$, in such a way that only a $(\cchi-1/2)^{-p}$ correction is introduced in the normalized force $F_d/F_{ds}$. The best fit with experimental and numerical data in the range $2\leq\cchi\leq10$ is obtained with $p=1.75$, a $\pm3\%$-difference on $p$ leading to a significantly poorer agreement. \color{black}

The full expression for the drag force incorporating this empirical term then reads
%\begin{equation}
%F_{d} = 2 \pi \mu L U_x \left( \frac{a^{(1)}}{\ln(2\cchi)} + \frac{a^{(2)}}{(\ln(2\cchi))^2} + \frac{a^{(3)}}{(\ln(2\cchi))^3} + \frac{a^{(4)}}{(\ln(2\cchi))^4} \right) - \mu (LD^2)^{1/3} U_x \frac{14.56 }{(\cchi -1/2)^{1.8}},
%\label{eq:slender_modified}
%\end{equation}
\begin{equation}
F_{d}^{Re=0} \approx 2 \pi \mu L U \left( \frac{a^{(1)}}{\ln(2\cchi)} + \frac{a^{(2)}}{(\ln(2\cchi))^2} + \frac{a^{(3)}}{(\ln(2\cchi))^3} + \frac{a^{(4)}}{(\ln(2\cchi))^4} - \frac{2.4}{\cchi ^{2/3}(\cchi -\frac{1}{2})^{1.75}} \right)\,,
\label{eq:slender_modified}
\end{equation}
%all the theoretical laws based on the slender-body theory does not match well the numerical results for $\cchi\leq 7$.
%due to the divergence of the slender-body based laws in $\chi=1/2$
 %where the relative mean square error $\epsilon\geq15\%$ ($\epsilon=\frac{numerical\_value - theoretical\_value}{numerical\_value}$). For $\raisebox{2pt}{$\chi$}=2$, the law \ref{eq:loi_keller} tends to diverge, for this reason it needs to be modified to be valid for small aspect ratio $\raisebox{2pt}{$\chi$}\leq10$ without influencing its accuracy at high aspect ratio.
%Deux choses a dire deja que l'accord nest pas forcemente excellent pour l'orrdre 3 et que lle truc diverge quand chi=1/2.
% The aim of this section is to measure the validity of this law at low Reynolds number for small aspect ratios. Here we choose to normalize \ref{eq:loi_keller} by the quantity $2\pi\mu DU_x$, $F_d= f_d/(2\pi\mu DU_x)$.\\
\noindent with $U=\bf{U}\cdot\bf{e_x}$ and the expressions for the $a ^{(i)}$ as provided in appendix \ref{app:slender}. As Fig. \ref{fig:forcere0} shows, the modified drag law fits the numerical and experimental results well for $\cchi \gtrsim2$. It still diverges for lower $\cchi$, but the drag is then close to that experienced by the equivalent sphere ($F_d/F_{ds}\leq1.05$). %{\color{red}{Moreover, the cylindrical pellets encountered in bubbling fluidized bed, the current application of interest, have most of the time a larger aspect ratio.}} 
Present numerical predictions in the range $3\leq\cchi\leq10$ are also seen to agree well with results available in the literature. %This good agreement strengthen our confidence in the numerical methodology. 
The slight deviation observed for $\cchi=20$ is not unexpected. Indeed, the Reynolds number based on the body length is of $\mathcal{O}(1)$ in this case, making inertial corrections to the drag significant (see below). %The appearance of inertia effects will be discussed further below.
%but this appear unimportant since the cylindrical pellets encountered in bubbling fluidized bed are larger than this aspect ratio. Moreover a correlation accurate for this aspect ratio can be found in \citet{clift1978}. 
%Note also than a simple minimal bound can be obtained for the drag force on a cylinder of aspect ratio 1 using the minimum dissipation theorem. 
% The relative error decreases drastically and varies from $1\%$ to $5\%$. We notice, that \ref{eq:loi_keller2} is only valid from $\raisebox{2pt}{$\chi$}\geq2$,
\begin{figure}[h]
\centering
\includegraphics[scale=0.4]{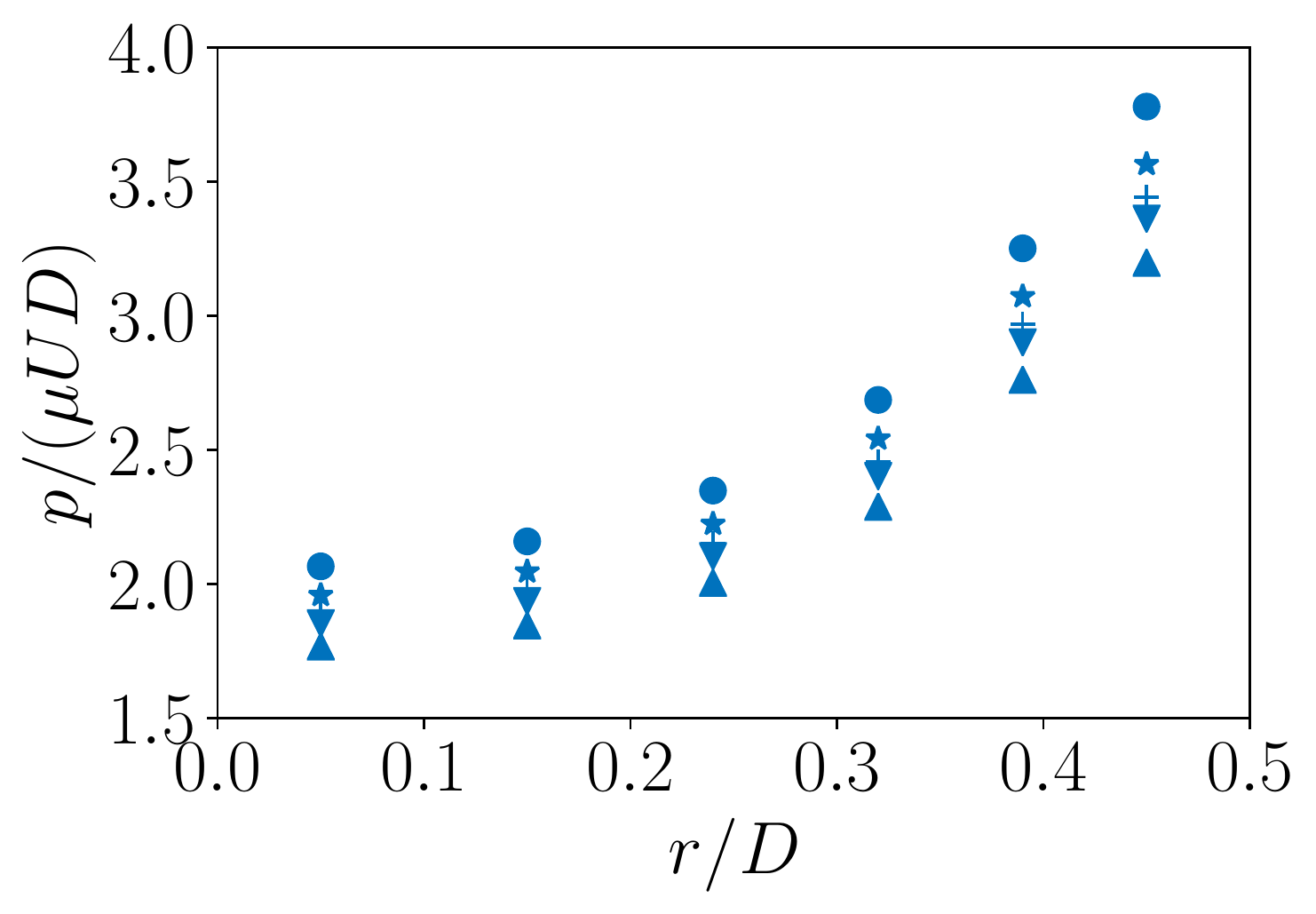}
\caption{Radial pressure distribution on the upstream end of the cylinder for $\text{Re}=0.05$. $\bullet$:  $\cchi=3$; $\star$:  $\cchi=5$; $+$:  $\cchi=7$; $\blacktriangledown$: $\cchi=10$; $\blacktriangle$: $\cchi=20$. The pressure is assumed to be zero in the far field.}
%\caption{Pressure as function of the distance along the perimeter of the cylinder for different values of $\chi$. The $s$ origin is chosen upstream of the cylinder, on the cylinder symmetry line. (left) pressure along the upstream disk, (middle) pressure along the lateral surface of the cylinder, (right) pressure along the downstream disk.}
\label{fig:pressure_re0}
\end{figure}
% Those figures evidenced the upstream-downstream symmetry of the pressure characteristic of the Stokes flow regime.
% We only note a slight decrease of the pressure peak on the cylinder edges as function of $\cchi$.
%One of the most noticeable features of the Stokes flow regime is the upstream - downstream symmetry of the flow. 
The pressure component to the drag, $F_{dp}$, results from the difference between the pressure distributions on the upstream and downstream ends of the cylinder. Owing to the fore-aft symmetry of the flow in the Stokes regime, these two distributions only differ by the sign of the corresponding pressure, since $p(r, x=L/2)=-p(r,x=-L/2)$. Figure \ref{fig:pressure_re0} displays the radial pressure distribution on the upstream end. For each $\cchi$, the pressure is seen to increase with the distance $r$ to the symmetry axis. At a given radial position, pressure variations are only mildly influenced by the aspect ratio, but this influence becomes larger as the lateral surface is approached.

\begin{figure}[h]
	\centering
	\includegraphics[scale=0.4]{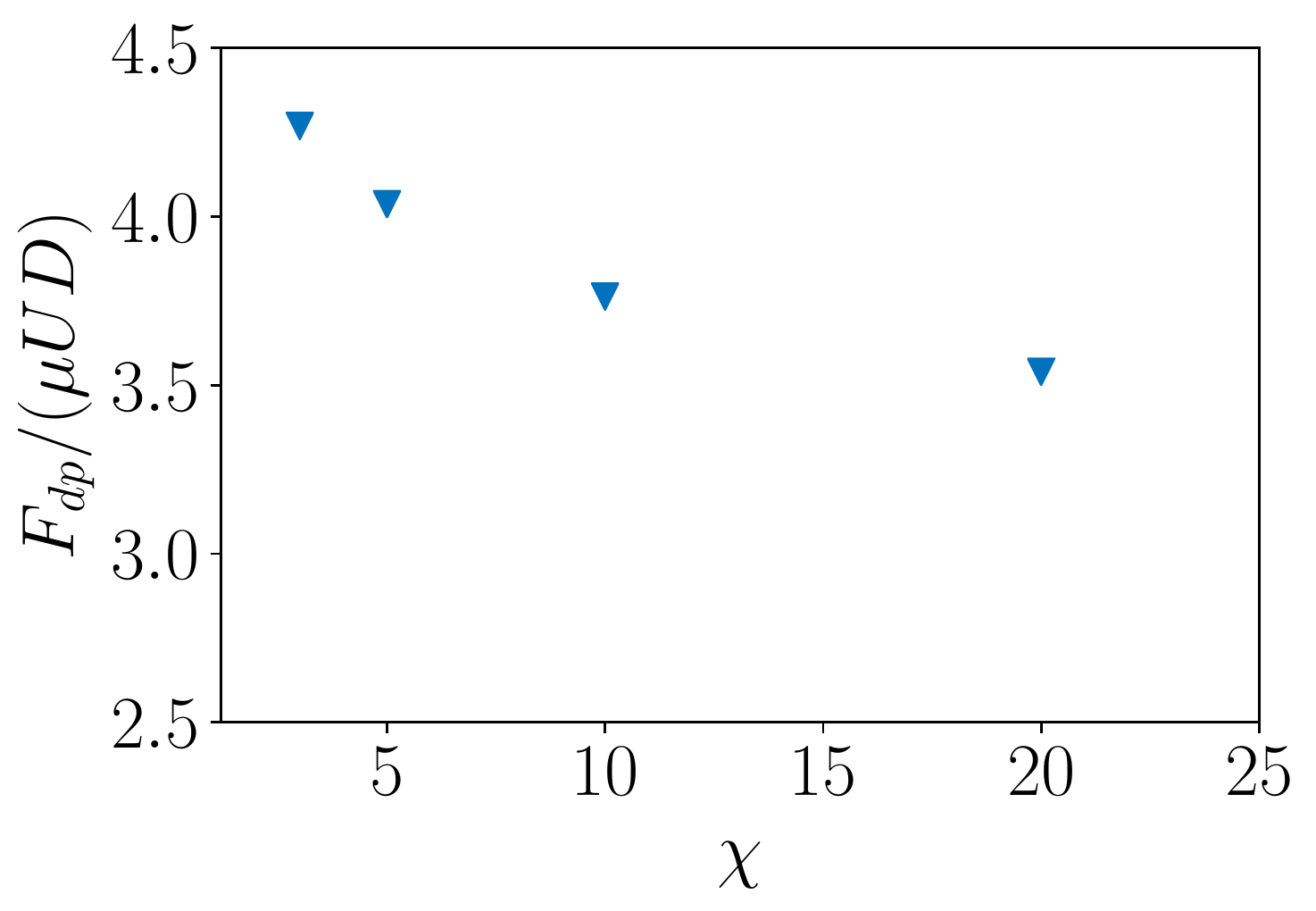}
	\includegraphics[scale=0.4]{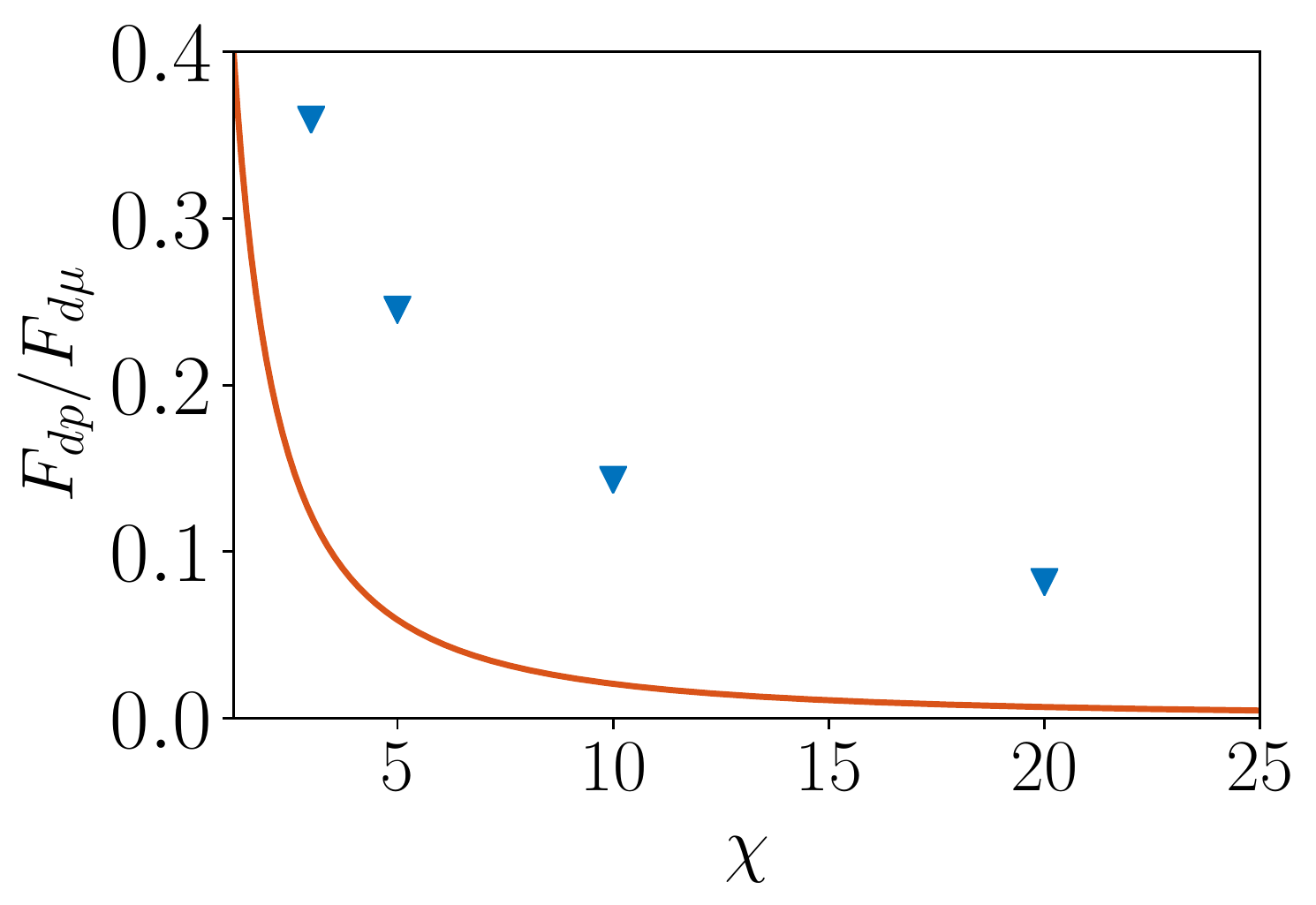}\\
	\vspace{-1mm}
	 \hspace{-1.5cm}{$(a)$}\hspace{6cm}{$(b)$}
	\caption{Contributions to the drag force. $(a)$: pressure contribution; $(b)$: ratio of the pressure to the shear stress contributions. $\blacktriangledown$: numerical results for $\text{Re}=0.05$; solid line: low-$\text{Re}$ prediction (\ref{eq:prolate_spheroid_Re0}) for a prolate spheroid.}
\label{fig:forces_mu_p_re0}	
\end{figure}
%The pressure distribution on the upstream disk has a noticeable effect on the pressure force $F_{dp}$ which is the part of the drag force due to pressure. Indeed 
Figure \ref{fig:forces_mu_p_re0}$(a)$ indicates that $F_{dp}$ only varies weakly with $\cchi$. It consistently decreases with the aspect ratio but its variation is less than 20\,\% from $\cchi=3$ to $\cchi=20$. Figure \ref{fig:forces_mu_p_re0}$(b)$ shows the ratio of the pressure contribution to the viscous stress contribution to the drag as a function of $\cchi$. This ratio is close to $1/3$ for $\cchi=3$. Then it decreases gradually as $\cchi$ increases, and becomes less than $1/10$ for $\cchi=20$. Hence, as expected from purely geometrical considerations, $F_{dp}$ becomes negligible with respect to $F_{d\mu}$ for large $\chi$. It is of some interest to compare this variation with that found for prolate spheroids, namely \citep{masliyah1970}
\begin{equation}
\displaystyle
\frac{F_{dp}}{F_{d\mu}} = \frac{\cchi \ln (\cchi + (\cchi ^2 -1)^{1/2})-(\cchi ^2 -1)^{1/2}}{\cchi ^2(\cchi ^2 -1)^{1/2}-\cchi \ln(\cchi + (\cchi ^2 -1)^{1/2})}\,.
\label{eq:prolate_spheroid_Re0}
\end{equation} 

\begin{figure}[h]
	\centering
		{\includegraphics[scale=0.35]{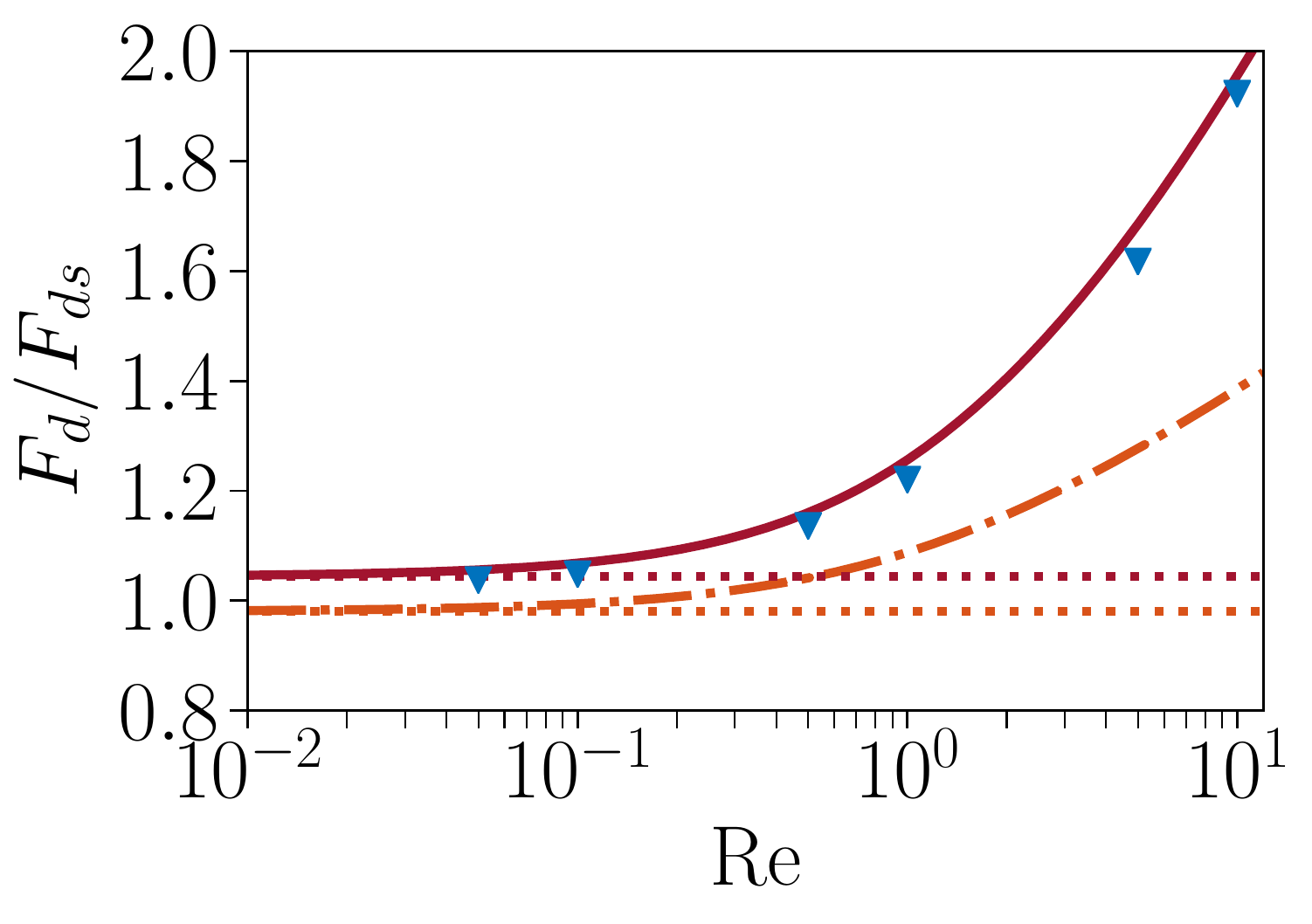}\label{fig:a}}
		{\includegraphics[scale=0.35]{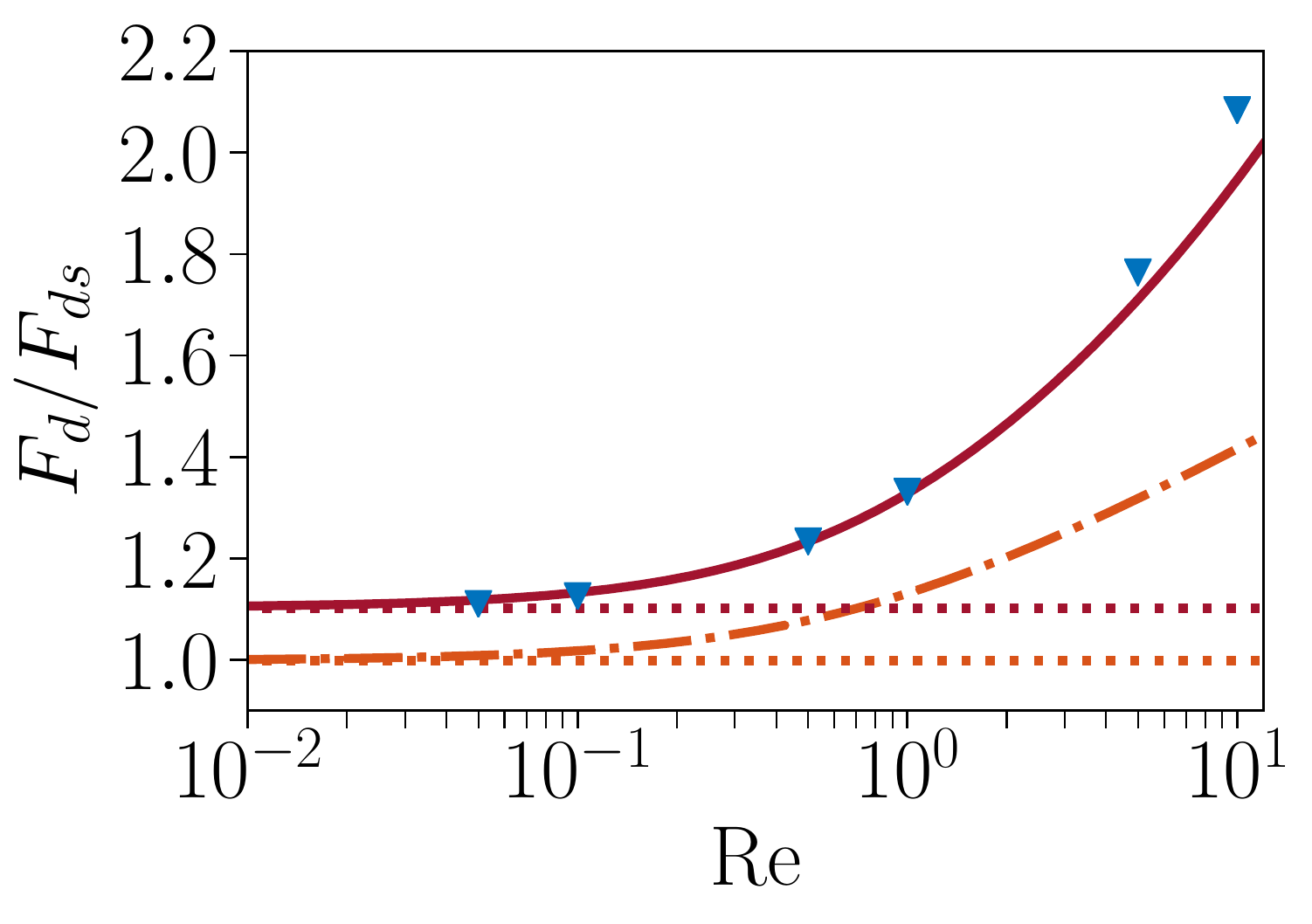}\label{fig:b}}\\
			\vspace{-4mm}
		 \hspace{-4cm}$(a)$\hspace{5cm}$(b)$\\
		 \vspace{2mm}
		{\includegraphics[scale=0.35]{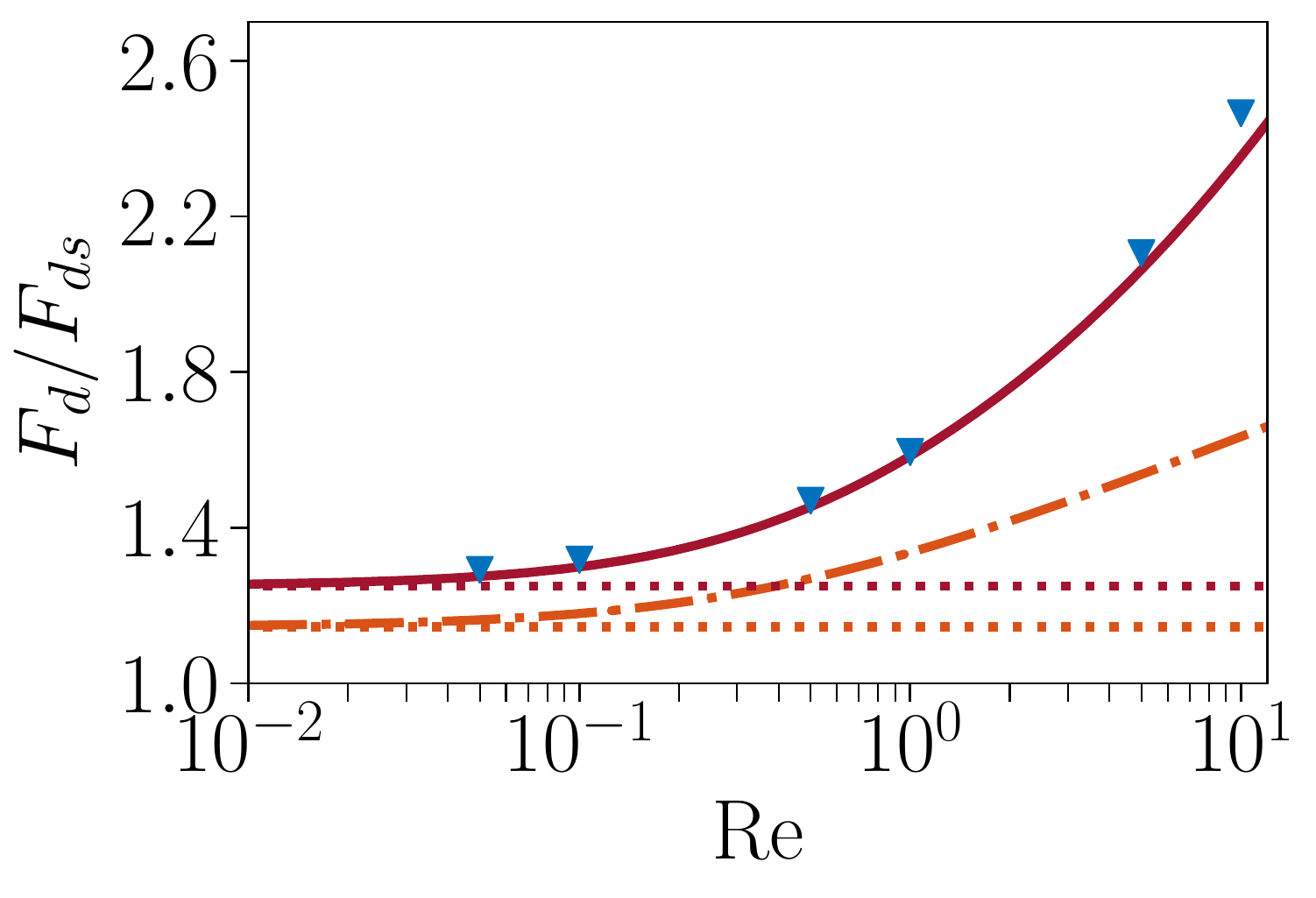}\label{fig:c}}
		{\includegraphics[scale=0.35]{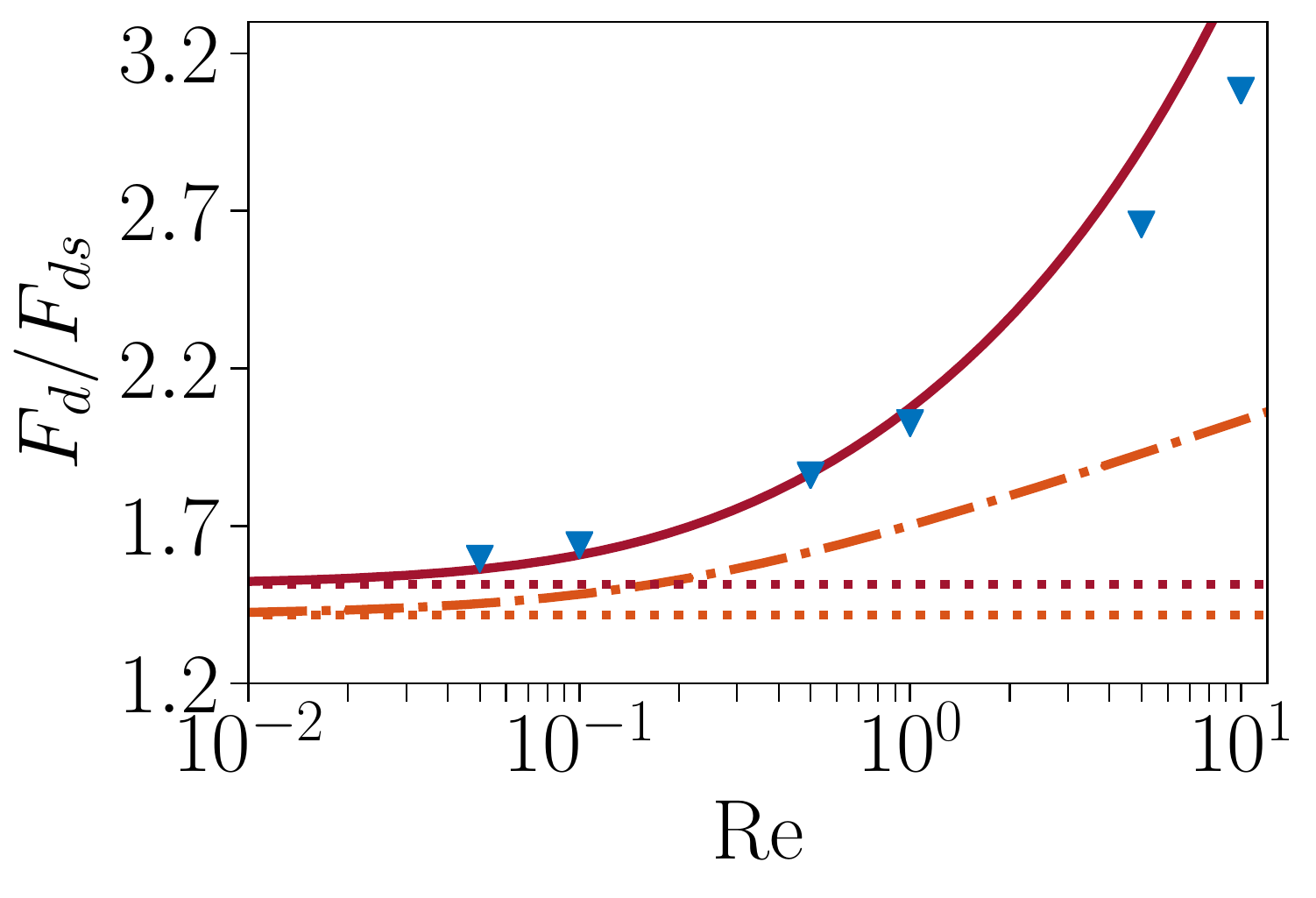}\label{fig:d}}\\
		\vspace{-4mm}
		 \hspace{-4cm}$(c)$\hspace{5cm}$(d)$\\
		 \vspace{-72mm}
		  \hspace{-14mm}$\cchi=3$\hspace{5cm}$\cchi=5$\\
		   \vspace{37mm}
		   \hspace{-12mm}$\cchi=10$\hspace{4.8cm}$\cchi=20$
		  \vspace{33mm}
%	\caption{ Drag force coefficient against $\chi$. - : Theoretical law \ref{eq:loi_keller} , $\star$ : numerical results of \cite{youngren1975},  $...$ :\cite{heiss1952}, $\bullet$ : Our numerical results.}
	\caption{Influence of inertial effects on the drag of finite-length cylinders aligned with the upstream flow. The drag is normalized by that of a sphere of same volume. % (a): $\cchi=3$, (b): $\cchi=7$, (c): $\cchi=10$, (d): $\cchi=20$. 
	$\blacktriangledown$: numerical results; solid line: semiempirical prediction (\ref{eq:slender_smallinertia2}); dash-dotted line: theoretical prediction (\ref{eq:slender_smallinertia}); dotted line: predictions (\ref{eq:slender_modified}) (upper line) and (\ref{eq:slender_smallinertia}) with $\text{Re}=0$ (lower line).}
		\label{fig:small_inertia}
\end{figure}

%A poor agreement is found between the numerical results and equation \ref{eq:prolate_spheroid_Re0}. 
\noindent As Fig. \ref{fig:forces_mu_p_re0}$(b)$ indicates, the pressure drag decreases much more slowly with the aspect ratio in the case of a cylindrical body with flat ends.  This difference underlines the importance of end effects even for long cylindrical bodies.\vspace{2mm}\\
\indent Next we consider the influence of inertial effects at low-to-moderate Reynolds number. %Figure \ref{fig:vort_theta0} illustrates these effects through the vorticity distribution around a  cylinder with $\cchi=5$. While the vorticity contours exhibit the expected fore-aft symmetry characteristic of the creeping-flow regime at $\text{Re}=0.05$, advection in the downstream direction makes these contours significantly asymmetric at $\text{Re}=10$. 
%%the flow is clearly asymmetric. Indeed as $\text{Re}$ increases, the vorticity advection becomes larger than the vorticity diffusion. For each Reynolds number, we observe a stronger magnitude of the vorticity near the edges of the body. It is one of the reason for which those regions are preferentially refined.
%as a consequence of the larger magnitude vorticity advection%of the wake The vorticity contours becomes more and more asymmetric as the Reynolds number increases. Indeed as $\text{Re}$ increases the vorticity advection becomes larger than with respect to vorcity fdiffusion.
The slender-body theory initially derived under Stokes flow conditions was extended to small-but-finite Reynolds numbers in \cite{khayat1989}, still assuming $\text{Re} \ll 1$ but considering that the Reynolds number based on the body length, ${\cchi\text{Re}}$, may be arbitrarily large. \color{black} \citet{lopez2017} and \citet{roy2019} provided experimental confirmations of the relevance of the finite-$\text{Re}$ corrections derived in \cite{khayat1989} by examining the settling of long fibers ($10\lesssim\cchi \lesssim35$) in a Taylor-Green type vortical flow and the sedimentation of isolated fibers with $\cchi=20$ and $100$ in a fluid at rest, respectively. \color{black} Nevertheless, we are not aware of any detailed validation of the finite-$\text{Re}$ theory of \cite{khayat1989} in the case of a body aligned with the upstream flow. According to this theory (see also \cite{lopez2017} and \cite{roy2019}), the drag force on a long cylindrical body aligned with the flow, disregarding terms of $\mathcal{O}((1/\ln\cchi)^3)$ and higher, reads
% \begin{equation}
%F_{d,Re \ll 1} = 2 \pi \mu L U_x \frac{1}{\ln(2\cchi) - F_\parallel},
%\label{eq:slender_smallinertia}
%\end{equation} 
 \begin{equation}
F_d^{\cchi\text{Re}=\mathcal{O}(1)}\approx  2 \pi \mu L U \left( \frac{a^{(1)}}{\ln\cchi} + \frac{a^{(2)}-\ln2+f_\parallel}{(\ln\cchi)^2} \right)\,,
\label{eq:slender_smallinertia}
\end{equation} 
 where  
% \begin{equation}
% f_\parallel = \frac{1}{2}\left[\frac{E_1(\cchi \text{Re} )+\ln(\cchi \text{Re} )-e^{-\cchi \text{Re} }+\gamma +1}{\cchi \text{Re} } + E_1(\cchi \text{Re} ) +\ln(\cchi/2 Re) +\gamma - \ln 2 +1 \right], \nonumber
% \end{equation}
  \begin{equation}
 f_\parallel =\frac{1}{2}\left( \frac{E_1(\cchi \text{Re} )+\ln(\cchi \text{Re} )-e^{-\cchi \text{Re} }+\gamma +1}{\cchi \text{Re} } + E_1(\cchi \text{Re} ) +\ln(\cchi \text{Re}) +\gamma  -2\right) \,,
 \label{fpKC}
 \end{equation}
\noindent with $E_1(X)=\int_X^{+\infty}{\frac{e^{-t}}{t}dt}$ (related to the exponential integral $\text{Ei}(X)$ through $E_1(X)=-\text{Ei}(-X)$), and $\gamma$ the Euler constant. In the limit of small $\cchi\text{Re}$, the inertial correction factor reduces to $f_\parallel\approx\frac{1}{8}\cchi\text{Re}$. Note that in (\ref{eq:slender_modified}) and throughout the rest of the paper, the slender-body solution is expanded with respect to $1/\ln(2\cchi)$, similar to \cite{batchelor1970}, \cite{tillett1970} and \cite{,keller1976}. In contrast, an expansion with respect to $1/\ln(\cchi)$ was used in \cite{khayat1989}. Once truncated at the same order, the two formulations are of course equivalent for $\cchi\gg2$, but significant differences exist for moderate aspect ratios. This is why we keep the original formulation in (\ref{eq:slender_smallinertia}) and in similar expressions of Sec. \ref{incl1} involving the inertial corrections derived in \cite{khayat1989}. \\%We observe that inertia effects do not change the leading order term but appear in the expansion at order 2. 
\indent The main shortcoming of (\ref{eq:slender_smallinertia}) is obviously the truncation at second order with respect to $1/\ln\cchi$. This limitation is confirmed in Fig. \ref{fig:small_inertia}, where the influence of inertial effects on the drag force is shown for cylindrical bodies with different aspect ratios. Significant deviations between numerical results and predictions of (\ref{eq:slender_smallinertia}) are observed whatever $\text{Re}$ and $\cchi$. %We also observe, as expected, that the drag force is an increasing function of the Reynolds number.
%\citet{lopez2017} studied experimentally the settling of long fibres ($\cchi \geq 10$) in vortical flows. They found a good agreement between their experimental results and the analytical prediction of \citet{khayat1989}. It seems however than the prediction of  \citet{khayat1989} has not been validated yet on the flow past an axisymmetric cylinder. Hence despite the limited applicability of \citet{khayat1989} results on the present range of aspect ratio of interest, it is nonetheless interesting to check the validity of their prediction by comparison to the present numerical results. %The drag force derived by \citet{khayat1989} reads :
It is thus desirable to include higher-order corrections with respect to $1/\ln(2\cchi)$ to improve the validity of (\ref{eq:slender_smallinertia}). The analysis of \cite{khayat1989} was recently extended to third order in \cite{khair2018} using the reciprocal theorem. However, the pre-factor of the third-order term is found in the form of a volume integral to be evaluated in Fourier space. To obtain a more straightforward formula accounting for small-but-finite inertial effects, we use present numerical results to empirically modify (\ref{eq:slender_smallinertia}) by taking advantage of the higher-order corrections present in (\ref{eq:slender_modified}). %For this, we just duplicate the inertial correction derived at order $2$ by \citet{khayat1989} in the higher-order terms of (\ref{eq:slender_modified}). %This modification ensures that inertia terms are effectively negligible at $\text{Re}=0$ while increasing with increasing Reynolds number. However, it has to be noted that 
%Obviously, this modification is only an approximation and the coefficient of the third-order term differs from that provided by the asymptotic expansion of \citet{khair2018}. 
The best agreement with the numerical results is obtained with the expression 
\begin{equation}
F_d^{\cchi\text{Re}=\mathcal{O}(1)} \approx 2 \pi \mu L U \left[ \frac{a^{(1)}}{\ln(2\cchi)} + \frac{a^{(2)}+ f_\parallel}{(\ln(2\cchi))^2} + \frac{a^{(3)}+ f_3f_\parallel}{(\ln(2\cchi))^3} + \frac{a^{(4)}+ f_4f_\parallel }{(\ln(2\cchi))^4} - \frac{2.4}{\cchi ^{2/3}(\cchi -\frac{1}{2})^{1.75}} \right]\,,
\label{eq:slender_smallinertia2}
\end{equation}
%Figure \ref{fig:small_inertia} displays the drag force on a cylinder of aspect ratio $10$ as function of the Reynolds number.  
%between the present numerical results and equation \ref{eq:slender_smallinertia2}
\noindent with $f_3=(\cchi\text{Re})^{0.07\chi^{0.5}}$ and $f_4=(\cchi\text{Re})^{0.03\chi^{0.9}}$. Predictions of (\ref{eq:slender_smallinertia2}) are compared with numerical results in Fig. \ref{fig:small_inertia}. The agreement is found to be good up to $\text{Re}\approx 1$ whatever the aspect ratio. This is quite remarkable since the derivation of (\ref{eq:slender_smallinertia2}) assumes $\text{Re}\ll1$ (but possibly $\cchi\text{Re}\gtrsim1$ since $\cchi\gg1$). \citet{khair2018} also found that their analytical prediction agrees well with the numerically computed drag on a long prolate spheroid up to $\text{Re} \approx 2$ for $\cchi=10$. %The extended slender-body theory for small but finite Reynolds number with the ad hoc modification proposes here is thus an accurate tool to compute the drag force on long body up to moderate Reynolds numbers. 
%\begin{equation}
%F_{d,Re \ll 1} = 2 \pi \mu L U_x \left(\ln(2\cchi) - \frac{1}{2}\left[\frac{E_1(\cchi \text{Re} )+\ln(\cchi \text{Re} )-e^{-\cchi \text{Re} }+\gamma +1}{\cchi \text{Re} } + E_1(\cchi \text{Re} ) +\ln(\cchi/2 Re) +\gamma - \ln(2) +1 \right] \right)^{-1}
%\label{eq:slender_smallinertia}
%\end{equation} 
  %Indeed \citet{khair2018} reported a remarkable agreement between the theory and numerical results 
% Hence despite their limited range of applicability for moderate aspect ratio of this study it is interesting to compare their prediction to the present numerical results, thing which has not been done yet in the literature.
% and \citet{khair2018} have extended the slender-body theory derived in the Stokes flow regime, to small but finite Reynolds number.  Also of limited applicability in practice due to the complexity of the analytical formula given by \citet{khayat1989} it is interesting to compare the analytical resutls of Khayat et Cox et ... have extended the results of slender-body theory for Stokes flow to the small inertia limit. 
%They developed a theory for moderate Reynolds number
%\subsubsection{Forces at high Reynolds numbers $\text{Re} \geq 1$}
\begin{figure}
	\centering
		\includegraphics[scale=0.213]{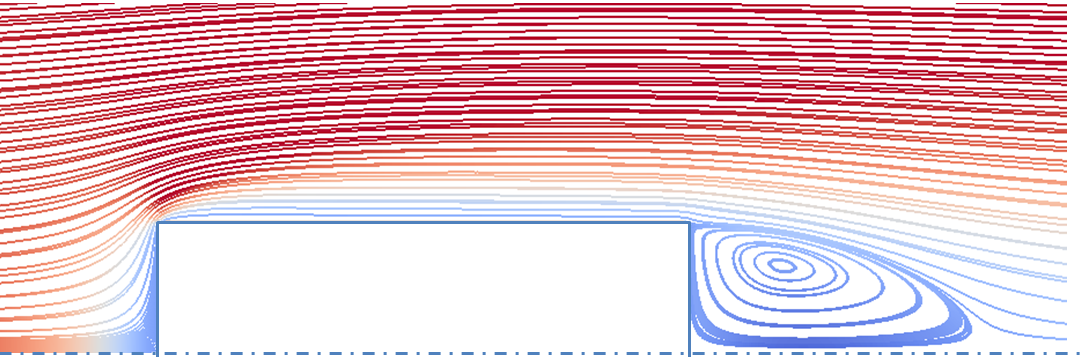}\hspace{15mm}
		\includegraphics[scale=0.275]{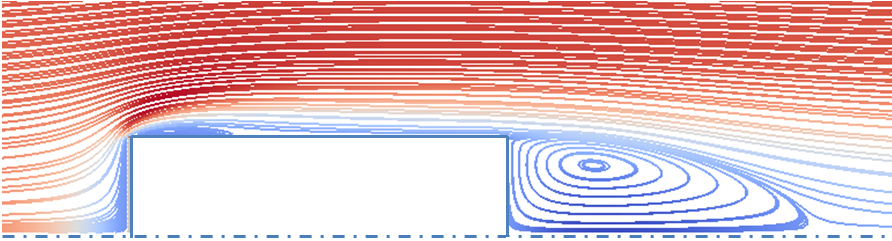}\\
		\vspace{1mm}\hspace{-55mm}$(a)$\hspace{75mm}$(b)$
	\caption{Streamlines pattern for $\cchi=2$ colored with the magnitude of the axial velocity (from $-0.25$ to $1$). $(a)$ $\text{Re}=140$; $(b)$ $\text{Re}=300$. %(\textcolor{red}{red $U_x=1$}, \textcolor{black}{blue $U_x=0$}) . 
		\label{fig:streamlines}}
\end{figure}
\subsection{From $\mathcal{O}(10)$ to $\mathcal{O}(4\times10^2)$ Reynolds numbers}
\label{moder}
%\begin{figure}
%\includegraphics[scale=0.2]{contour_w_x5R10}
%\end{figure}
%One of the most noticeable feature of high Reynolds number flows is the appearance of a recirculation eddy in the cylinder wake. 
\color{black}Increasing the Reynolds number, we considered six aspect ratios ranging from $2$ to $10$ and twenty Reynolds numbers from $\text{Re}=20$ to $\text{Re}=400$. \color{black} The flow past the cylinder was found to reach a steady state in all cases. Nevertheless, the flow structure revealed new features as the Reynolds number increases.\\
Figure \ref{fig:streamlines}$(a)$ displays the streamlines around a cylinder with $\cchi=2$ for $\text{Re}=140$. In this regime, the flow is attached to the body all along the lateral surface but the separation of the boundary layer at the downstream edge results in the generation of a toroidal eddy in the near wake. Simulations indicate that this standing eddy sets in for $\text{Re} \approx 10$ for $\cchi=2$ and $\text{Re} \approx 20$ for $\cchi=7$. %and $\cchi=10$.
\begin{figure}[h]
	\centering
		\includegraphics[scale=0.4]{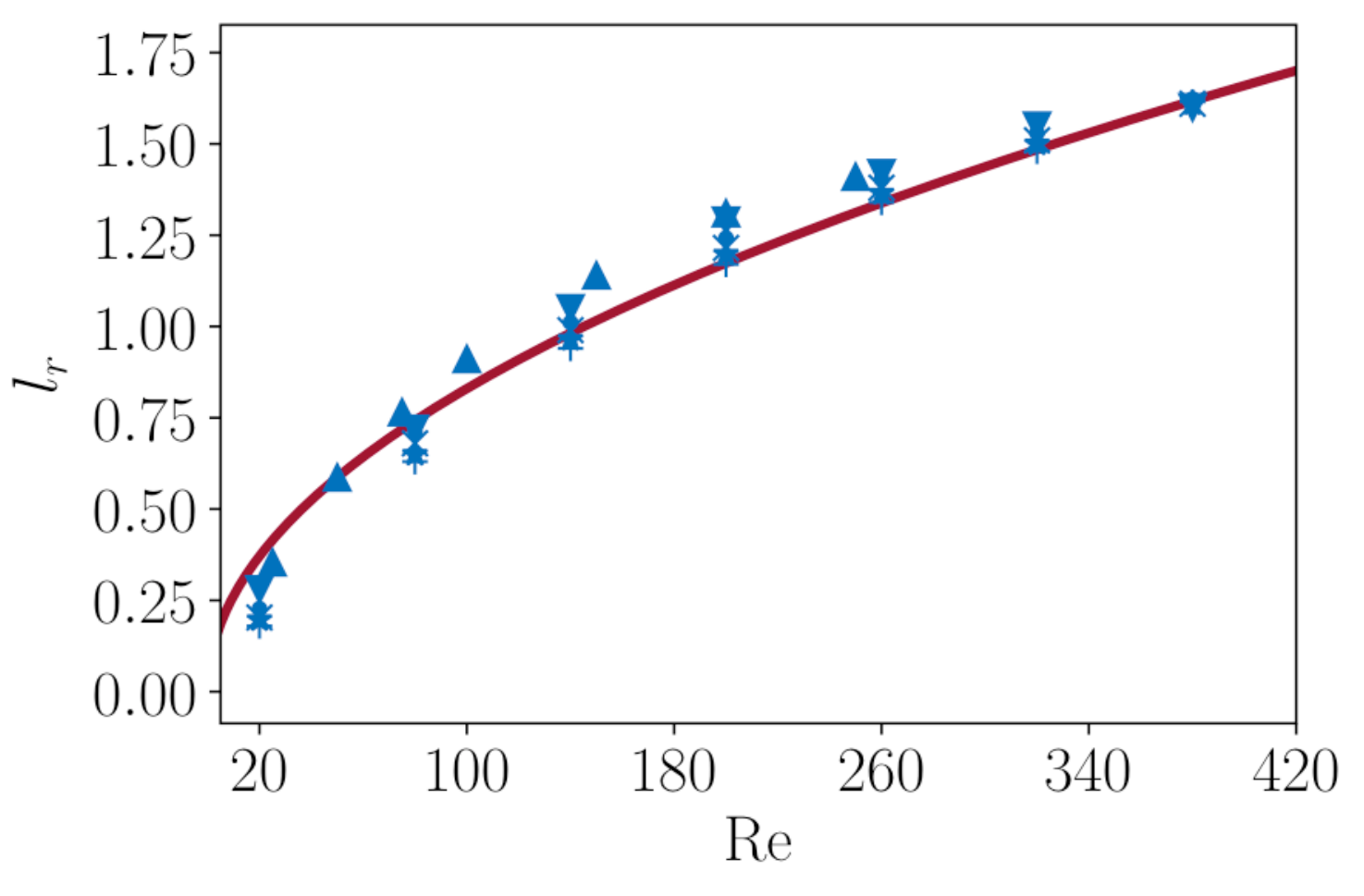}
	\caption{Variation of the length $l_r$ of the toroidal eddy vs. the Reynolds number past cylinders with different aspect ratios; $l_r$ is measured along the symmetry axis, from the downstream end of the cylinder to the point at which the axial velocity changes sign. $\blacktriangle$: $\cchi=1$ from \cite{auguste2010}; $\blacktriangledown$:  $\cchi=2$; $\bullet$:  $\cchi=3$; $\times$:  $\cchi=4$; $\star$:  $\cchi=5$; $+$:  $\cchi=7$; solid line: $l_r(\text{Re})=0.083\text{Re}^{1/2}$.
		\label{fig:toroidal}}
		\vspace{-3mm}
\end{figure}
\noindent The length of the standing eddy is plotted in Fig. \ref{fig:toroidal} as a function of the Reynolds number. Remarkably, it is found to be almost independent of $\cchi$ and grows approximately as the square root of the Reynolds number. %$l_r$ depends strongly on the shape of the obstacle since follows logarithmic law for a sphere as predicted by \cite{tomboulides2000}. 
%The length is quasi-identical for all aspect ratios because the boundary layer thickness is the same for all $\raisebox{2pt}{$\chi$}$ sand it peels away at the same point in the cylinder boundary for all $\raisebox{2pt}{$\chi$}$. We can observe a small distinction for $\raisebox{2pt}{$\chi$=2}$ where the vorticity generated at its surface is more intense than the others $\raisebox{2pt}{$\chi$}$, that is why the length of the recirculation which is linked to that vorticity is a little bit higher compared to high aspect ratio.
%\color{black}

%\color{red}
%\huge
%ICI il faudra discuter des sillages non axisymetrique (stationnaire / instationnaire, etc)...
%\normalsize
%\color{black}
As the Reynolds number increases, a thin secondary annular eddy sets in along the upstream part of the lateral surface of the body (Fig. \ref{fig:streamlines}$(b)$), owing to the detachment of the boundary layer along the upstream edge. The critical Reynolds number $\text{Re}_{c0}$ beyond which this flow pattern is detected is a slowly increasing function of the aspect ratio: from $\text{Re}_{c0}\approx180$ for $\cchi=2$, to $\text{Re}_{c0}\approx200$ for $\cchi=5$, $\text{Re}_{c0}\approx220$ for $\cchi=7$, until $\text{Re}_{c0}\approx240$ for  $\cchi=10$. It will be seen later that this secondary eddy has a strong influence on the viscous friction experienced by the cylinder

%\begin{figure}[h]
%\centering
%\includegraphics[scale=0.3]{ldc_t0x2R300.png}
%\caption{Streamlines pattern for  $\text{Re} = 300$ and $\cchi=2$, colored with the magnitude of the axial velocity (from $-0.25$ to $1$).}
%\label{secondeddy}
%\end{figure}
\begin{figure}[h]
		\centering
	    \hspace{-10mm}\includegraphics[height=4cm]{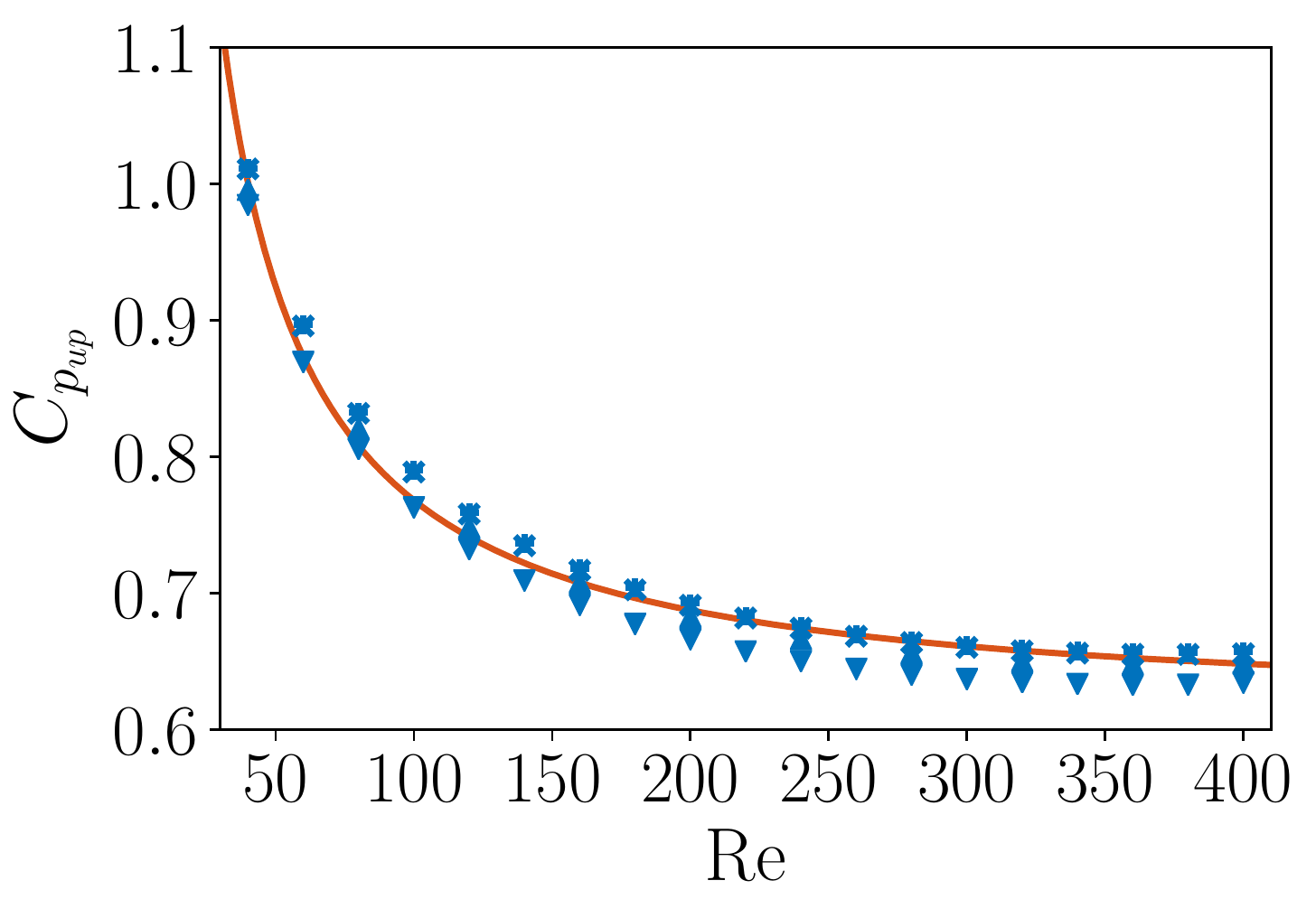} %\hspace{0.1cm}
		\includegraphics[height=4cm]{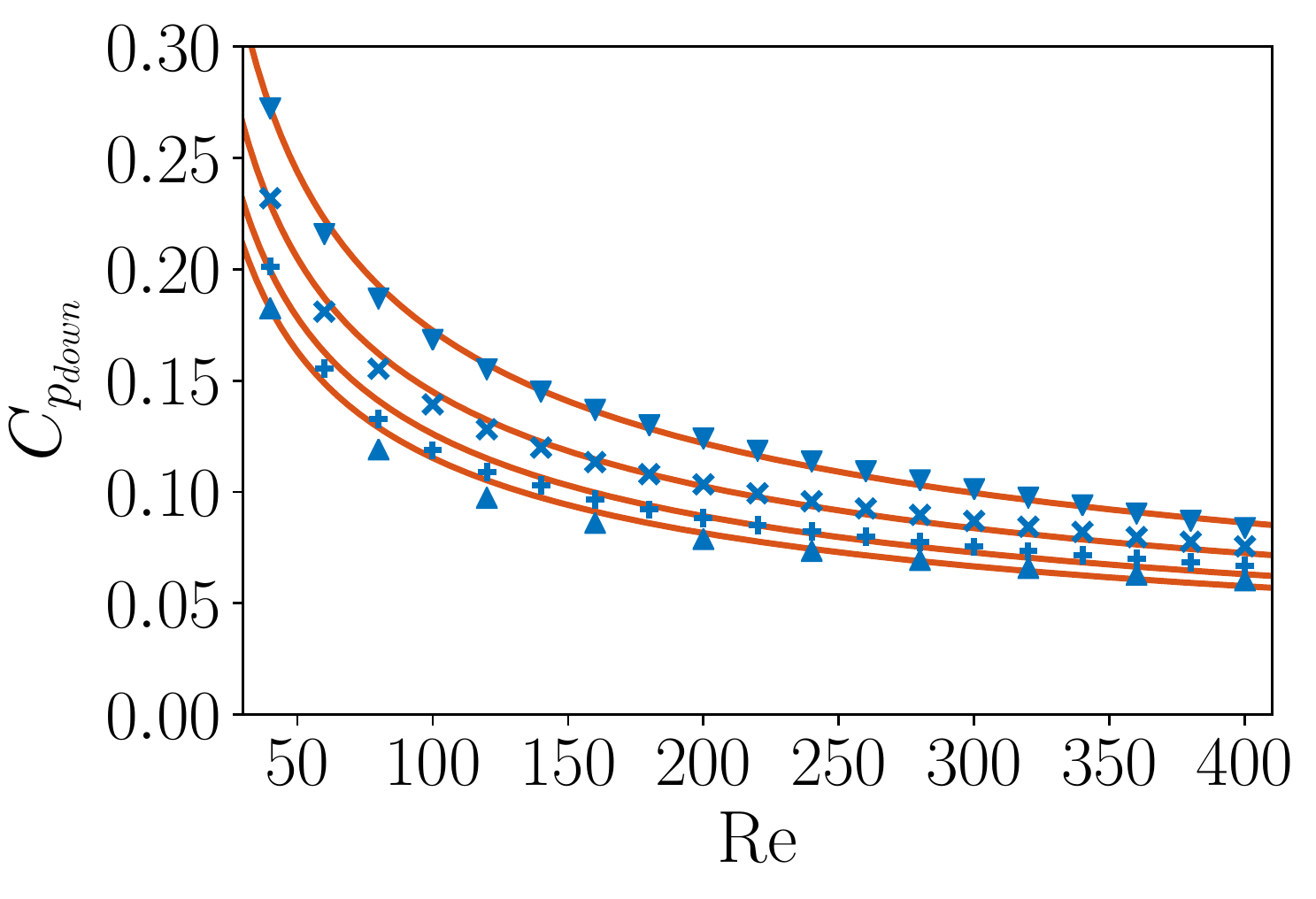}
		\includegraphics[height=4cm]{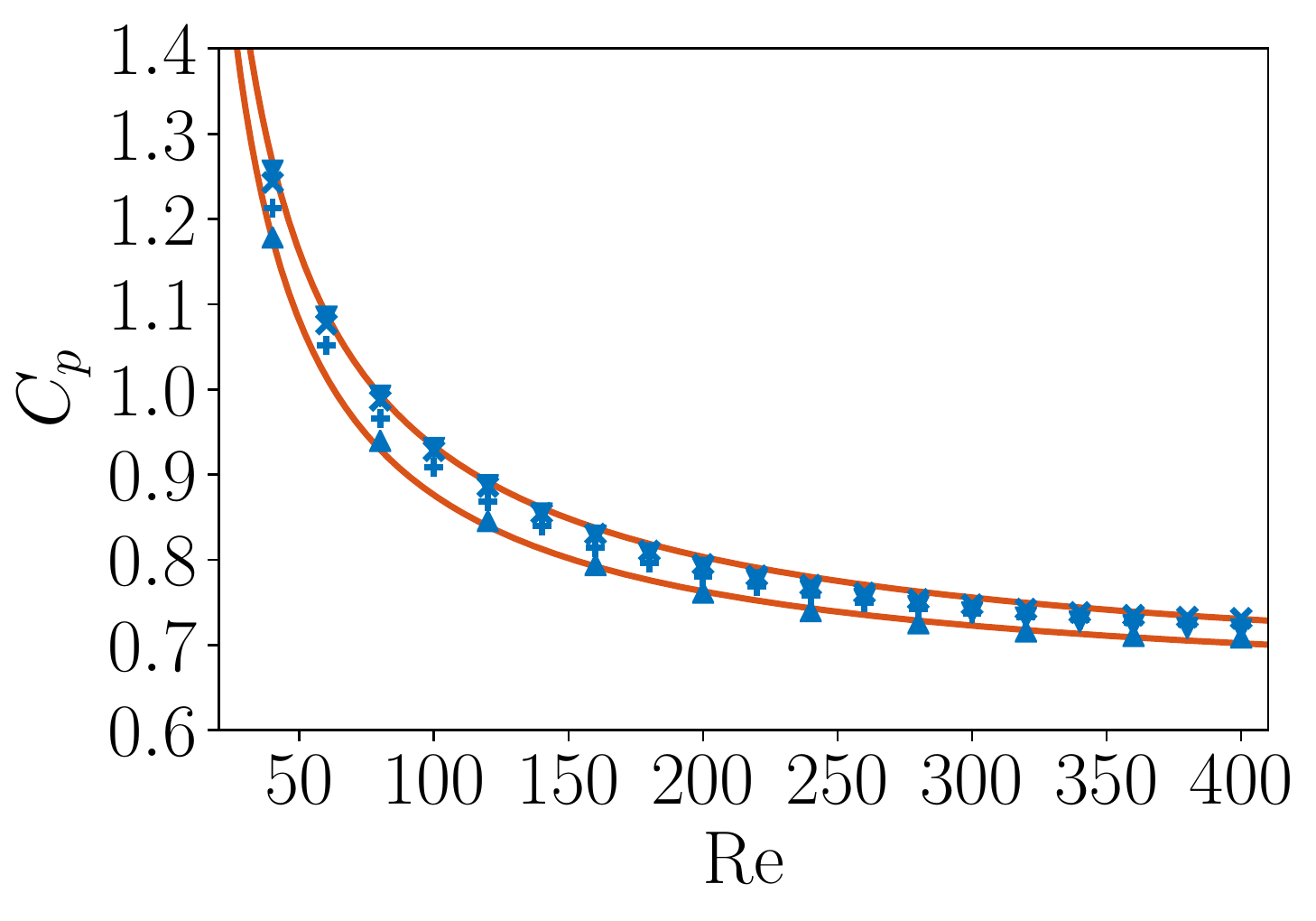} %\hspace{0.1cm}
		\\\vspace{-3mm}
		 \hspace{-3cm}$(a)$\hspace{55mm}$(b)$\hspace{55mm}$(c)$
	\caption{Pressure drag coefficient $C_{p}$ as a function of $\text{Re}$ for different aspect ratios. $(a)$: upstream end; $(b)$: downstream end; $(c)$: sum of the two contributions. $\blacktriangledown$:  $\cchi=2$; $\times$:  $\cchi=4$; $+$:  $\cchi=7$; $\blacktriangle$:  $\cchi=10$; solid line in $(a)$: empirical fit (\ref{eq:cp_theta0_up}); solid line in $(b)$: empirical fit (\ref{eq:cp_theta0_down}); solid line in $(c)$: sum (\ref{eq:cp_theta0_up})+(\ref{eq:cp_theta0_down}) for $\cchi=2$ (upper line) and $\cchi=10$ (lower line). 
\label{fig:Cp_theta0_relarge}}
\end{figure}
%We now come back to the main purpose of this section which is the derivation of a drag law valid for a wide range of aspect ratios and Reynolds numbers. In the following, the force coefficients are normalized by the quantity $\pi/8\rho U^2D^2$. 
\noindent Figures \ref{fig:Cp_theta0_relarge}$(a)-(b)$ display the variation with $\text{Re}$ of the pressure drag coefficient $C_p=8F_{dp}/(\pi\rho D^2U^2)$ on the upstream and downstream ends of the body for various $\cchi$. For each aspect ratio, both contributions decrease monotonically with $\text{Re}$. The pressure coefficient on the downstream end ($C_{p_{down}}$) is $3-4$ times smaller than that on the upstream end ($C_{p_{up}}$) for $\text{Re}\approx100$ and becomes only a small fraction of the latter for $\text{Re}\approx400$. %This difference is explained by the asymmetry of the wake as well as the appearance of a toroidal eddy downstream. 
Moreover, $C_{p_{up}}$ is almost independent of $\cchi$ (except for the shortest cylinder), while $C_{p_{down}}$ gradually decreases as the aspect ratio increases. Last, for $\cchi>2$ and $\text{Re}\gtrsim300$, $C_{p_{up}}$ is seen to tend toward an almost constant value slightly larger than $0.6$. Based on these remarks, approximate expressions for the two contributions take the form %Owing to these combined features, the total pressure drag coefficient depends only marginally on the aspect ratio. Indeed, for $\text{Re} \gtrsim100$, all pressure drag coefficients almost collapse on a master curve. As the Reynolds number increases, $C_p$ gets closer and closer to a constant value approximatively equal to $0.7$. A least-square fit of the difference $C_p(Re,\cchi)-0.7$ in the range $50\lesssim \text{Re}\lesssim400$ provides the simple approximation
%Since we observed in the previous section that the pressure coefficient is almost independent of the aspect ratio we can write it as $C_p = F_{dp,Re=0} / F_N + 0.72 - F(\text{Re})$, where F is a positive function of the Reynolds number which is unknown a priori. This function has to respect two important properties, it has to be negligible at low and high Reynolds number. It appears that $1.49 / \text{Re}^{1/2}$ works pretty well.
%This mean that the distribution of pressure on the upstream and downstream disk for a given Reynolds number is independent of $\cchi$....
%Figure \ref{fig:Cp_theta0_relarge} displays the evolution of the pressure force coefficient $C_p$ as function of the Reynolds number for various $\cchi$.
%The most noticeable feature shown by figure \ref{fig:Cp_theta0_relarge} is the independence of the pressure coefficient on the aspect ratio
%The pressure coefficient finally reads :
\begin{eqnarray}
%C_p \approx 0.7 + \frac{16.3}{\text{Re}^{0.9}(1+0.004\text{Re}^{0.95})}
\label{eq:cp_theta0_up}
C_{p_{up}}(\cchi,\text{Re}) &\approx& 0.62 + \frac{11.7}{\text{Re}^{0.9}(1+0.004\text{Re}^{0.9})}\\
\label{eq:cp_theta0_down}
C_{p_{down}}(\cchi,\text{Re}) &\approx& 2.05\cchi^{-1/4}\text{Re}^{-1/2}
%\label{eq:cp_theta0}
\end{eqnarray} 

\noindent As Fig. \ref{fig:Cp_theta0_relarge}$(c)$ confirms, numerical data for the total pressure drag coefficient $C_p=C_{p_{up}}+C_{p_{down}}$ are accurately fitted by the sum of (\ref{eq:cp_theta0_up}) and (\ref{eq:cp_theta0_down}). %Note that the decrease of the pressure drag coefficient with the Reynolds number is steeper than the usual $\text{Re}^{-1}$-law corresponding to the creeping-flow regime. %However due to the  this stonger decreases broke the minimum dissipation theorem   
\begin{figure}[h]
\centering
\includegraphics[width=6cm]{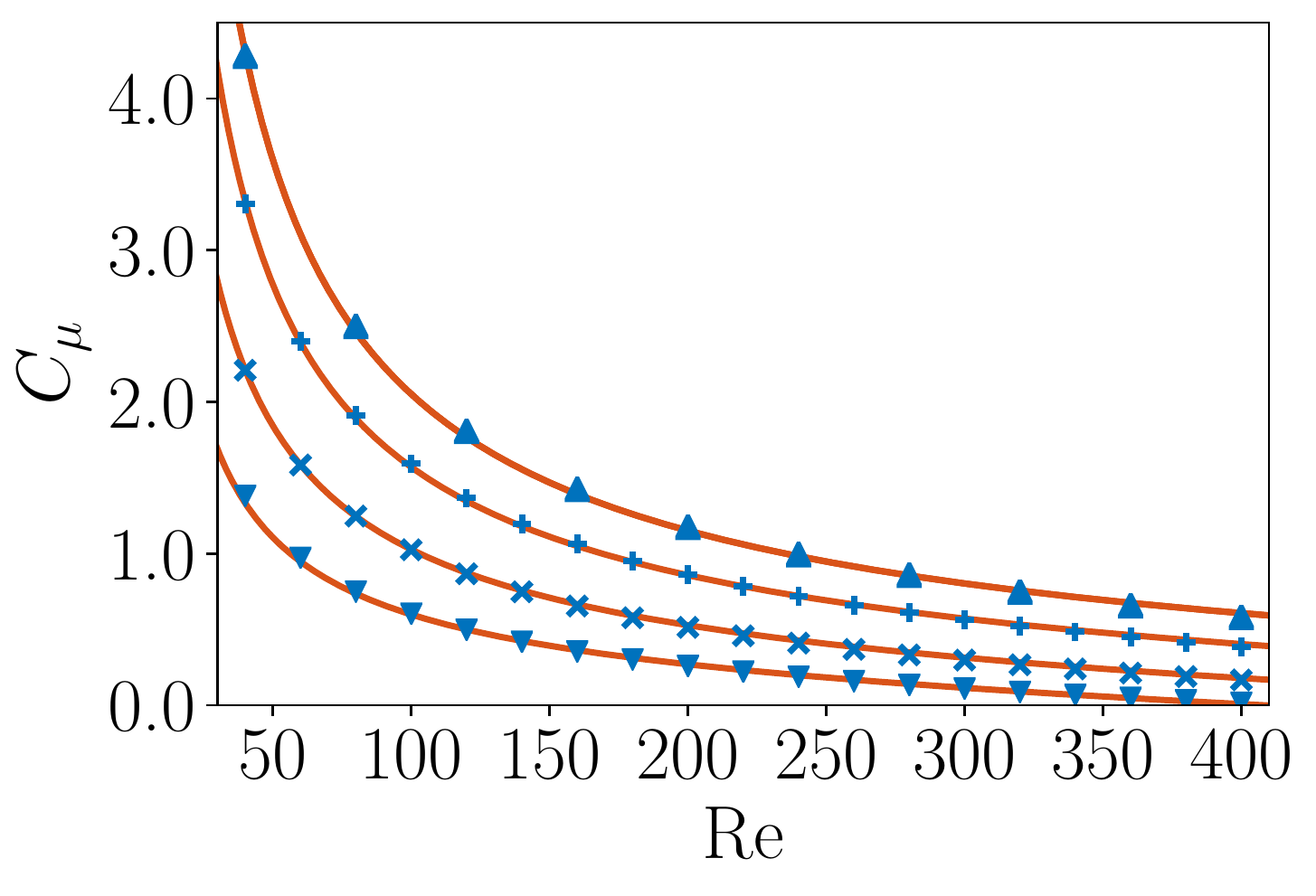}\hspace{15mm}\includegraphics[width=6cm]{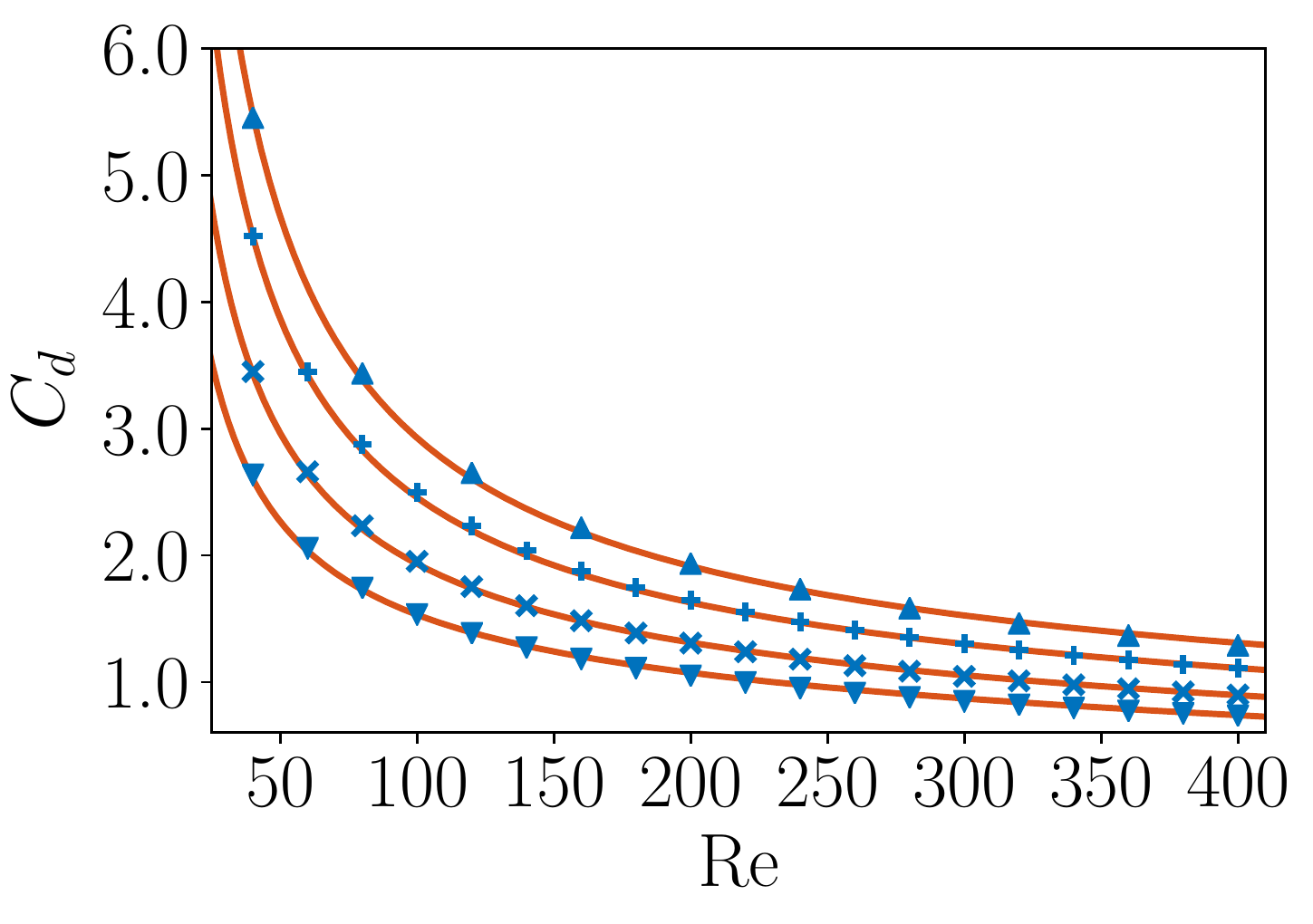}
 \vspace{-3mm}\\
 \hspace{-5cm}$(a)$\hspace{7cm}$(b)$\\
	\caption{Drag coefficient as a function of $\text{Re}$ for different aspect ratios. $(a)$: viscous contribution $C_{\mu}$; $(b)$: total drag coefficient $C_d$. $\blacktriangledown$:  $\cchi=2$; $\times$:  $\cchi=4$; $+$:  $\cchi=7$; $\blacktriangle$:  $\cchi=10$; for each aspect ratio, the solid lines in $(a)$ and $(b)$ represent the fits (\ref{eq:cmu_theta0}) and (\ref{eq:cd_theta0}), respectively. 
\label{fig:Cmu_theta0_relarge}}
\end{figure}
%\begin{figure}[h]
%\centering
%\includegraphics[width=6cm]{Cd.pdf}
%\caption{Drag coefficient as function of $\text{Re}$ for different for different aspect ratios. $\blacktriangledown$:  $\cchi=2$; $\times$:  $\cchi=4$; $+$:  $\cchi=7$; $\blacktriangle$: $\cchi=10$; $-$ : fit (\ref{eq:cd_theta0}) for the corresponding four aspect ratios. 
%\label{fig:Cd_theta0_relarge}}
%\end{figure}
Figure \ref{fig:Cmu_theta0_relarge}$(a)$ displays the variation of the viscous drag coefficient, $C_{\mu}(\cchi, \text{Re})$. This contribution is seen to be an increasing function of $\cchi$. Indeed, the area of the lateral surface of the body may be expressed in the form $\pi D^2\cchi$. Hence the normalized viscous force $C_{\mu}= 8F_{d\mu}/(\pi\rho D^2U^2)$ is expected to increase almost linearly with the aspect ratio. %Hence we can assume that $C_\mu$ is proportional to $\cchi$. This simple scaling proves himself to work well in the range of parameters studied here. 
 As usual, $C_\mu$ is also a decreasing function of the Reynolds number. However, the observed decrease is much steeper than the classical $\text{Re}^{-1/2}$ behavior expected on the basis of the boundary layer theory. The reason for this stands in the presence of the secondary annular eddy along the lateral surface. The corresponding backflow generates a negative local shear stress which lowers the overall viscous drag. For sufficiently short cylinders and large enough Reynolds numbers, this negative contribution may exceed the positive contribution of the shear stress on the rest of the lateral surface, yielding an overall negative viscous drag. This change of sign takes place at $\text{Re}\approx 420$ for $\cchi=2$. Based on the previous findings, a simple fit for $C_\mu$ is found to be
%Hence by integrating the viscous stress tensor over the lateral area the viscous friction becomes higher.
\begin{equation}
C _\mu(\cchi,\text{Re}) \approx16.45 \cchi^{0.7}\text{Re}^{-0.8}+(a_1\cchi-a_2)\text{Re}\,,\quad\mbox{with}\quad a_1=4.1\times10^{-5}\quad\mbox{and}\quad\,a_2=6\times10^{-4}\,.
\label{eq:cmu_theta0}
\end{equation} 
\noindent As Fig. \ref{fig:Cmu_theta0_relarge}$(a)$ shows, the above fit describes the variations of $C_\mu$ well throughout the entire range of aspect ratios and Reynolds numbers explored numerically. The last term in the right-hand side of (\ref{eq:cmu_theta0}) accounts for the influence of the annular eddy. Note that according to (\ref{eq:cmu_theta0}), the dependence of $C _\mu$ with respect to $\cchi$ is slightly weaker than expected on the basis of the above simple geometrical argument.\\%ormula matches reasonably with the numerical results up to an upper Reynolds number in the range $[350,\,400]$, depending on $\chi$. Beyond this limit, the effect of the lateral toroidal vortex on the shear friction becomes significant and (\ref{eq:cmu_theta0}) starts to deviate significantly from the numerical results. 
As Figs. \ref{fig:Cp_theta0_relarge} and \ref{fig:Cmu_theta0_relarge} evidence, pressure effects contribute less to the drag than viscous friction for $\text{Re} \lesssim 100$. For Reynolds numbers in the range $100\lesssim \text{Re}\lesssim300$ and short cylinders ($\cchi \leq 4$), both contributions have a comparable magnitude. However, due to the sharp decrease of the viscous contribution for $\text{Re}\gtrsim400$ for such short cylinders, the latter eventually becomes smaller than the pressure contribution in the upper part of the $\text{Re}$-range covered by present computations. Adding the approximate expressions (\ref{eq:cp_theta0_up}), (\ref{eq:cp_theta0_down}) and (\ref{eq:cmu_theta0}), the total drag coefficient is approached as %. \textit{Hence the small deviation observed between equation \ref{eq:cmu_theta0} and the numerical results do not have a strong effect on the drag law based on the sum of viscous and pressure contribution. This drag law reads}
% The more the cylinder is large more the pressure is greater. Increasing Reynolds number until $\text{Re}=400$, the inertial effects dominate in this regime so the pressure contributions are very higher than the viscous ones. It reaches $90\%$ for  $\raisebox{2pt}{$\chi$}=3$, which means that the pressure controls the dynamic of the flow. For a slender cylinder ( $\raisebox{2pt}{$\chi$}=7$), the fraction $\frac{C_p}{C_\mu}$ moves from $1/2$ to $3/2$ by increasing $\text{Re}$, which is a kind of balance between viscous and inertial regimes.
%due to a lower contribution of viscous effect at the higher inertial regime
\begin{equation}
%C_d \approx 0.7+ \frac{15.7 \cchi^{0.72}}{\text{Re}^{0.8}}+ \frac{16.3}{\text{Re}^{0.9}(1+0.004\text{Re}^{0.95})}+(a_1\cchi-a_2)Re
C_d(\cchi,\text{Re})\approx C_{p_{up}}(\cchi,\text{Re})+C_{p_{down}}(\cchi,\text{Re})+C_\mu(\cchi,\text{Re})\,.
\label{eq:cd_theta0}
\end{equation}
As Fig. \ref{fig:Cmu_theta0_relarge}$(b)$ indicates, this fit is in good agreement with the numerically predicted drag throughout the range of $\cchi$ and $\text{Re}$ covered by the simulations. Note that the upper limit of validity of this fit is presumably $\text{Re}_{max}\approx400$. In particular, the linear variation of $C_\mu$ with $\text{Re}$ predicted by (\ref{eq:cmu_theta0}) cannot continue at very large Reynolds number, as the drag coefficient is expected to decrease with $\text{Re}$ for $\text{Re}>\text{Re}_{max}$ and become $\text{Re}$-independent in the limit $\text{Re}\rightarrow\infty$. Conversely, it may be checked that the above fit properly matches the modified low-but-finite $Re$ prediction (\ref{eq:slender_smallinertia2}). Consider for instance a cylinder with $\cchi=7$ and $Re=5$. On the one hand, (\ref{eq:cp_theta0_up})-(\ref{eq:cd_theta0}) predict $C_d\approx21.6$ for this set of parameters. On the other hand, according to Fig. \ref{fig:small_inertia}, the ratio $F_d/F_{ds}$ predicted by  (\ref{eq:slender_smallinertia2}) is approximately $1.85$. Keeping in mind that the diameter $\mathcal{D}$ of the equivalent sphere is related to $D$ through $\mathcal{D}/D=(\frac{3}{2}\cchi)^{1/3}$, one has $C_d=24\text{Re}^{-1}(\frac{3}{2}\cchi)^{1/3}F_d/F_{ds}\approx21.9$. Similarly, for the same Reynolds number but $\cchi=20$, one has $C_d\approx40.7$ from (\ref{eq:cp_theta0_up})-(\ref{eq:cd_theta0}) and $C_d\approx40.3$  from Fig. \ref{fig:small_inertia} where $F_d/F_{ds}\approx2.70$.
This agreement, which is confirmed with other sets of parameters, allows us to conclude that combining (\ref{eq:slender_smallinertia2}) for Reynolds numbers less than a few units with (\ref{eq:cp_theta0_up})-(\ref{eq:cd_theta0}) for larger $\text{Re}$ provides an accurate description of drag variations from $\text{Re}=0$ up to $\text{Re}=400$.% We only notice small deviations of our numerical data at moderate $\text{Re}$ for  and at very high Reynolds number for $\cchi =2$.
% The mean square error varies from $2\%$ to $10\%$. %This model is used in the following section to build a general expression for drag force exerted on a yawed cylinder when the yawed angle is taking into account.

\color{black}
\section{Forces and torque on a moderately inclined cylinder at low-to-moderate Reynolds number}
\label{incl1}
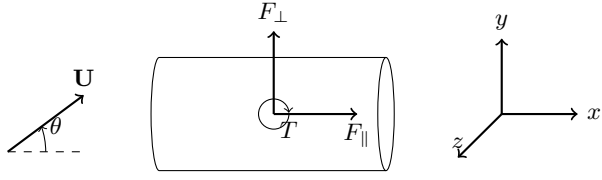
\begin{figure}[h]
		\centering
	\begin{tabular}
	{>{\centering\arraybackslash} m{2cm} >{\centering\arraybackslash} m{4cm}}
	   \begin{tikzpicture}[scale=1.]

\node [cylinder, rotate=0, draw, minimum height=3.2cm, minimum width=1.5cm]  at (-0.1, 0, 0) (c) {}; %ne sera pas a lechelle mais pas grave
%\draw[dashed] (-1.5,-0.75) rectangle +(3,1.5);

\draw[->, thick] (3,0)--(4,0) node[right]{$x$};
\draw[->, thick] (3,0)--(3,1) node[above]{$y$};
\draw[->, thick] (3,0,0)--(3,0,1.5) node[above]{$z$};

\draw[->,black] (-3,-0.5)  arc (0:20:1)node[right]{$\theta$}; %draws an arc of radius 1 starting from (0,0) 
\draw[dashed] (-3.5,-0.5)--(-2.5,-0.5);
\draw[->,thick] (-3.5,-0.5)--(-2.5,0.25)node[above]{$\mathbf{U}$};

\draw[->,black] (0.2,0)  arc (0:-355:0.2)node[below]{$T$}; 

\draw[->,thick] (0,0)--(1.1,0)node[below]{$F_{\parallel}$};
\draw[->,thick] (0,0)--(0,1.1)node[above]{$F_{\perp}$};

%\draw[red,thick] (-1.5,0)--(-1.5,0.5);
%\draw[red,thick] (1.5,0)--(1.5,-0.5);

%\draw[blue,thick] (-1.5,0)--(-1.5,0.5);
%\draw[blue,thick] (1.5,0)--(1.5,-0.5);

%\draw[dashed] (1.45,1.5,2)--(2.5,1.94,2);
%\draw (0.5,1.5,2) node[]{$\rho, \mu$};
\end{tikzpicture}

%	\begin{tikzpicture}[scale=0.04]

%	\node [cylinder, black, rotate=10, draw, minimum height=3cm, minimum width=1cm]  at (3, 2 ,2) (c) {}; %ne sera pas a lechelle mais pas grave

%	\end{tikzpicture}
 &
	\end{tabular} 
	\caption{ Schematic of the force components on an inclined cylinder. In this configuration, the inertial torque is negative, tending to rotate the cylinder clockwise.
\label{fig:cylindre_incline}}
\end{figure}

We now move to the more general configuration in which the cylinder is inclined with respect to the incoming flow by a nonzero angle. For reasons discussed in Sec. \ref{intro}, we limit ourselves to maximum inclinations of $30^\circ$. The present section focuses on the low-to-moderate Reynolds number range $[ 0.1-5]$. Higher Reynolds numbers are considered in the next section.\\\color{black} %The meshes used in the computations are the same as the one used in section \ref{creep} with the addition of 32 azimuthal planes.  
\indent The nonzero inclination breaks the flow axial symmetry. Therefore, in addition to the force component $F_{\parallel}$ parallel to the cylinder axis, a perpendicular force component ($F_\perp$) takes place, together with a spanwise torque ($T$) (Fig.  \ref{fig:cylindre_incline}). 
These components are linearly related to the drag and lift forces $F_d$ and $F_l$, respectively parallel and perpendicular to the incoming velocity $\textbf{U}$, via the geometric relations
\begin{eqnarray}
\label{eq:geom1}
%C_d(\theta)&=&C_{\parallel}(\theta)cos\theta+C_{\perp}(\theta)sin\theta \\
F_\parallel(\cchi,\theta, \text{Re})&=&F_d(\cchi,\theta, \text{Re})\cos\theta-F_l(\cchi,\theta, \text{Re})\sin\theta\,,\\
\label{eq:geom2}
%C_l(\theta)&=&C_{\parallel}(\theta)sin\theta-C_{\perp}(\theta)cos\theta
F_\perp(\cchi,\theta, \text{Re})&=&F_l(\cchi,\theta, \text{Re})\cos\theta+F_d(\cchi,\theta, \text{Re})\sin\theta\,.
\end{eqnarray}

In the Stokes regime, the linearity of the loads with respect to the boundary conditions implies that the force acting on the inclined cylinder is linearly related to the drag acting on the same body in the two extreme configurations $\theta=0^\circ$ and $\theta=90^\circ$ through 
\begin{eqnarray}
\label{eq:Cpara_stokes}
F_{\parallel}(\cchi,\theta)&=&F_\parallel^{ \theta=0^\circ}(\cchi)\cos\theta\,, \\
\label{eq:Cperp_stokes}
F_{\perp}(\cchi,\theta)&=&F_\perp^{ \theta=90^\circ}(\cchi)\sin\theta\,.
\end{eqnarray}
\color{black}
These simple `Stokes laws' are not expected to remain valid when inertial effects become significant. To assess and possibly extend their validity, an approximate expression for $F_\perp^{ \theta=90^\circ}$, similar to (\ref{eq:slender_smallinertia2}) for $F_\parallel$, is required. In appendix \ref{perps}, we establish the fourth-order slender-body approximation of $F_\perp^{ \theta=90^\circ}(\cchi,\text{Re}=0)$, and modify it empirically to extend its validity toward aspect ratios and Reynolds numbers of $\mathcal{O}(1)$. To get some insight into the way inertial effects alter (\ref{eq:Cpara_stokes})-(\ref{eq:Cperp_stokes}), it is informative to consider the finite-Reynolds-number approximate expressions for $F_d$ and $F_l$ established for arbitrary inclinations by \citet{khayat1989}. % \ref{eq:slender_smallinertia}) and (\ref{eq:slender_perp_smallinertia}) for the drag force on a cylinder aligned or perpendicular to the flow, respectively. 
Evaluating these expressions in the limit $\cchi\text{Re}\ll1$ and making use of (\ref{eq:geom1})-(\ref{eq:geom2}) yields 
\begin{eqnarray}
\label{eq:Cpara_stokes1}
\frac{F_{\parallel}(\cchi,\theta,\cchi\text{Re}\ll1)}{2\pi\mu UL}&\approx&\left(F_\parallel^{* \theta=0^\circ}(\cchi,\cchi\text{Re}\ll1)-\frac{1}{16}\sin^2\theta\frac{\cchi Re}{(\ln\cchi)^2}\right)\cos\theta\,, \\
\label{eq:Cperp_stokes1}
\frac{F_{\perp}(\cchi,\theta,\cchi\text{Re}\ll1)}{4\pi\mu UL}&\approx&\left(F_\perp^{* \theta=90^\circ}(\cchi,\cchi\text{Re}\ll1)+\frac{1}{16}\cos^2\theta\frac{\cchi Re}{(\ln\cchi)^2}\right)\sin\theta\,,
\end{eqnarray}
where $F_\parallel^{*}$ and $F_\perp^{*}$ stand for the dimensionless second-order expansion of the corresponding force with respect to $1/\ln\cchi$ in the limit of low-but-finite $\cchi\text{Re}$, as provided by (\ref{eq:slender_smallinertia}) and (\ref{eq:slender_perp_smallinertia}), respectively. According to (\ref{eq:slender_smallinertia}) and the asymptotic form of the inertial correction $f_\parallel$ in the limit $\cchi\text{Re}\rightarrow0$, one has $F_\parallel^{* \theta=0^\circ}(\cchi,\cchi\text{Re}\ll1)=[\text{ln}\cchi]^{-1}+(\frac{3}{2}-2\text{ln}2+\frac{1}{8}\cchi\text{Re})[\text{ln}\cchi]^{-2}$. Similarly, (\ref{eq:slender_perp_smallinertia}) and the asymptotic form of $f_\perp$ yield $F_\perp^{* \theta=0^\circ}(\cchi,\cchi\text{Re}\ll1)=[\text{ln}\cchi]^{-1}+(\frac{1}{2}-2\text{ln}2+\frac{1}{4}\cchi\text{Re})[\text{ln}\cchi]^{-2}$. Expressions (\ref{eq:Cpara_stokes1})-(\ref{eq:Cperp_stokes1}) indicate that the angular dependence of $F_\parallel$ and $F_\perp$ becomes more complex in the presence of inertial effects, involving higher-order harmonics of $\theta$. Moreover they suggest that the $\theta$-dependent inertial corrections tend to decrease $F_\parallel$ and increase $F_\perp$, compared to the prediction of the extrapolated `Stokes law' based on the finite-$\text{Re}$ drag forces $F_\parallel^{* \theta=0^\circ}$ and $F_\perp^{* \theta=90^\circ}$. % found in the two limit configurations $\theta=0^\circ$ and $\theta=90^\circ$.

%\vspace{0mm}\\
%Combining the purely geometric relations (\ref{eq:geom1})-(\ref{eq:geom2}) with (\ref{eq:Cpara_stokes})-(\ref{eq:Cperp_stokes}) yields in this regime
%\begin{eqnarray}
%\label{eq:Cd_stokes}
%C_{d}(\cchi,\theta)&=&C_\parallel^{ \theta=0^\circ}(\cchi) + (C_\perp^{ \theta=90^\circ}(\cchi)-C_\parallel^{ \theta=0^\circ}(\cchi))\sin^2\theta\,,  \\
%\label{eq:Cl_stokes}
%C_l(\cchi,\theta)&=&\frac{1}{2}(C_\parallel^{ \theta=0^\circ}(\cchi) - C_\perp^{ \theta=90^\circ}(\cchi))\sin2\theta \,.
%\end{eqnarray}

%Combining the purely geometric relations (\ref{eq:geom1})-(\ref{eq:geom2}) with `Stokes laws' (\ref{eq:Cpara_stokes})-(\ref{eq:Cperp_stokes}) yields under creeping flow conditions
%\begin{eqnarray}
%\label{eq:Cd_stokes}
%F_{d}(\cchi,\theta)&=&F_\parallel^{ \theta=0^\circ}(\cchi) + (F_\perp^{ \theta=90^\circ}(\cchi)-F_\parallel^{ \theta=0^\circ}(\cchi))\sin^2\theta\,,  \\
%\label{eq:Cl_stokes}
%F_l(\cchi,\theta)&=&\frac{1}{2}(F_\perp^{ \theta=90^\circ}(\cchi)-F_\parallel^{ \theta=0^\circ}(\cchi))\sin2\theta \,.
%\end{eqnarray}
\indent Still in the Stokes regime, the spanwise torque is zero whatever $\theta$, owing to the geometrical symmetries of the cylinder and the reversibility of Stokes equations. However, nonlinearities inherent to inertial effects result in a finite torque. In the limit $Re \cchi\ll1$, the finite-Reynolds-number expression for this torque obtained in \cite{khayat1989} reduces to
\begin{equation}
\frac{T(\cchi,\theta,\cchi\text{Re}\ll1)}{\mu U(L/2)^2}\approx-\frac{5\pi}{12}\frac{\cchi Re}{(\ln\cchi)^2}\sin2\theta\,.
\label{KCt0}
\end{equation}
This negative torque tends to rotate the cylinder perpendicular to the flow direction. \vspace{1mm}\\
\indent In what follows we make use of the numerical results to examine the validity of `Stokes laws' (\ref{eq:Cpara_stokes})-(\ref{eq:Cperp_stokes}) and of the predictions of \cite{khayat1989} for low-to-moderate Reynolds numbers.
%\vspace{2mm}
\\  %The low-to-moderate Reynolds number range $1\lesssim\text{Re}\lesssim10$ is of special interest to check how theoretical predictions from \cite{khayat1989}  compare with numerical results for the drag and lift force components and the spanwise torque. 
%\subsection{Approximate laws for the parallel force}
\begin{figure}[h]
\centering
\includegraphics[width=4.9cm]{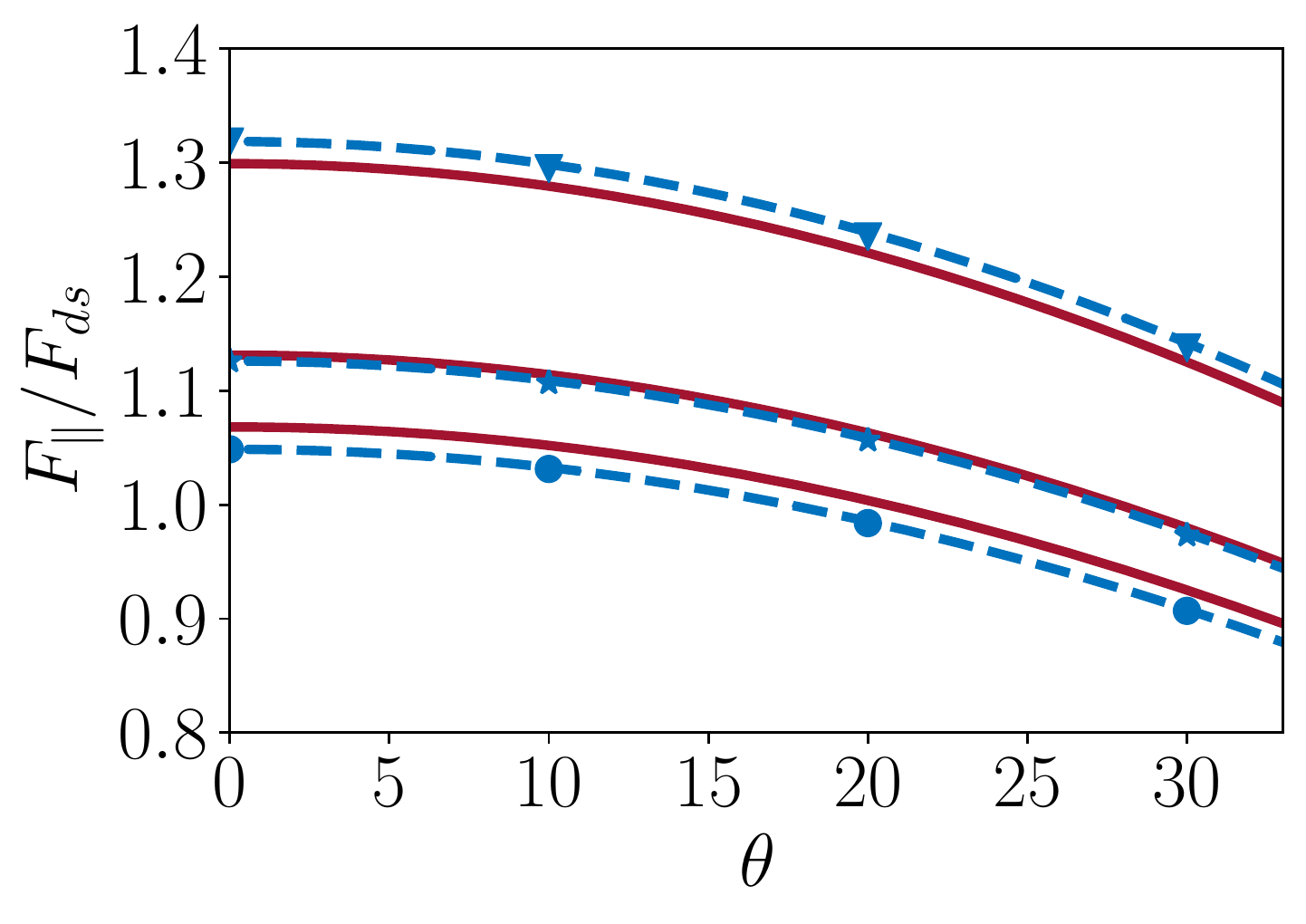}\includegraphics[width=5cm]{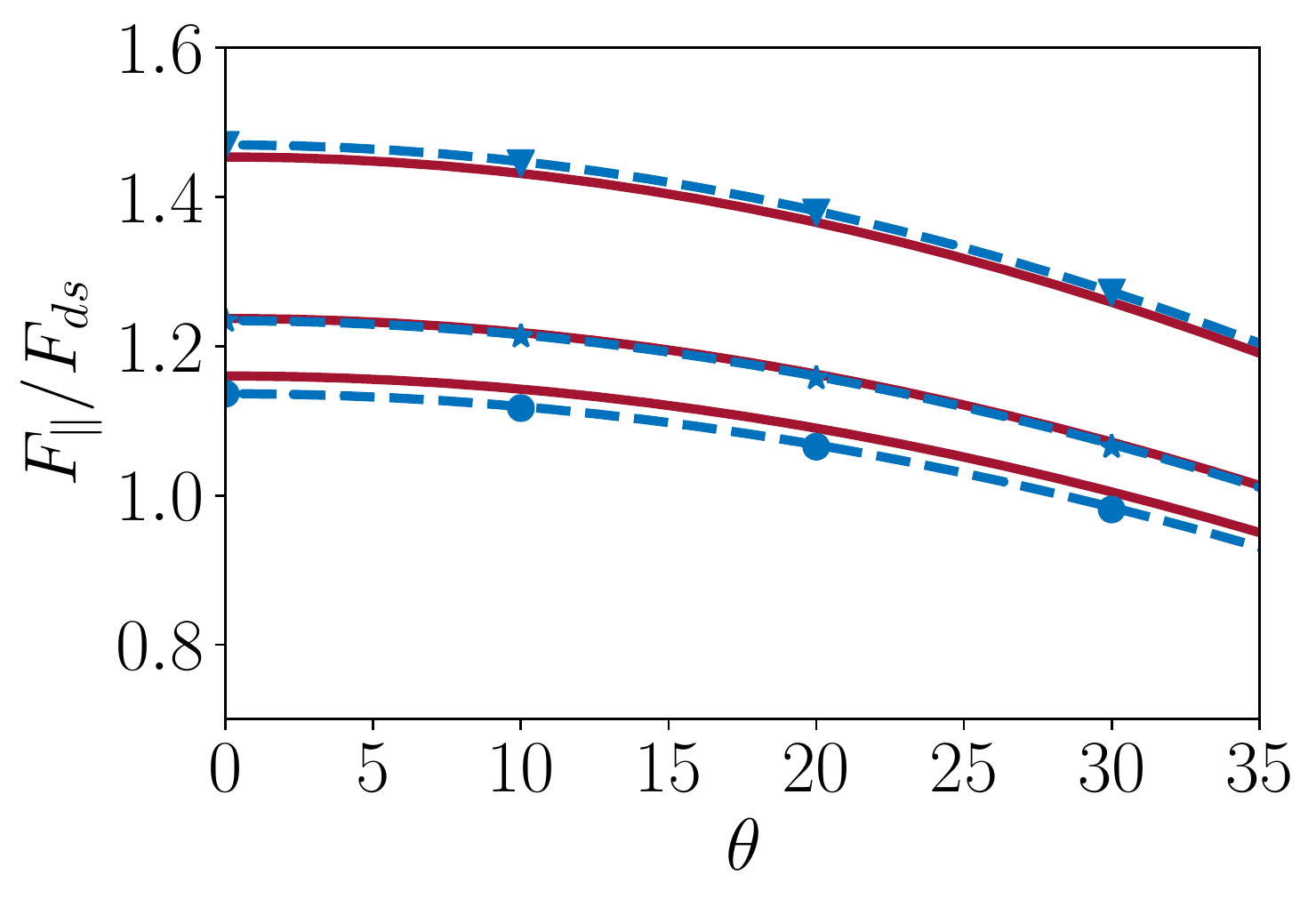}\includegraphics[width=5cm]{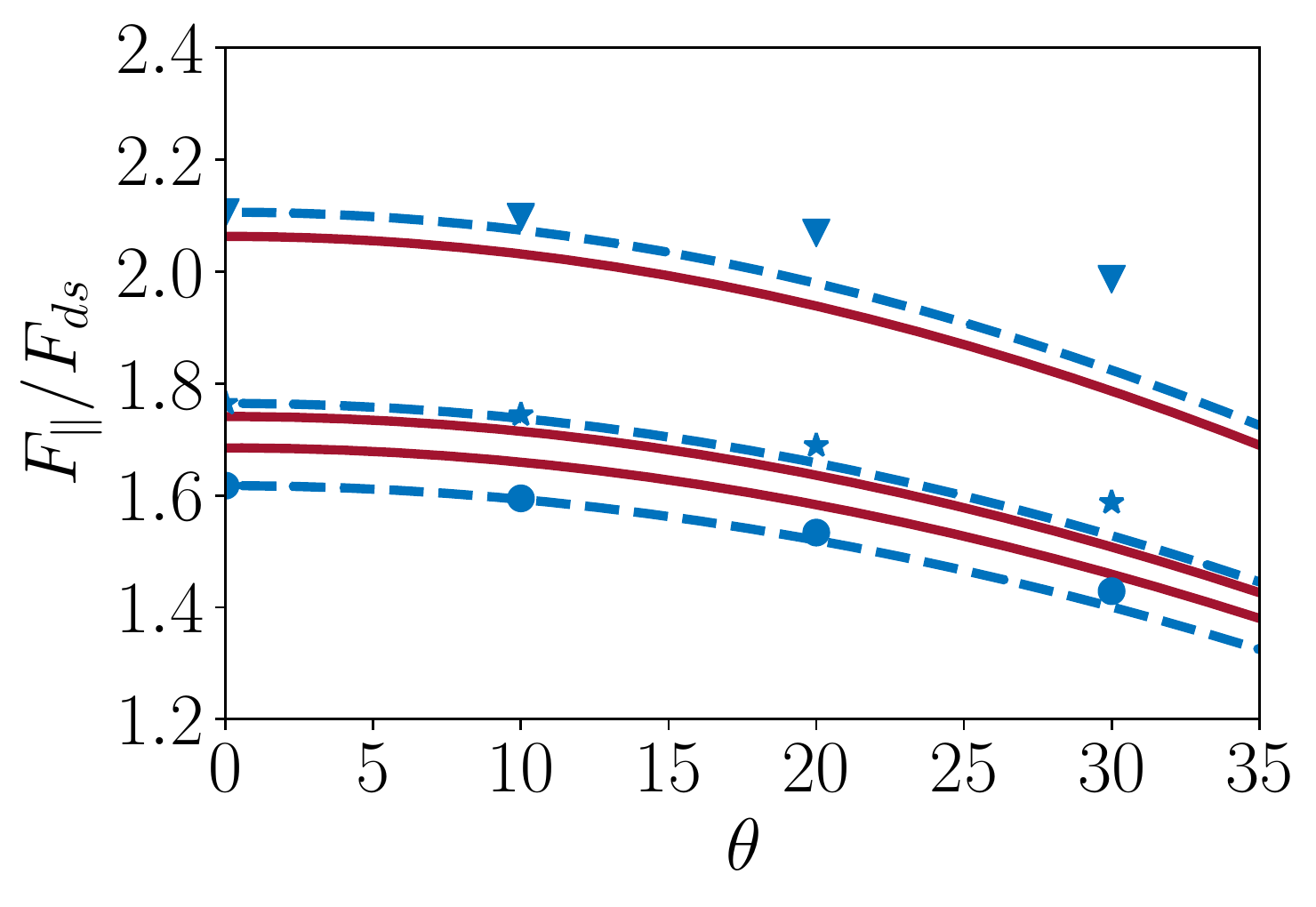}
\vspace{-3mm}\\
\hspace{-35mm}$(a)$\hspace{47mm}$(b)$\hspace{47mm}$(c)$\\
\vspace{-13mm}
\hspace{-13mm}$\text{Re}=0.1$\hspace{38mm}$\text{Re}=0.5$\hspace{38mm}$\text{Re}=5$
\vspace{11mm}
%\vspace{1mm}
\caption{Parallel force component $F_\parallel$ vs. the inclination angle $\theta$ for $\bullet$: $\cchi=3$, $\star$: $\cchi=5$, $\blacktriangledown$: $\cchi=10$. %$(a)$: $\text{Re}=0.5$, $(b)$: $\text{Re}=5$. 
Dashed line: prediction (\ref{eq:Cpara_stokes}) based on the numerical value of $F_\parallel^{ \theta=0^\circ}$; solid line: prediction (\ref{eq:Cpara_stokes}) based on the semiempirical approximation (\ref{eq:slender_smallinertia2}) for $F_\parallel^{ \theta=0^\circ}$.}%, dashed dotted line : equation (\ref{eq:cpp_theta0_up})-(\ref{eq:cpmu_theta0_lat})}
\label{fig:c_par_moderate}
\vspace{-6mm}
\end{figure}\indent
Figure \ref{fig:c_par_moderate} displays the parallel force component $F_\parallel$ for three Reynolds numbers and aspect ratios $\cchi=3,\,5$ and $10$, respectively. At the lower two $\text{Re}$, variations of $F_\parallel$ closely follow the $\cos\theta$-behavior predicted by (\ref{eq:Cpara_stokes}) whatever the aspect ratio. This implies that the $\sin^2\theta$-inertial correction in (\ref{eq:Cpara_stokes1}) has only a marginal influence at moderate inclinations for $\cchi\text{Re}\lesssim5$. Indeed, for $\theta=30^\circ$, this correction is less that $2\%$ of $F_\parallel^{* \theta=0^\circ}$ whatever the aspect ratio. However, the inertial correction included in $F_\parallel^{*}$ reaches $16\%$ for $\cchi=10$, indicating that inertial effects are already significant. In other words, in the range of moderate inclinations considered here, the `Stokes law' still accurately predicts $F_\parallel(\cchi,\theta,\text{Re})$ for $\text{Re}\lesssim1$, provided the prediction makes use of the inertia-corrected drag $F_\parallel^{\theta=0^\circ}(\cchi,\text{Re})$.
%Note that, as (\ref{eq:slender_smallinertia2}) slightly overestimates $F_\parallel^{ \theta=0^\circ}(\cchi,\text{Re})$ for small-to-moderate aspect ratios (see Fig. \ref{fig:small_inertia}$(a)$), the same happens for the prediction of $F_\parallel(\cchi,\theta, \text{Re})$ through the $\cos\theta$-law based on (\ref{eq:slender_smallinertia2}) for $\chi=3$ and $\chi=5$. However the overestimate quickly decreases as $\cchi$ increases and a good agreement is reached for $\chi=10$. 
\\ %Despite the complicated behaviour of equations (\ref{eq:cpp_theta0_up})-(\ref{eq:cpmu_theta0_lat}) as function of $\theta$ a good agreement is found with the numerical results. Suprisingly the agreement is worse for Re=5  (figure \ref{fig:c_par_moderate}  (right)). In this configuration equations (\ref{eq:cpp_theta0_up})-(\ref{eq:cpmu_theta0_lat}) overestimate the drag force for all aspect ratios. 
\indent At $\text{Re}=5$, influence of inertial effects has become dominant. %This is why (\ref{eq:Cpara_stokes}) based on the prediction (\ref{eq:slender_smallinertia2}) for $F_\parallel^{ \theta=0^\circ}$ better agrees with numerical results up to $\theta\approx10^\circ$, a slight overestimate ($\approx7\%$) only remaining for  $\cchi=3$. 
Since the magnitude of $\theta$-dependent inertial correction also increases with $\text{Re}\cchi$, the larger the aspect ratio the stronger these effects at a given $\text{Re}$. This may be appreciated in Fig. \ref{fig:c_par_moderate}$(c)$, where the difference between the computed force and the prediction of the `Stokes law' is seen to increase significantly with $\cchi$ for $\theta>10^\circ$, from less than $2\%$ for $\cchi=3$ at $\theta=30^\circ$ to $9\%$ for $\cchi=10$ at the same inclination. Interestingly, the under-prediction of $F_\parallel$ by the Stokes law is at odds with the low-$\cchi\text{Re}$ prediction (\ref{eq:Cpara_stokes1}) which suggests that this `law' should overestimate $F_\parallel$, since it ignores the negative $\sin^2\theta$-inertial contribution. This contradiction emphasizes the fact that the $\theta$-dependences of inertial effects in the low-$\cchi\text{Re}$ range and in the range $10\lesssim\cchi\text{Re}\lesssim10^2$ are drastically different. 
%therefore larger forFor $\chi=10$ the parallel force do not follow the $\cos\theta$ law. This is not surprinsing since the Reynolds based on the the lentgh of the body is 50 in this case.
%\subsection{Approximate laws for the perpendicular force}
\begin{figure}[h]
\centering
\includegraphics[width=5cm]{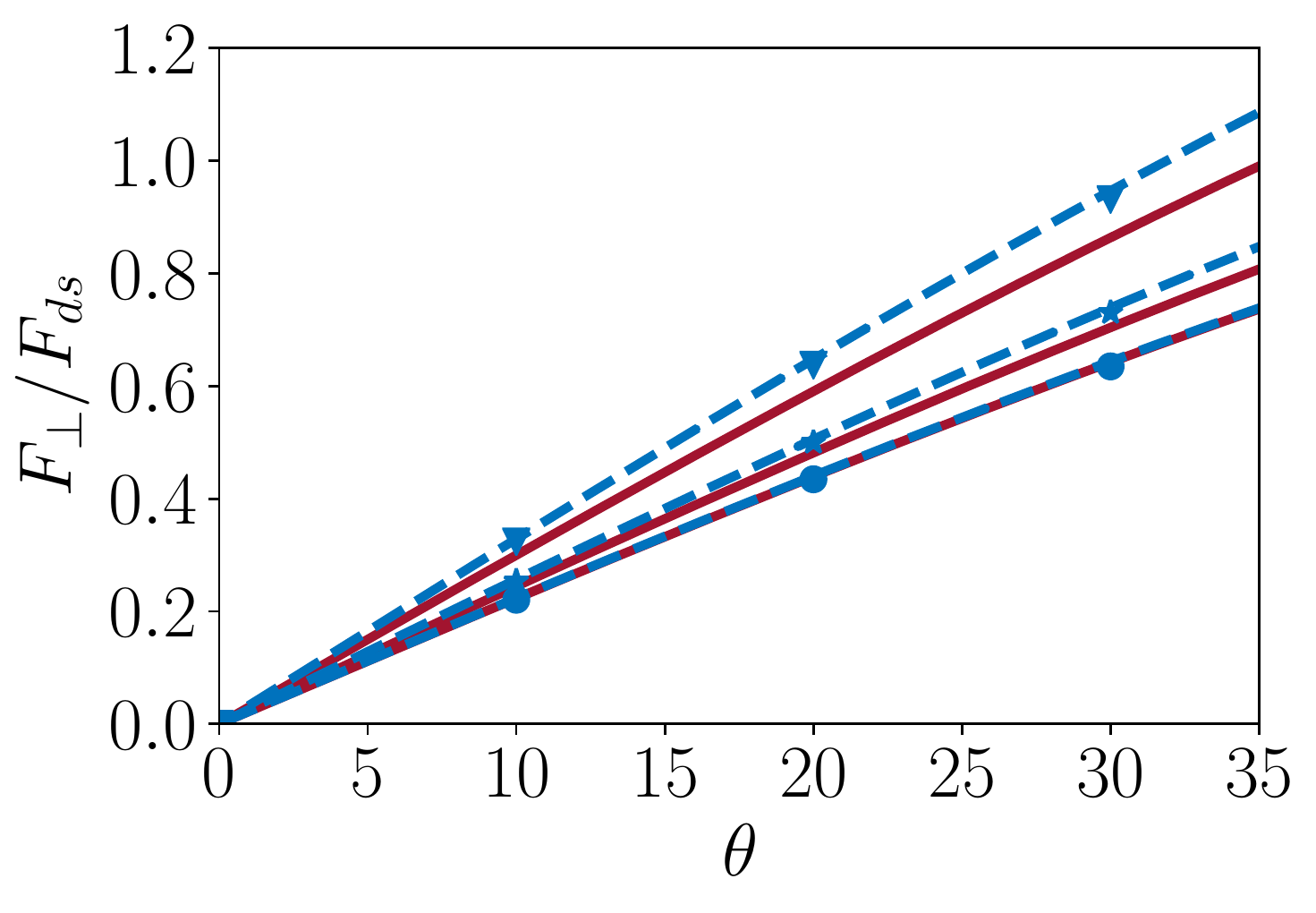}\includegraphics[width=5cm]{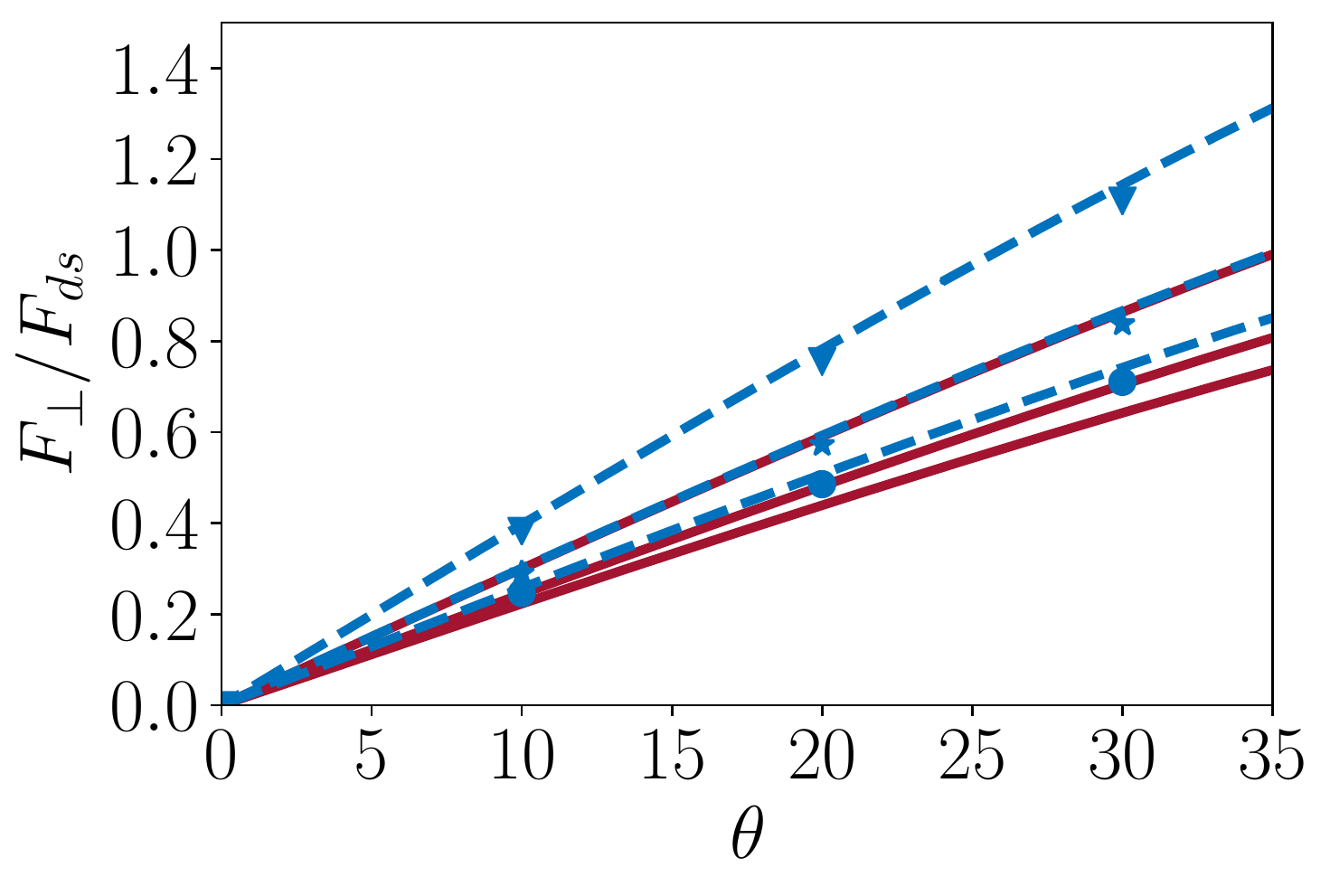}\includegraphics[width=5cm]{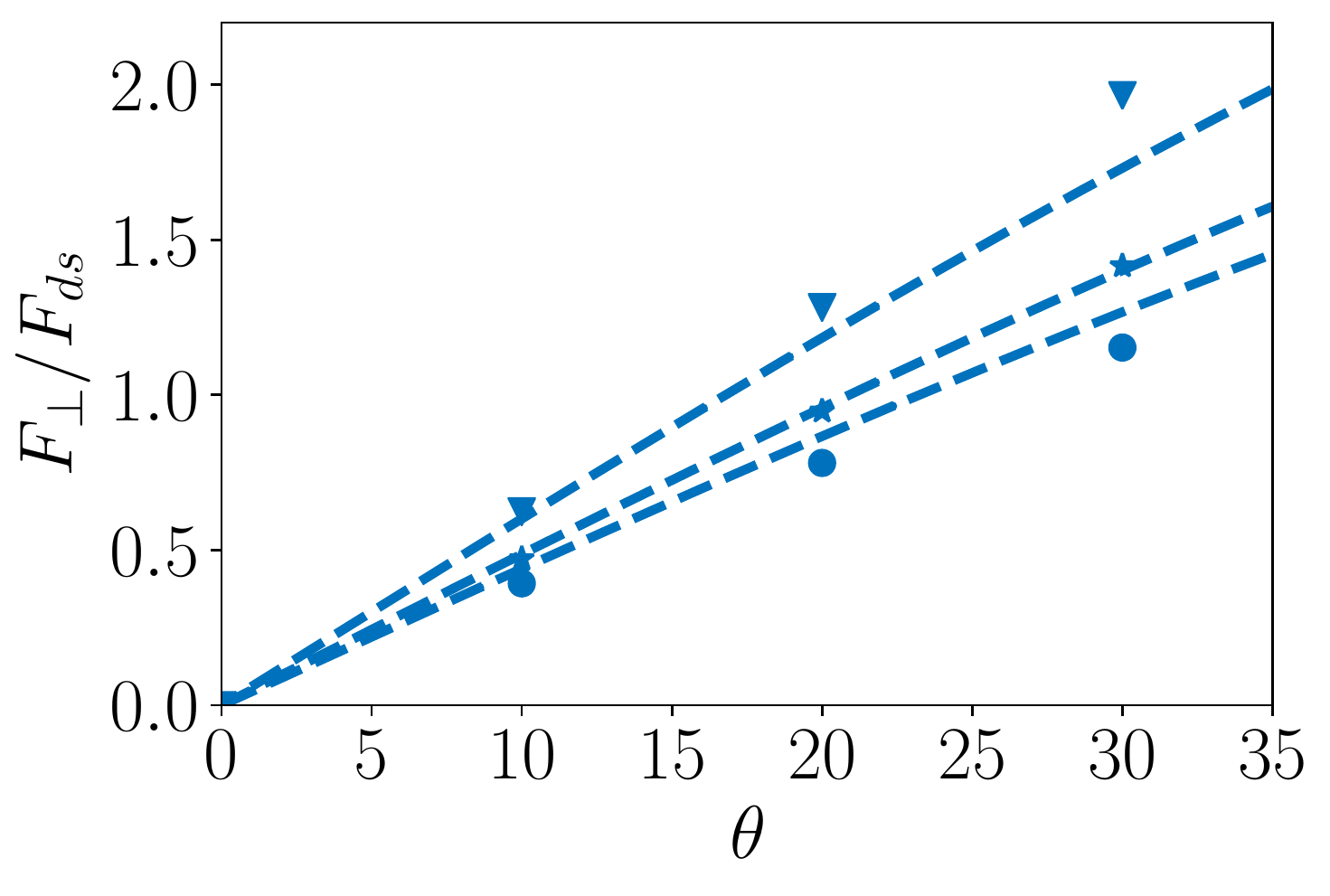}
\vspace{-3mm}\\
\hspace{-35mm}$(a)$\hspace{47mm}$(b)$\hspace{47mm}$(c)$\\
\vspace{-32mm}
\hspace{-13mm}$\text{Re}=0.1$\hspace{38mm}$\text{Re}=0.5$\hspace{38mm}$\text{Re}=5$
\vspace{30mm}\\
\caption{Perpendicular force component $F_\perp$ as a function of $\theta$ for $\bullet$: $\cchi=3$, $\star$: $\cchi=5$, $\blacktriangledown$: $\cchi=10$. %$(a)$: $\text{Re}=0.1$, $(b)$: $\text{Re}=0.5$, $(c)$: $\text{Re}=5$. 
 Solid line: `Stokes law' (\ref{eq:Cperp_stokes}) based on the semiempirical creeping-flow estimate (\ref{eq:fperp4m}) for $F_\perp^{ \theta=90^\circ}(\cchi)$; dashed line: same with $F_\perp^{ \theta=90^\circ}$ based on the semiempirical finite-$\text{Re}$ estimate (\ref{eq:slender_perp_smallinertia_m}). In $(a)$, the two estimates overlap for $\cchi=3$; in $(b)$, the prediction for $\cchi=10$ based on (\ref{eq:fperp4m}) overlaps with that for $\cchi=5$ based on  (\ref{eq:slender_perp_smallinertia_m}).%dash-dotted line in $(c)$: empirical extension (\ref{perp}) of the `Stokes law' to strongly inertial regimes.
 }
\label{fig:c_perp_moderate}
\vspace{-2mm}
\end{figure}

\begin{figure}[h]
\centering
\includegraphics[width=5.5cm]{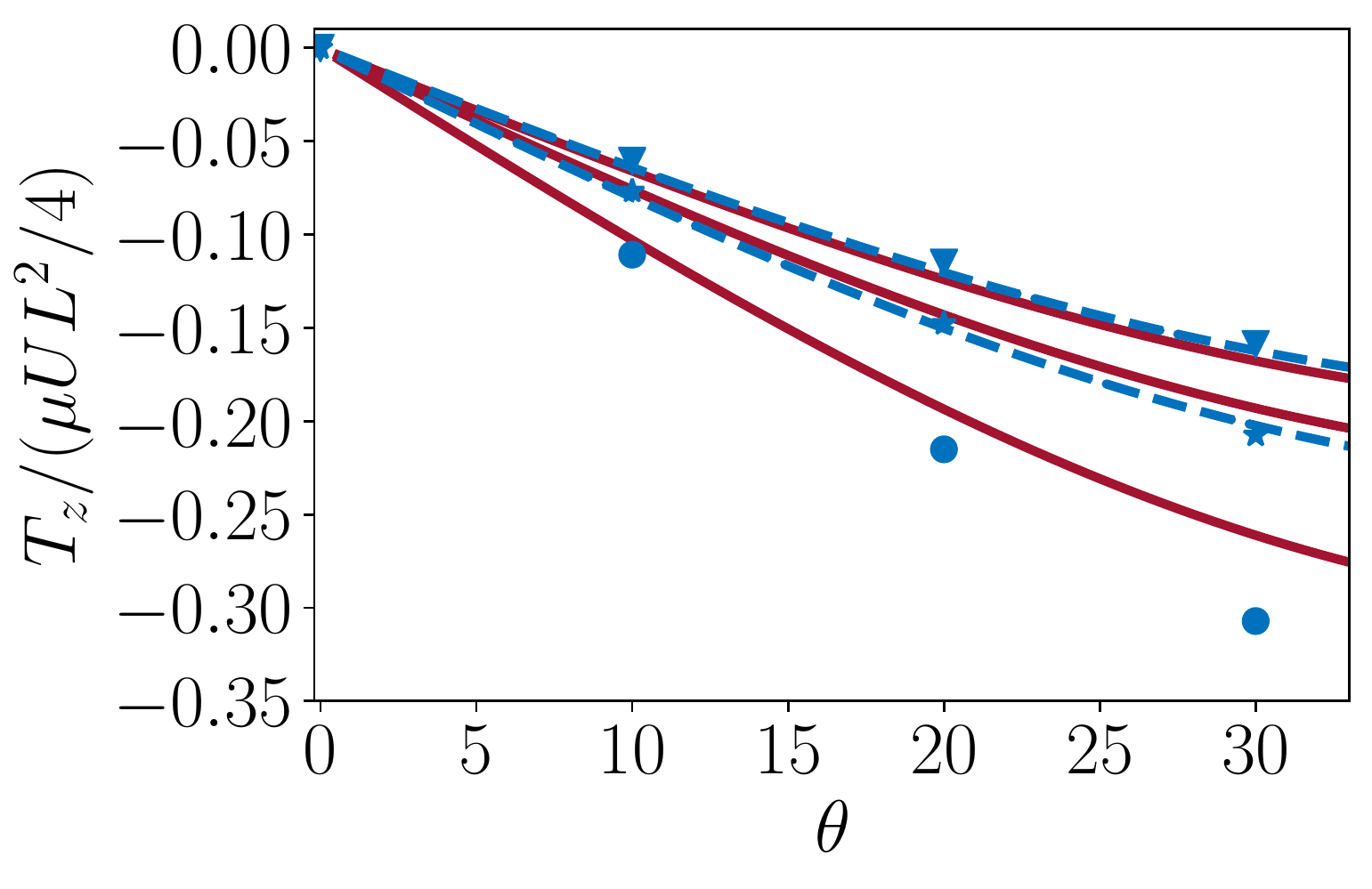}\includegraphics[width=5.5cm]{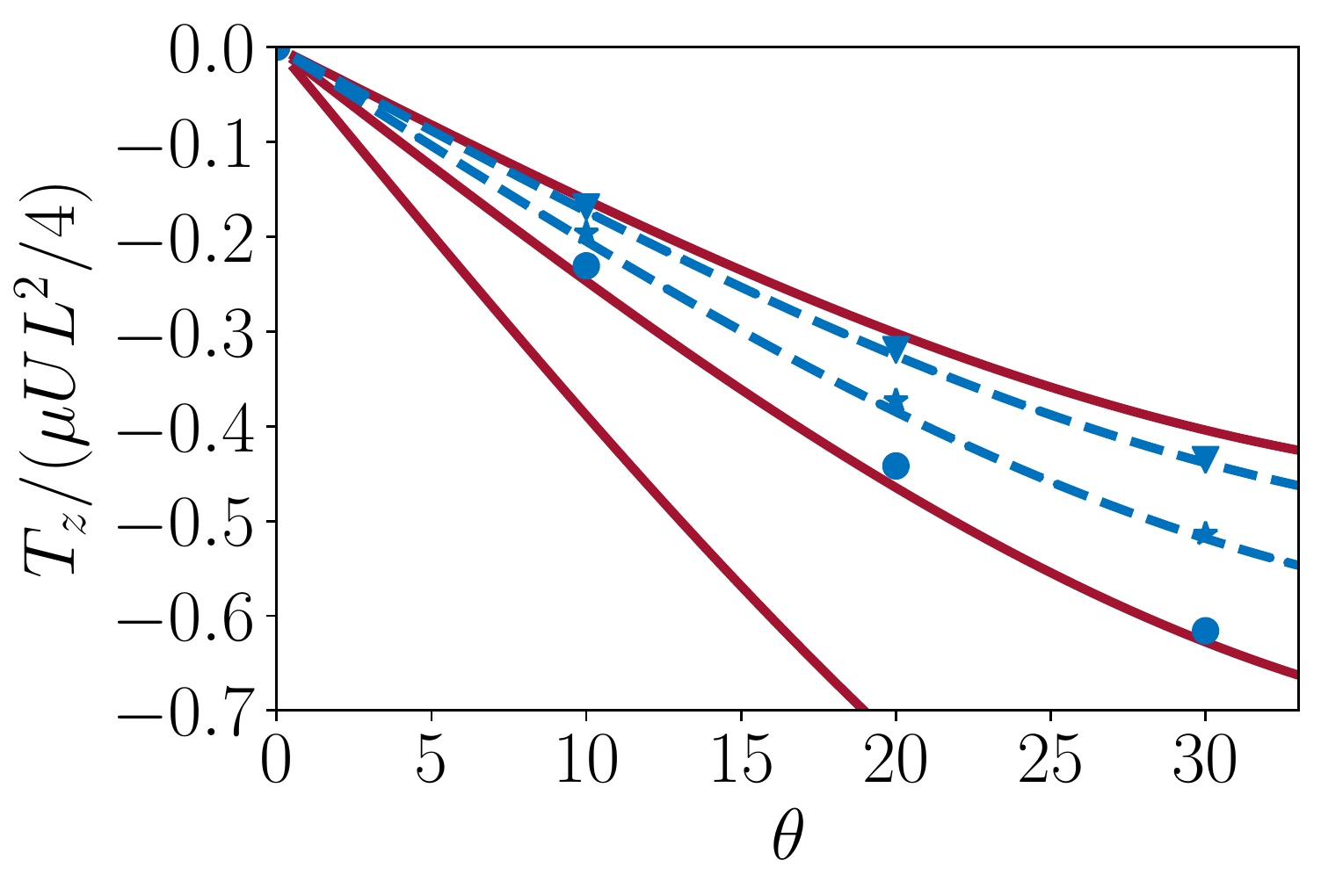}\includegraphics[width=5.5cm]{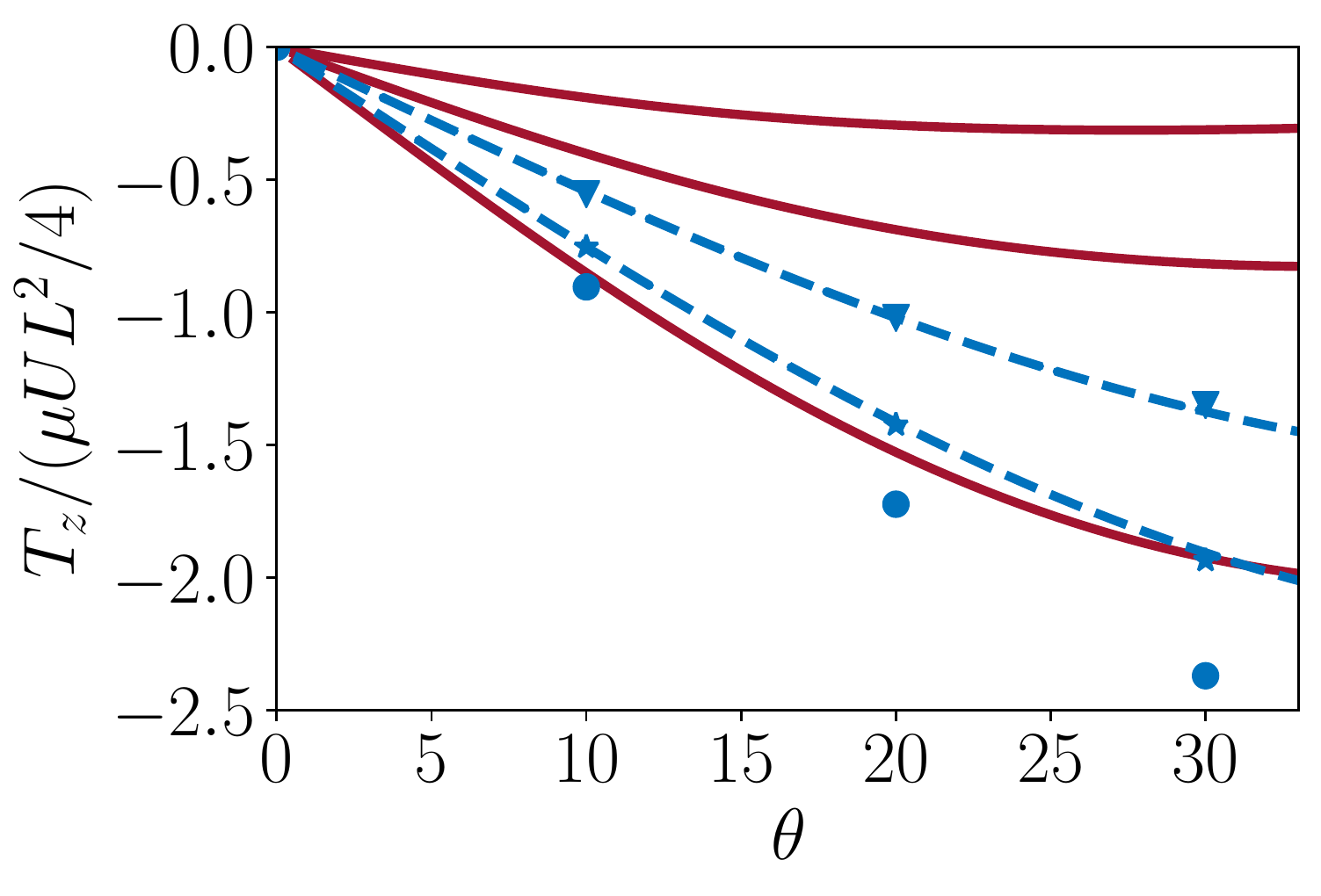}
\vspace{-3mm}\\
\hspace{-35mm}$(a)$\hspace{47mm}$(b)$\hspace{47mm}$(c)$\\
\vspace{-15mm}
\hspace{-14mm}$\text{Re}=0.1$\hspace{42mm}$\text{Re}=0.5$\hspace{42mm}$\text{Re}=5$
\vspace{12mm}\\
\caption{Torque as a function of $\theta$ for $\bullet$: $\cchi=3$, $\star$: $\cchi=5$, $\blacktriangledown$: $\cchi=10$. %$(a)$: $\text{Re}=0.1$, $(b)$: $\text{Re}=0.5$, $(c)$: $\text{Re}=5$. 
 Solid line: asymptotic prediction (\ref{KCcomplet}) for low-but-finite $\text{Re}$; dashed line: approximate fit (\ref{KCJM}) for $\cchi=5$ and $10$.%dash-dotted line in $(c)$: empirical extension (\ref{perp}) of the `Stokes law' to strongly inertial regimes.
 }
\label{fig:T_moderate}
\vspace{-2mm}
\end{figure}

Figure \ref{fig:c_perp_moderate} shows the variations of the perpendicular force component in the same low-to-moderate $\text{Re}$-range. At $\text{Re}=0.1$ (Fig. \ref{fig:c_perp_moderate}$(a)$), $F_\perp$ is accurately approximated for the two cylinders with $\cchi=3$ and $\cchi=5$ by the `Stokes law' based on the creeping-flow prediction (\ref{eq:fperp4m}) for $F_\perp^{ \theta=90^\circ}(\cchi)$. %More precisely, the Stokes law based on the fourth-order creeping-flow prediction (\ref{eq:fperp4m}) is seen to work well up to $\cchi\text{Re}\approx1$, 
This is no longer the case for $\cchi=10$, where a significant underestimate may be noticed. In this case, the length-based Reynolds number $\cchi\text{Re}$ is unity, implying that inertial effects are already significant. This is why the Stokes law based on the composite expression (\ref{eq:slender_perp_smallinertia_m}) for $F_\perp^{ \theta=90^\circ}$, which incorporates  the inertial correction derived in \cite{khayat1989}, closely approaches the numerical data set. The validity of the Stokes law based on (\ref{eq:slender_perp_smallinertia_m}) is maintained for $\text{Re}=0.5$ whatever the aspect ratio.
%For such larger $\cchi\text{Re}$ ((Fig. \ref{fig:c_perp_moderate}$(b)$), the Stokes law based on the composite fourth-order expression (\ref{eq:slender_perp_smallinertia_m}) incorporating  the inertial correction derived in \cite{khayat1989} is observed to \color{red}???? TBC with new post-processed data... \color{black}
%\begin{figure}[h]
%\centering
%\includegraphics[width=5cm]{cy_teta_xR5.pdf}
%\caption{Perpendicular force as function of $\theta$ for $\bullet$: $\cchi=3$, $\star$: $\cchi=5$, $\blacktriangledown$: $\cchi=10$ and Re=5. Dashed dotted line : equation \ref{perp}.}
%\label{fig:c_perp_moderate5}
%\end{figure}
For $\text{Re}=5$ (Fig. \ref{fig:c_perp_moderate}$(c)$), predictions of the same law are found to deviate significantly from numerical results, overestimating $F_\perp$ by more than $10\%$ for $\cchi=3$ and underestimating it by a similar percentage for $\cchi=10$. This is no surprise, since the inertial corrections involved in (\ref{eq:slender_perp_smallinertia_m}) are based on the finite-$\text{Re}$ theory of \cite{khayat1989} which assumes $\text{Re}\ll1$. In this respect, the deviations observed in Fig. \ref{fig:c_perp_moderate}$(c)$ may even be considered as surprisingly small.\vspace{1mm}\\
\indent Variations of the spanwise torque with $\cchi,\,\theta$ and $\text{Re}$ are displayed in Fig. \ref{fig:T_moderate}. As expected, the torque is negative, tending to orient the cylinder broadside on. Variations with the cylinder inclination closely follow the $\sin2\theta$-dependence predicted in the low-$\cchi\text{Re}$ limit by (\ref{KCt0}). The magnitude of the torque increases with the Reynolds number, in line with its inertial nature. For a given $\text{Re}$, the shorter the cylinder the stronger the normalized torque. Numerical results are compared with the full theoretical prediction of \citet{khayat1989}, namely
 \begin{eqnarray}
 \nonumber
 \frac{T(\cchi,\theta,\text{Re}\ll1)}{\mu U(L/2)^2}&=&\frac{2\pi}{(\ln{\cchi})^2}\left\{\cos\theta\left[\mathcal{Z}(X)-\frac{E_1(X)+\ln X+\gamma}{X}+\mathcal{Z}(Y)-\frac{E_1(Y)+\ln Y+\gamma}{Y}\right]\right.\\
&& \bigg.\frac{}{}+\mathcal{Z}(Y)-\mathcal{Z}(X)\bigg\}\sin\theta+\mathcal{O}\left(\frac{1}{(\ln{\cchi})^3}\right)\,,
 \label{KCcomplet}
 \end{eqnarray}
 where $\mathcal{Z}(X)=\frac{2}{X}\left(1+\frac{e^{-X}-1}{X}\right)$, $X=\frac{1}{2}\cchi\text{Re}(1-\cos\theta)$, $Y=\frac{1}{2}\cchi\text{Re}(1+\cos\theta)$ and $E_1$ and $\gamma$ as defined in (\ref{fpKC}). According to Fig. \ref{fig:T_moderate}, this prediction closely approaches the numerical results for the longest cylinder up to $\text{Re=0.5}$, together with those for the intermediate cylinder at $\text{Re}=0.1$. It is no surprise that the low-but-finite $\text{Re}$ theory is unable to provide a reasonable prediction for any of the three cylinders at $\text{Re}=5$. In their experiments,  \citet{roy2019} considered cylindrical fibers with $\text{Re}\approx0.15$. With $\cchi=20$ they found the theory of \cite{khayat1989} to over-predict the torque by more than $20\%$ for $\cchi=20$ at $\theta=30^\circ$, and to slightly under-predict it (by $7-8\%$) for $\cchi=100$. Present results provide a stronger support to the theory, since the difference observed in Fig.  \ref{fig:T_moderate}$(a)$ is only of the order of $5\%$ for $\cchi=10$ and clearly decreases with increasing $\cchi$. The fact that the asymptotic prediction, in which terms of $\mathcal{O}\left((\ln{\cchi})^{-3}\right)$ are neglected, correctly estimates the torque on a $\cchi=5$ cylinder at $\text{Re}=0.1$ but not at $\text{Re}=0.5$, while it still provides an accurate prediction at the same Reynolds number for $\cchi=10$ is noticeable. It suggests that the conditions $\text{Re}\ll1$ and $\cchi\gg1$ on which the asymptotic prediction is grounded must rather been understood as $\text{Re}\ll1$ and $\text{Re}/\cchi^2\ll1$. Indeed, the ratio $\text{Re}/\cchi^2$ stands below $0.005$ in all three configurations correctly predicted by (\ref{KCcomplet}), while it is beyond $0.01$ in all other cases. 
 For practical purposes, we sought an empirical fit of the torque valid for long enough cylinders and reducing to (\ref{KCt0}) (the limit form of (\ref{KCcomplet})) for $\cchi\text{Re}\rightarrow0$. As all three panels in Fig. \ref{fig:T_moderate} indicate, numerical data corresponding to $\cchi=10$ and $\cchi=5$ are accurately approached by the formula
%In the next section we shall show that an empirically modified Stokes law (\ref{perp}) incorporating a boundary-layer type correction works well in the fully inertial regimes corresponding to $\text{Re}\gtrsim10$. As shown in Fig. \ref{fig:c_perp_moderate}$(c)$, (\ref{perp}) actually correctly approximates $F_\perp$ down to $\text{Re}=5$, at least for $\cchi\geq 5$, \textit{i.e.} for $\cchi\text{Re}\gtrsim25$.
\begin{equation}
%\frac{T(\cchi,\theta,\text{Re})}{\mu U(L/2)^2}\approx-\frac{5\pi}{12}\frac{\text{Re}}{(1+10.5\text{Re})^{0.55}}\left\{\frac{\cchi}{(\ln\cchi)^2}+\frac{3.85e^{-6.35\text{Re}}+0.28\text{Re}}{(\ln\cchi)^3}\right\}\sin2\theta\,.
%\frac{T(\cchi,\theta,\text{Re})}{\mu U(L/2)^2}\approx-\frac{5\pi}{12}\frac{\text{Re}}{(1+\cchi\text{Re})^{0.5}}\left\{\frac{\cchi}{(\ln\cchi)^2}+\frac{1.7e^{-10\text{Re}/\cchi}}{(\ln\cchi)^3}\right\}\sin2\theta\,.
\frac{T(\cchi\gg1,\theta,\text{Re})}{\mu U(L/2)^2}\approx-\frac{5\pi}{12}\frac{\text{Re}}{(1+\cchi\text{Re}^{1.1})^{0.5}}\left\{\frac{\cchi}{(\ln\cchi)^2}+\frac{13.5-30\,\text{Re}^{1/2}}{\cchi (\ln\cchi)^3}e^{-0.7\text{Re}}\right\}\sin2\theta\,.
\label{KCJM}
\end{equation}
The $\mathcal{O}\left(\cchi^{-1}(\ln{\cchi})^{-3}\right)$-term only provides a marginal contribution  for $\cchi=10$. Consequently the leading-order $\mathcal{O}\left(\cchi(\ln{\cchi})^{-2}\right)$-truncation of (\ref{KCJM}) is sufficient to correctly estimate the torque on cylindrical fibers with $\cchi\gtrsim10$ up to $\text{Re}=5$. A correction proportional to $(\ln{\cchi})^{-4}$ may certainly be incorporated in (\ref{KCJM}) to properly estimate the torque on very short cylinders.

%\subsection{Approximate laws for the torque} 

\section{Fully inertial stationary regime}
\label{incl2}
\color{black}
As pointed out in Sec. \ref{moder}, the flow past a cylinder aligned with the incoming velocity is stationary and axisymmetric within the full range of $\text{Re}$ investigated here. %Three different stationary patterns are encountered : steady toroidal vortex, steady shedding of one counter-rotating vortex pair with planar symmetry, steady shedding of single-sided hairpin vortices. Two unsteady regimes are also encountered : the periodic shedding of single-sided and double-sided of hairpin vortices. Those periodic regimes can be either governed by a unique frequency or by two or more frequencies. 
 Although the axial symmetry breaks down when $\theta$ is nonzero, the flow remains stationary up to \color{black} a critical Reynolds number, $\text{Re}_c(\cchi,\,\theta)$, larger than $300$. Details on the first unsteady regimes that take place beyond this threshold are provided as Supplemental Material. Here we concentrate on the stationary non-axisymmetric regime extending from Reynolds numbers of $\mathcal{O}(10)$ up to $\text{Re}_c$. \color{black}
To limit the computational cost, simulations in this regime were only carried out for cylinders with aspect ratios $\cchi=3,5$ and $7$. However, the results to be discussed hereinafter suggest that the flow structure and loads are only weakly affected by finite-length effects beyond $\cchi\approx5$ in this range of Reynolds number, although %By this, we mean for instance that effects taking place close to the ends of the cylinder only marginally affect the loads for $\cchi\approx7$. However, 
the inertial load coefficients defined below may well continue to depend on $\cchi$, even for $\cchi\gg1$. Consequently, the physical phenomena involved and the simulation-based approximate expressions for the loads provided below are expected to apply without significant changes to cylinders with $\cchi\gg1$, which makes them relevant to analyze the motion of long cylindrical fibers in inertia-dominated regimes.\\
\indent \color{black} In Sec. \ref{moder}, we showed that, for Reynolds numbers of $\mathcal{O}(10^2)$, the flow structure past a cylinder aligned with the incoming velocity exhibits the presence of a thin annular standing eddy along the upstream part of the lateral surface. This feature was found to significantly influence the drag force, being able to change the sign of the viscous drag for sufficiently short cylinders and large Reynolds numbers. The situation is qualitatively similar in the $(\theta,\,\text{Re})$-range considered hereinafter since, for large enough inclinations and Reynolds numbers, the flow field exhibits specific features which directly affect the loads on the body. Consequently, it is relevant to examine first how the flow close to the body varies with the control parameters, which is the purpose of Sec. \ref{separ}. Then, the possibility to use numerical data to derive simple laws for the force components is considered in Sec. \ref{stokeslaw}, before empirical fits for the axial force and spanwise torque are built in Sec. \ref{approxpt}.\color{black}
\begin{figure}[H]
\centering
\raisebox{0.4in}{}\includegraphics[scale=0.05]{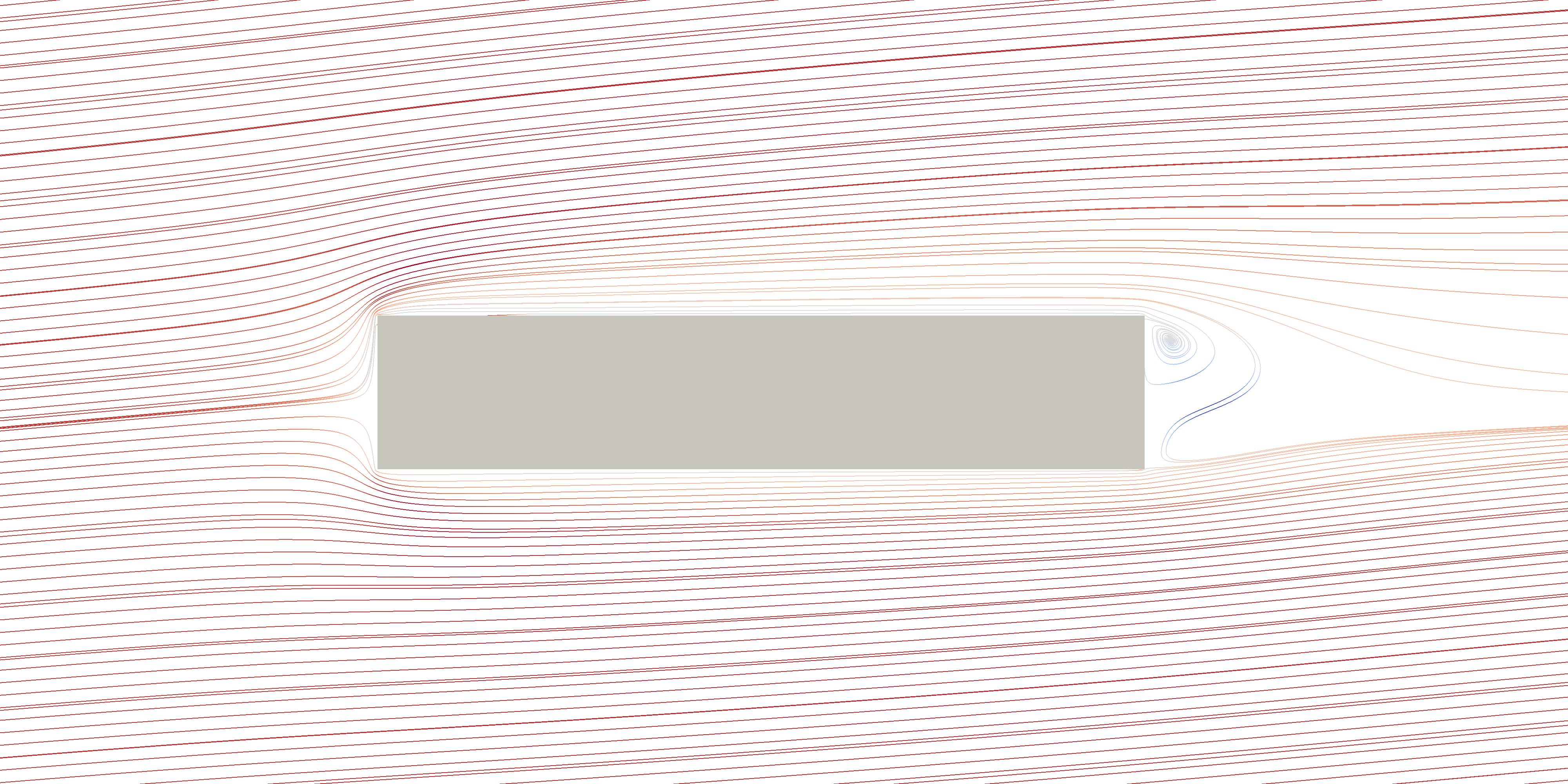}\quad
 \raisebox{0.4in}{}\includegraphics[scale=0.05]{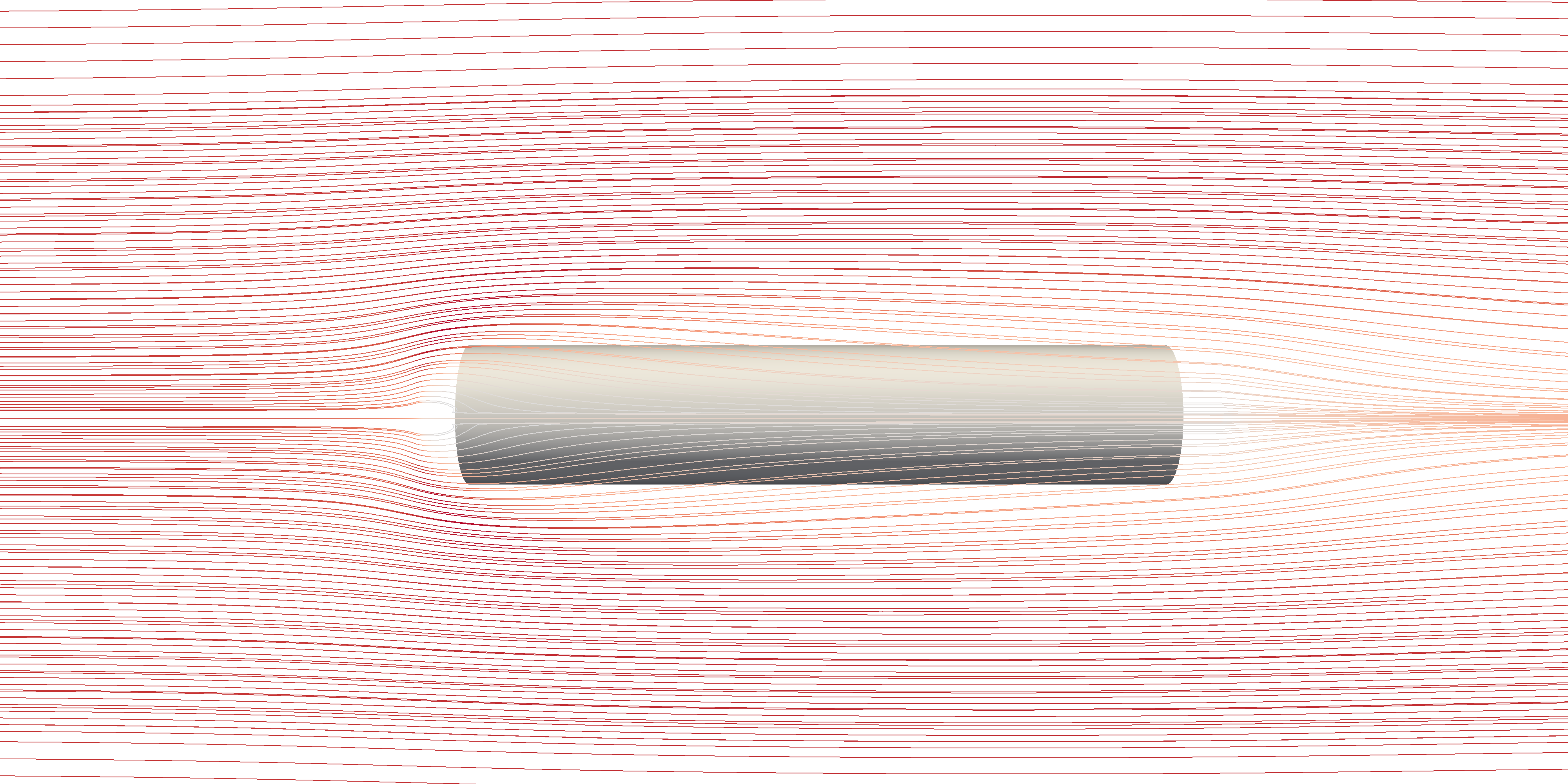}\\
 \vspace{-18mm}\hspace{-3.6cm}{\small{$S^-$}}\\
 \vspace{-6mm} \hspace{-1.2cm}{\small{$fs^-$}}\\
 \vspace{15mm} \hspace{-4cm}{$(a)$}\hspace{7cm}{$(b)$}\\
 \vspace{2mm}
\raisebox{0.4in}{}\includegraphics[scale=0.05]{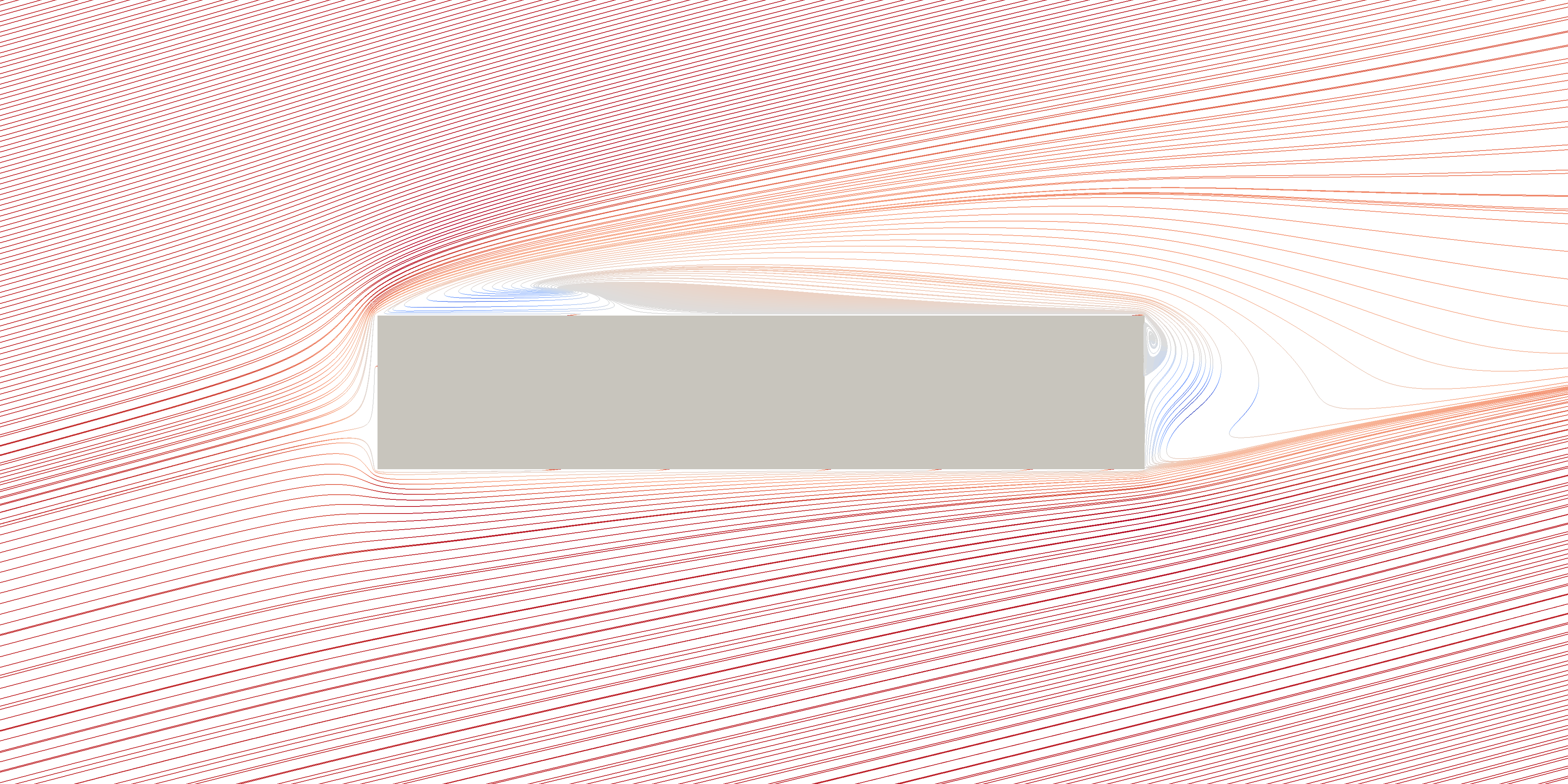}\quad
\raisebox{0.4in}{}\includegraphics[scale=0.05]{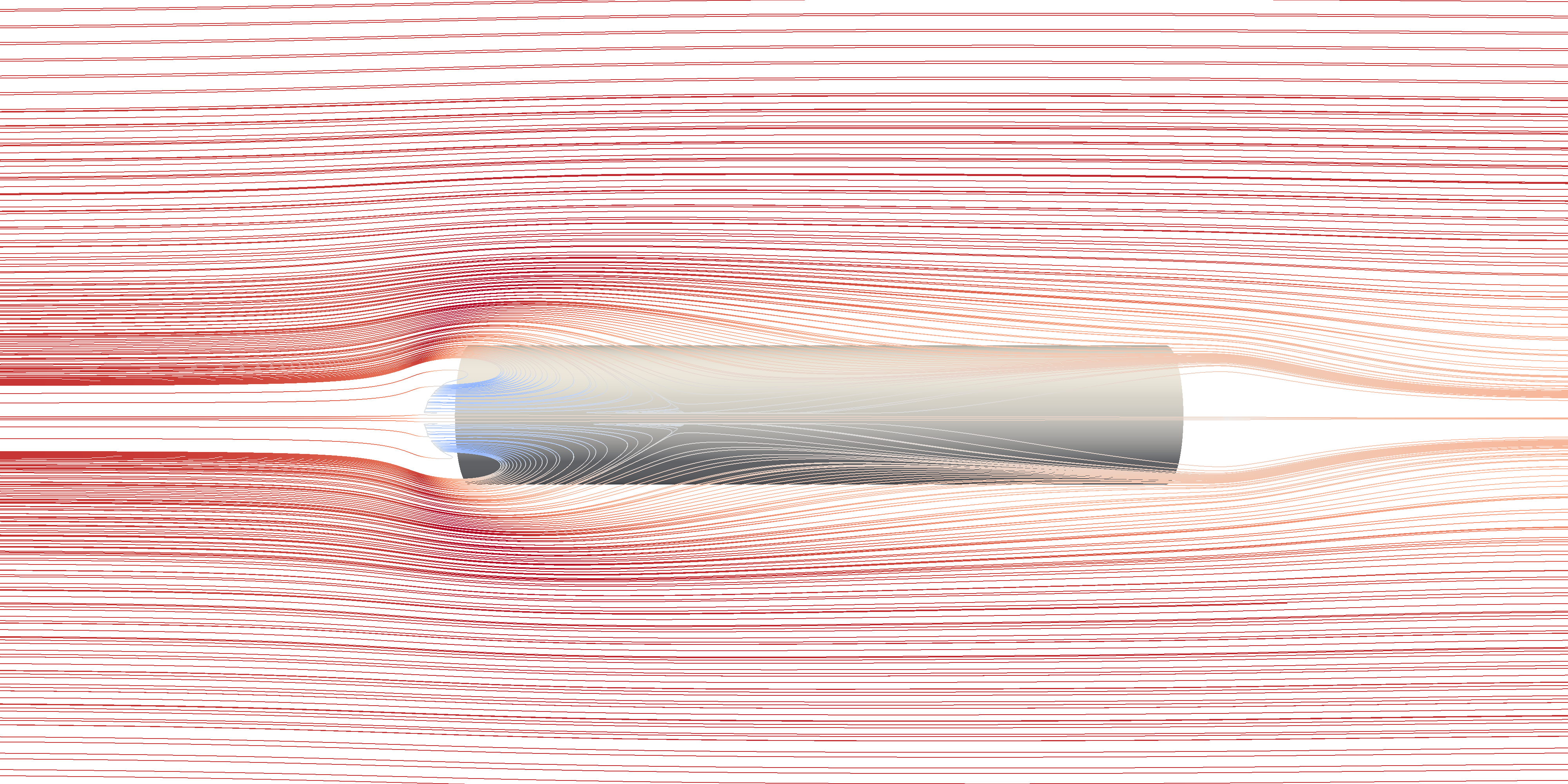}\\
\vspace{-27.2mm}\hspace{-1.2cm}{\small{$fs^+$}}\\
\vspace{-3.7mm}\hspace{-9.8cm}{\small{$E$}}\\
\vspace{-1.5mm}\hspace{-10cm}{\small{$S^+$}}\hspace{0.8cm}{\small{A}}\\
\vspace{1mm}\hspace{-1.2cm}{\small{$fs^-$}}\\
\vspace{-9.2mm}\hspace{5cm}{$\mathcal{R}_b$}\\
\vspace{3.2mm}\hspace{5.2cm}{\small{$S^+$}}\hspace{0.9cm}{\small{A}}\\
\vspace{2.5mm}\hspace{5cm}{$\mathcal{R}_f$}\\
\vspace{8.5mm}
 \hspace{-4cm}{$(c)$}\hspace{7cm}{$(d)$}\\
 \vspace{2mm}
\raisebox{0.4in}{}\includegraphics[scale=0.05]{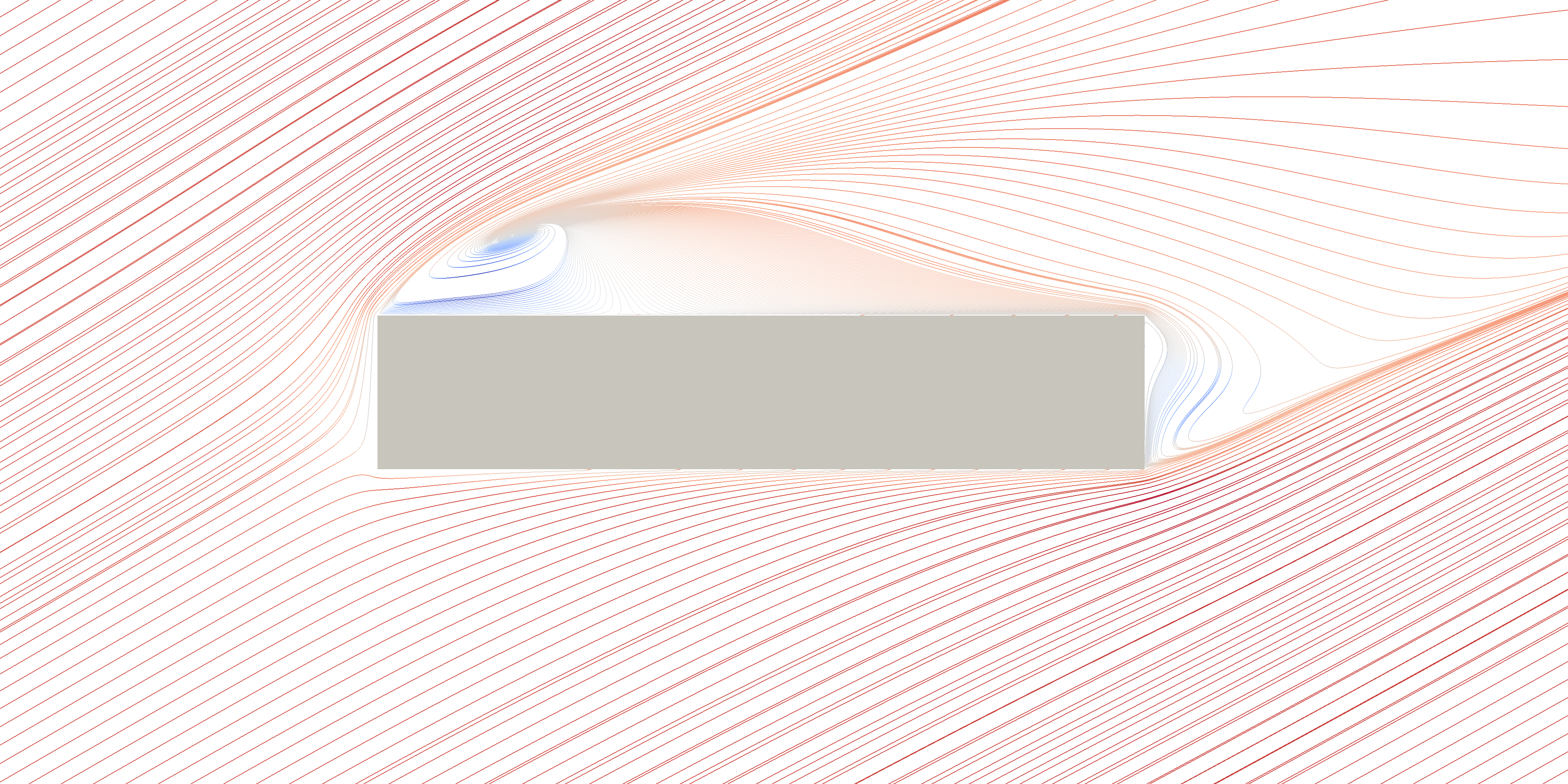}\quad
\raisebox{0.4in}{}\includegraphics[scale=0.065]{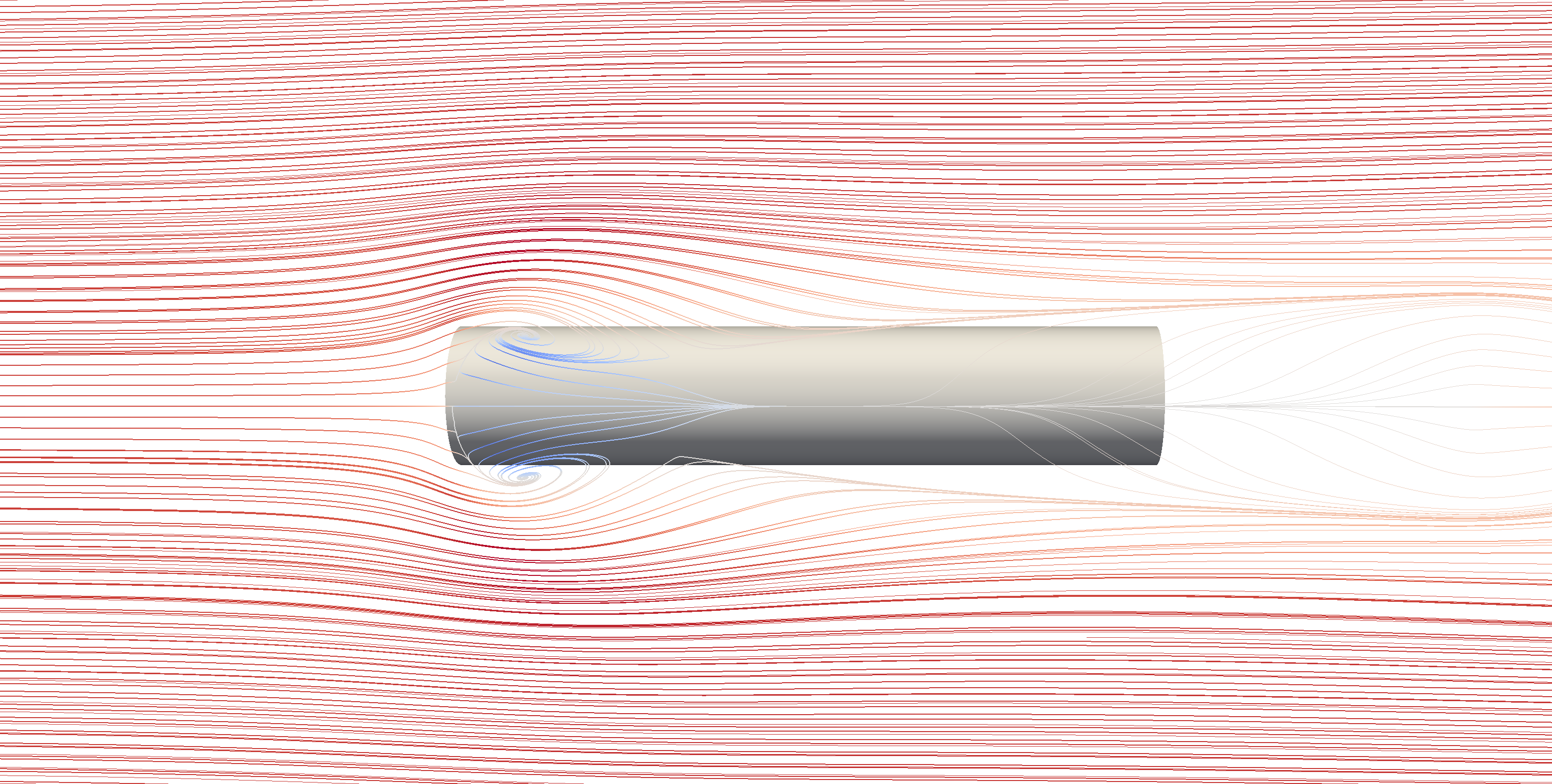}\\
\vspace{-36mm}\hspace{-5.5cm}{\small{$fs^+$}}\\
\vspace{3mm}\hspace{-9.8cm}{\small{$E$}}\\
\vspace{0.25mm}\hspace{-10cm}{\small{$S^+$}}\hspace{0.9cm}{\small{A}}\\
\vspace{-2.5mm}\hspace{-1.2cm}{\small{$fs^-$}}\\
\vspace{-7.5mm}\hspace{5cm}{$\mathcal{R}_b$}\\
\vspace{3.8mm}\hspace{5.5cm}{\small{$S^+$}}\hspace{1.2cm}{\small{A}}\\
\vspace{2.9mm}\hspace{5cm}{$\mathcal{R}_f$}\\
\vspace{9.4mm}
 \hspace{-4cm}{$(e)$}\hspace{7cm}{$(f)$}\\
 \vspace{2mm}
%\raisebox{0.4in}{}\includegraphics[scale=0.035]{ldc_x5t15R140}\quad
 %\hspace{-1cm}{(a)}\hspace{5cm}{(b)}\hspace{4.5cm}{(c)}\\
\vspace{0mm}
% \hspace{-1cm}{(d)}\hspace{5cm}{(e)}\hspace{4.5cm}{(f)}
	\caption{Streamlines in the symmetry plane $z=0$ (left) and in the plane $y=D/2$ tangent to the lateral surface along the top generatrix (right) for a cylinder with $\cchi=5$ at $\text{Re}=140$. Streamlines are colored according to the magnitude and sign of the axial velocity from $-0.25$ (deep blue) to $+1$ (deep red). % $(a)$: $\theta=5^\circ$; $(b)$: $\theta=15^\circ$; $(c)$: $\theta=30^\circ$. 
	$(a)-(b)$: $\theta=5^\circ$; $(c)-(d)$: $\theta=15^\circ$; $(e)-(f)$: $\theta=30^\circ$. 
\label{fig:ldc140}}
\end{figure}
\subsection{Flow structure}
\label{separ}
\begin{figure}[h]
%\vspace{-3mm}
\centering
\raisebox{0.4in}{}\includegraphics[scale=0.07]{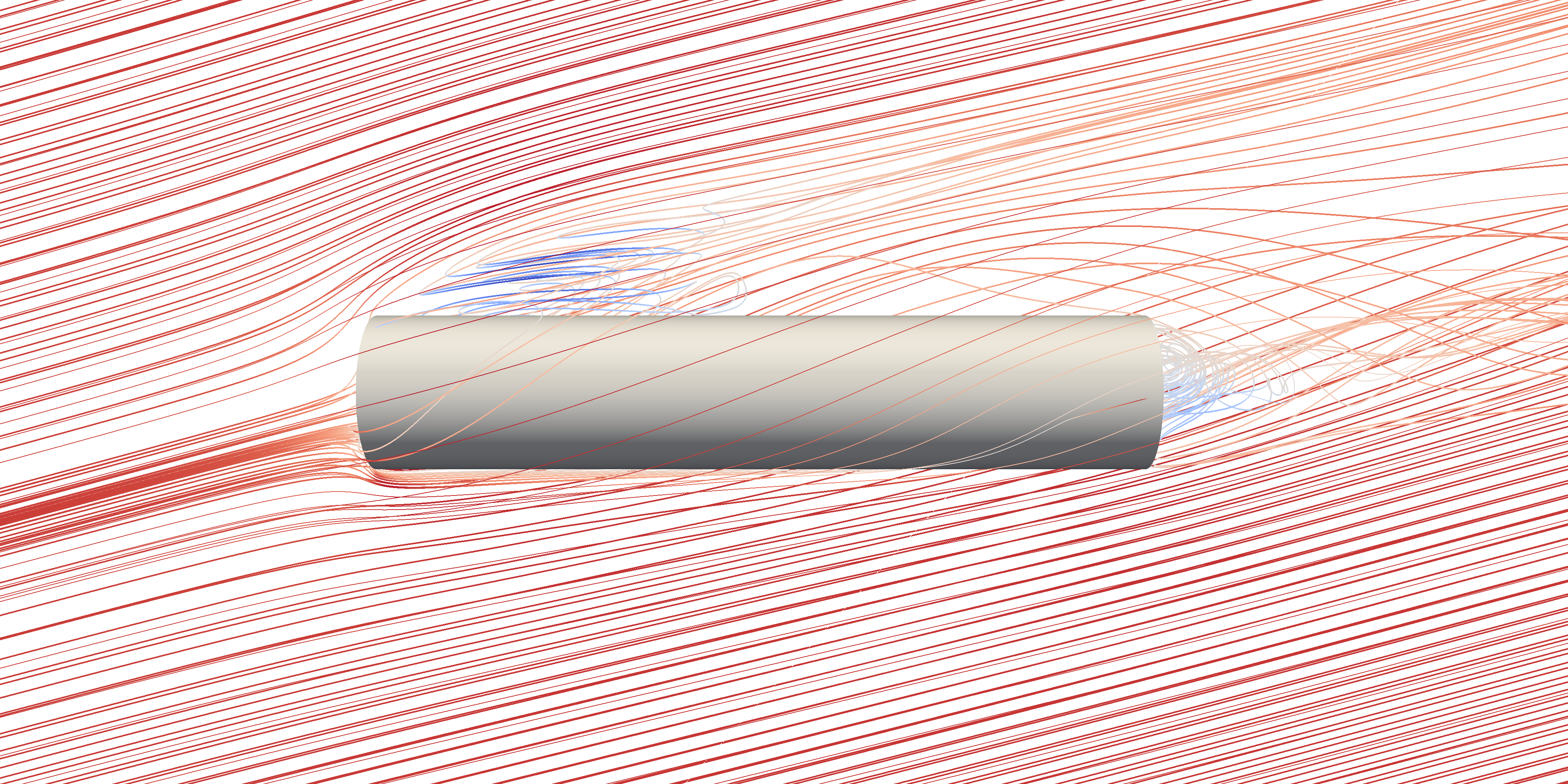}
\caption{Three-dimensional streamlines past an inclined cylinder with $\cchi=5$, $\theta=15^\circ$ and $\text{Re}=300$. 
\label{fig:ldc3d}}
\vspace{-3mm}
\end{figure}

Throughout the regime under consideration, the flow field exhibits a mirror symmetry with respect to the $(x,y)$-plane which contains both the body axis and the incoming velocity. %Consequently, the force on the cylinder involves a nonzero steady lift component perpendicular to the upstream flow and standing in the symmetry plane, and a torque in the spanwise $z$-direction. 
Figure \ref{fig:ldc140} helps to understand how the flow structure past the cylinder varies with the inclination angle.
% encountered under most conditions in the present study (crosses in Fig. \ref{fig:teta_Re}). 
%Lower- and higher-$\text{Re}$ configurations exhibit qualitatively similar behaviors, with of course a shift in the typical inclination at which  
Several generic features emerge. First, the front stagnation point standing in the symmetry plane $z=0$ is seen to move downward (\textit{i.e.} toward negative $y$) as $\theta$ increases, almost reaching the bottom generatrix ($y=-D/2)$ for $\theta=30^\circ$ whatever $\text{Re}$. At the back of the body, the fluid that has passed over the upper part of the lateral surface is entrained downwards. It recirculates toward the downstream end within a region whose length in the streamwise $(x)$ direction is typically of the order of the cylinder radius for $\text{Re}=140$. Examination of this recirculating region at other Reynolds numbers (not shown) indicates that its length increases gradually with $\text{Re}$, becoming of the order of $D$ for $\text{Re}=300$. Unlike the standing eddy existing in the axisymmetric configuration (Sec. \ref{moder}), this recirculating zone is no longer a closed toroid. Indeed, once the fluid entrained downward gets very close to the lowest generatrix (point $S^-$ in Fig. \ref{fig:ldc140}$(a)$), it is expelled downstream in the main flow, just above the open streamline $fs^-$ emanating from  $S^-$. Out of the symmetry plane, the three-dimensional streamlines displayed in Fig. \ref{fig:ldc3d} (for $\text{Re}=300$, to magnify the regions in which the fluid recirculates) reveal that fluid particles entrapped in the recirculating region describe successive loops before being sucked into the wake and advected downstream. This scenario is similar to that observed in \cite{johnson1999} at the back of a sphere in the first (steady) non-axisymmetric wake regime.

For $\text{Re}=140$ and $\theta=5^\circ$ (Fig. \ref{fig:ldc140}$(a)$), the flow along the body remains attached everywhere to the lateral surface. The lower (resp. higher) the Reynolds number, the larger (resp. smaller) the critical inclination at which separation occurs along the upper part of this surface. For instance, the flow is still unseparated at $\theta=15^\circ$ for $\text{Re}=40$ but is already separated at $\theta=5^\circ$ for $\text{Re}=300$. Separation starts at the intersection of the upstream end and the upper generatrix (point $S^+$ in Figs. \ref{fig:ldc140}$(c)-(f)$). Beyond the corresponding critical $\theta$ and/or $\text{Re}$, separation takes place over an open surface of finite extent, the most downstream point of which (point $A$ in Figs. \ref{fig:ldc140}$(c)-(f)$) stands on the symmetry plane. Still in the plane $y=D/2$, the separating line starting at $A$ develops upstream on the sides of the cylinder (regions $\mathcal{R}_b$ and $\mathcal{R}_f$ in Figs. \ref{fig:ldc140}$(d)$ and $(f)$) before joining the incoming flow. Not visible in Fig. \ref{fig:ldc140}, the lower part of the separation surface ($y<D/2$) on the sides of the cylinder exhibits a structure qualitatively similar to that of regions $\mathcal{R}_b$ and $\mathcal{R}_f$ and ends in some intermediate plane $0<y<D/2$, the position of which depends on $\theta$ and $\text{Re}$. Moving away from the lateral surface above the upper generatrix ($y>D/2$), the extent of the separation surface in the spanwise direction decreases gradually until its trace  reduces to a single point some distance above the cylinder. This apex (point $E$ in Figs. \ref{fig:ldc140}$(c)$ and $(e)$) looks like the `eye of the storm' of the separated region. The larger $\theta$ and $\text{Re}$, the larger the distance between points $S^+$ and $E$ in both the $x$- and $y$-directions. Fluid is brought in the neighborhood of $E$ from both sides of the cylinder in a way similar to that observed in Figs. \ref{fig:ldc140}$(d)$ and $(f)$ in the plane $y=D/2$ (see Fig. \ref{fig:ldc3d}). Then it is sent back toward the upstream end in between the `eye' and the cylinder. The dividing streamline joining the region of the `eye' to point $A$ gets very close to the uppermost point $S^+$ of the upstream end. There, fluid particles are deviated by the `fresh' fluid flowing along the free streamline $fs^+$ and advected downstream, just below this free streamline. Streamlines that pass closer to to the `eye' stay further away from the cylinder surface within the recirculating region. Consequently they also stay further away from $fs^+$ once they escape this region, the corresponding fluid filling the intermediate region in between $fs^+$ and the cylinder at the back of the separation surface. Overall, the flow past the cylinder looks massively separated in between the free streamlines $fs^-$ and $fs^+$, the position of the `eye' governing the flow structure in the intermediate region. A similar open separation configuration was recently observed over inclined prolate spheroids in the range $5\leq \text{Re}\leq100$ in \cite{frohlich2020}.
%\begin{figure}[H]
%\centering
%\raisebox{0.4in}{}\includegraphics[scale=0.035]{ldc_x5t5R40}\quad
%\raisebox{0.4in}{}\includegraphics[scale=0.05]{ldc_x5t15R40}\quad
%\raisebox{0.4in}{}\includegraphics[scale=0.05]{ldc_x5t30R40}\\
% \hspace{-1cm}{(a)}\hspace{5cm}{(b)}\hspace{4.5cm}{(c)}\\
 %\hspace{1cm}{(a)}\hspace{7cm}{(b)}\\
 %\vspace{2mm}
% \raisebox{0.4in}{}\includegraphics[scale=0.035]{ldcxz_x5t5R40}\quad
%\raisebox{0.4in}{}\includegraphics[scale=0.05]{ldcxz_x5t15R40}\quad
%\raisebox{0.4in}{}\includegraphics[scale=0.05]{ldcxz_x5t30R40}\\
%\vspace{1mm}
% \hspace{-1cm}{(d)}\hspace{5cm}{(e)}\hspace{4.5cm}{(f)}
 %\hspace{1cm}{(a)}\hspace{7cm}{(b)}\\
%	\caption{Streamlines in the symmetry plane $z=0$ (top) and in the plane $y=D/2$ tangent to the lateral surface along the top generatrix (bottom) for $\cchi=5$ and $\text{Re}=40$. Streamlines are colored according to the magnitude and sign of the axial velocity from $-0.25$ (deep blue) to $+1$ (deep red).% $(a)$: $\theta=5^\circ$; $(b)$: $\theta=15^\circ$; $(c)$: $\theta=30^\circ$. 
%	$(a)$: $\theta=15^\circ$; $(b)$: $\theta=30^\circ$.
%			\label{fig:ldc40}}
%\end{figure}
\subsection{Are `Stokes laws' valid for inclined cylinders at moderate-to-large Reynolds number?}
\label{stokeslaw}
\color{black} We now make use of the numerical results to build approximate force and torque laws valid for $|\theta|\lesssim30^\circ$ and $\text{Re}\gtrsim10$ in the stationary non-axisymmetric flow regime.  For this purpose, we characterize the force components through the parallel and perpendicular force coefficients defined \textit{via} the identify $(F_{\perp},F_{\parallel})\equiv(C_{\perp},C_{\parallel})\rho U^2LD/2$. Similarly, we introduce the torque coefficient $C_{t}$ related to the spanwise torque $T$ through $T\equiv C_{t}\rho U^2L^2D$. \color{black}
%\noindent In what follows, force coefficients are defined by normalizing the corresponding force component by the quantity ${\frac{1}{2}LD\rho U^2}$ (not by ${\frac{\pi}{8}D^2\rho U^2}$ any more), while the torque coefficient is obtained by normalizing the spanwise torque by ${\frac{1}{2}L^2D\rho U^2}$.

\begin{figure}
	\centering
	\begin{tabular}{cccc}
		\includegraphics[width=5cm]{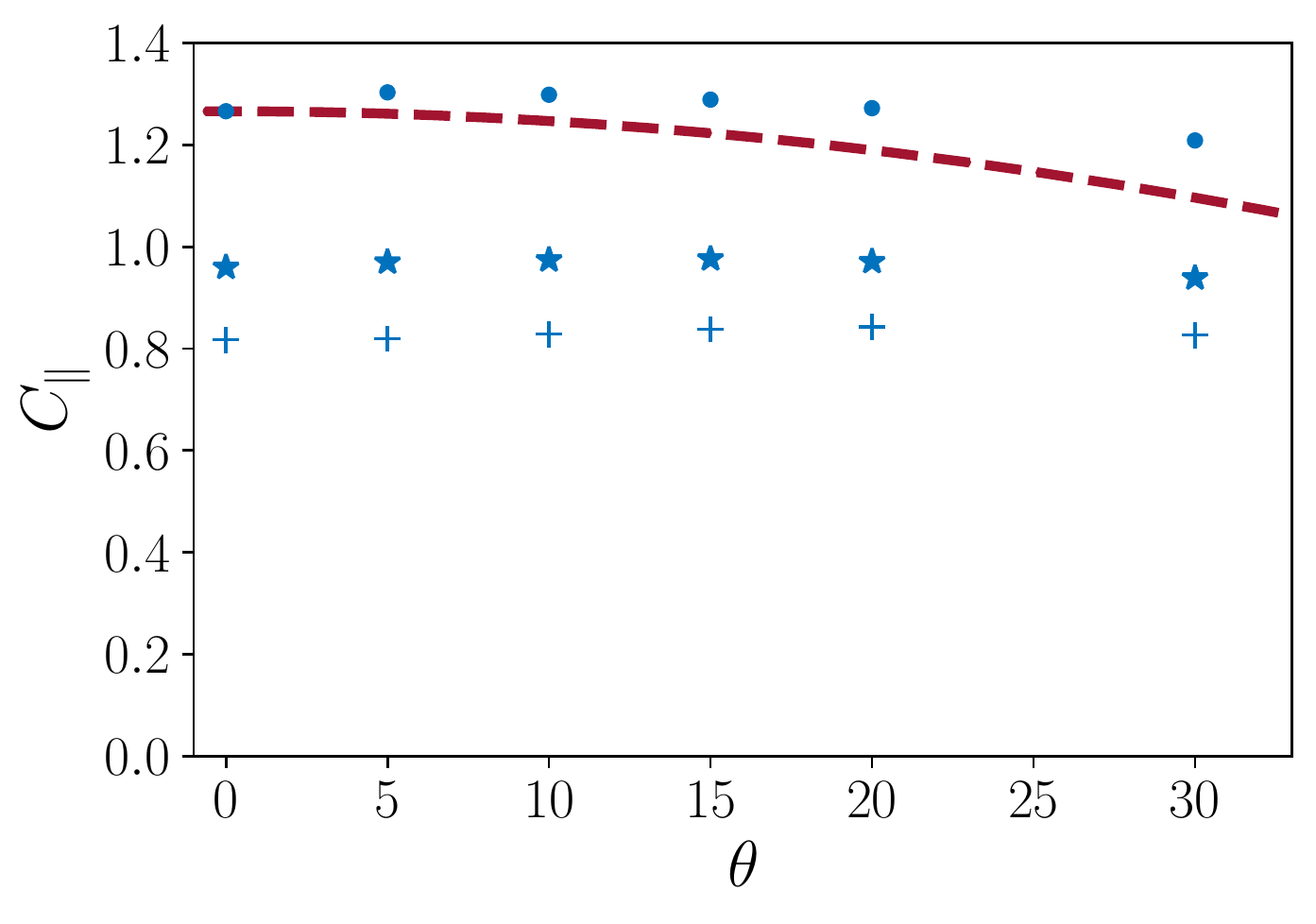}&
		\includegraphics[width=5cm]{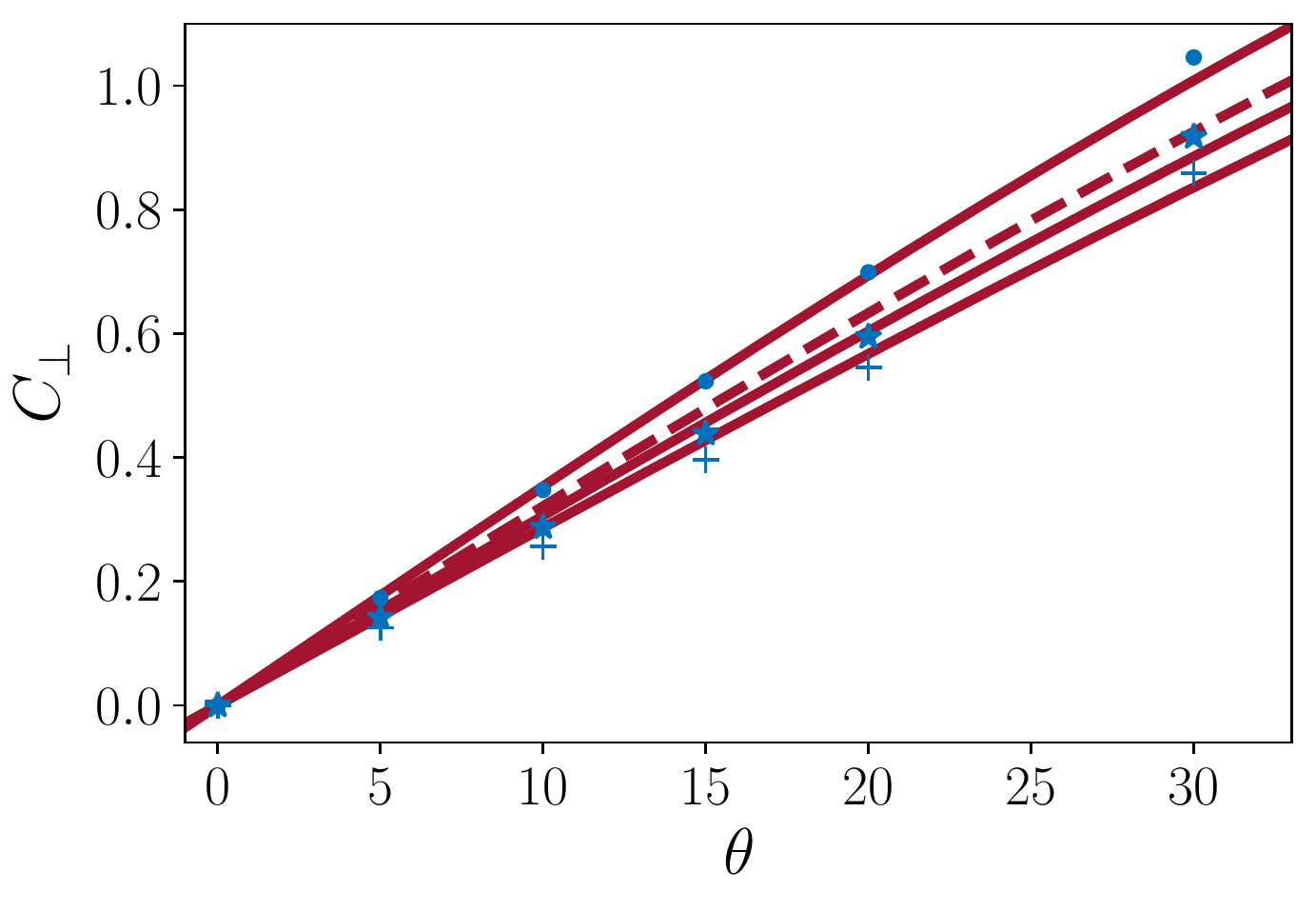}&
		&
				\vspace{-3mm}
		\\
		\hspace{-30mm}$(a)$ & \hspace{-30mm}$(b)$ \\
		\vspace{-18mm}\\
	\hspace{28mm}$\text{Re}=20$&
	\hspace{33mm}$\text{Re}=20$
	\vspace{12mm}\\
	\end{tabular}
	\centering
	\begin{tabular}{cccc}
		\includegraphics[width=5cm]{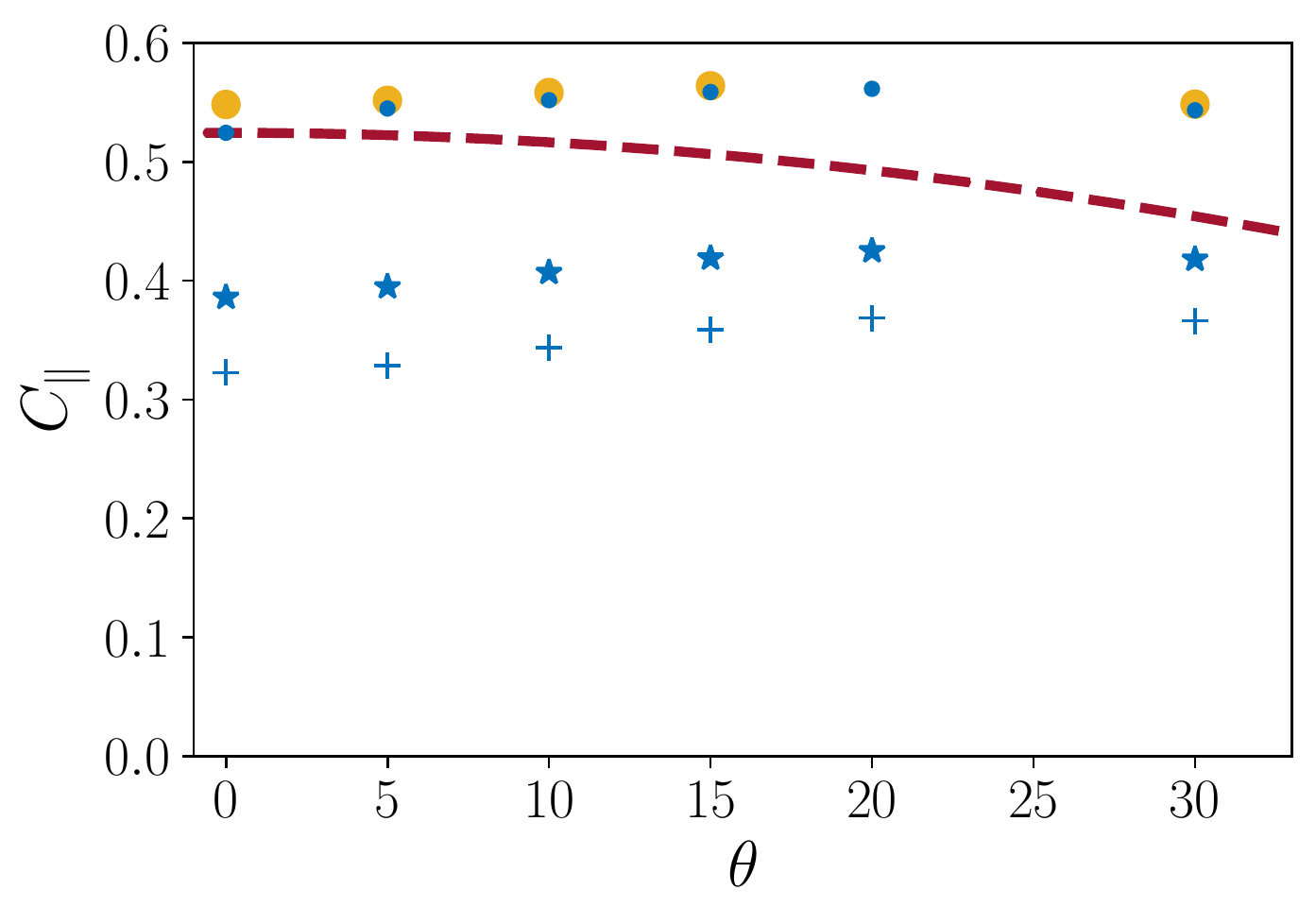}&
		\includegraphics[width=5cm]{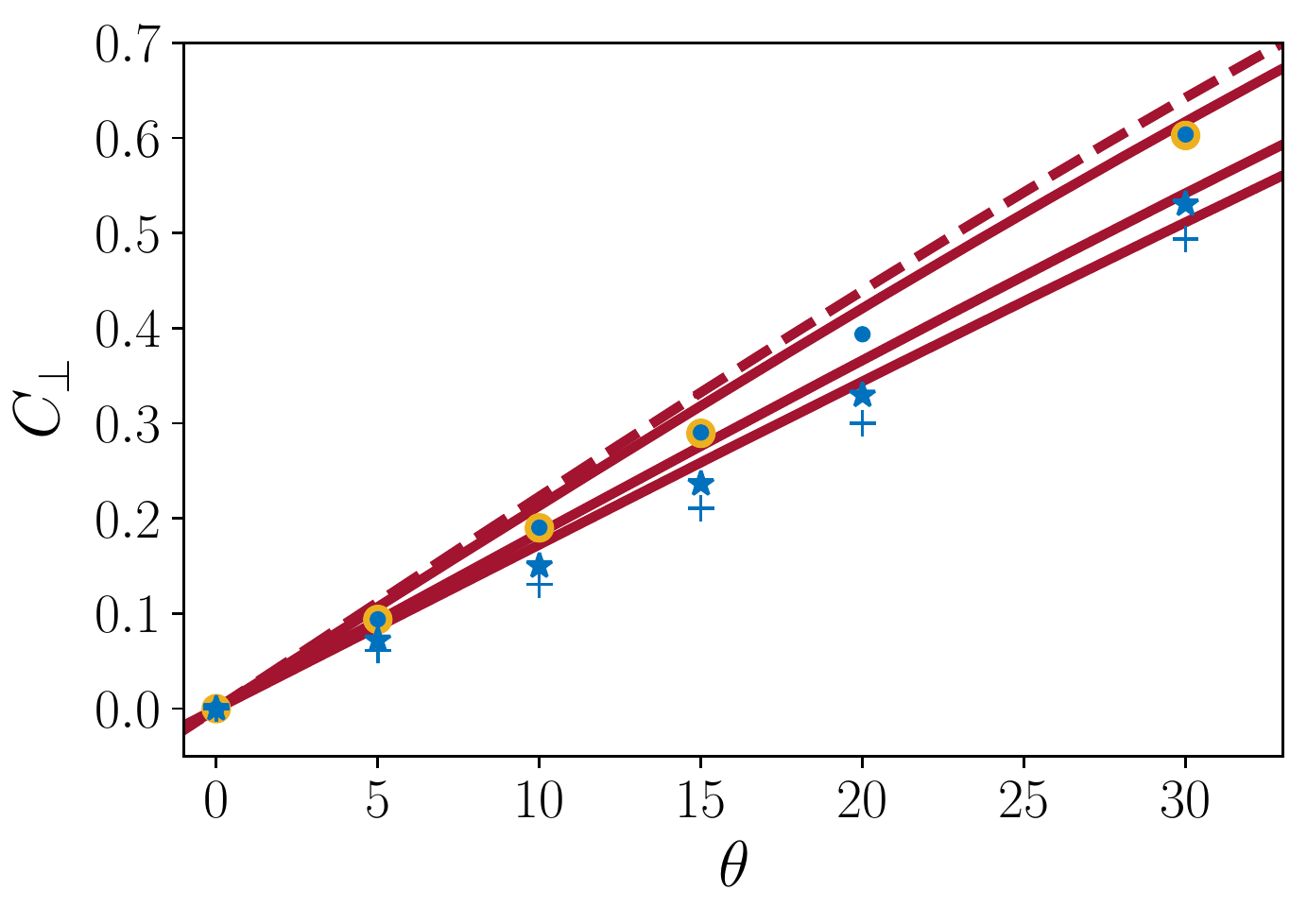}&
		&
		\vspace{-3mm}
		\\
		\hspace{-30mm}$(c)$ & \hspace{-30mm}$(d)$ \\
		\vspace{-18mm}\\
	\hspace{28mm}$\text{Re}=80$&
	\hspace{33mm}$\text{Re}=80$
	\vspace{12mm}\\	
	\end{tabular}
	\centering
	\begin{tabular}{cccc}
		\includegraphics[width=5cm]{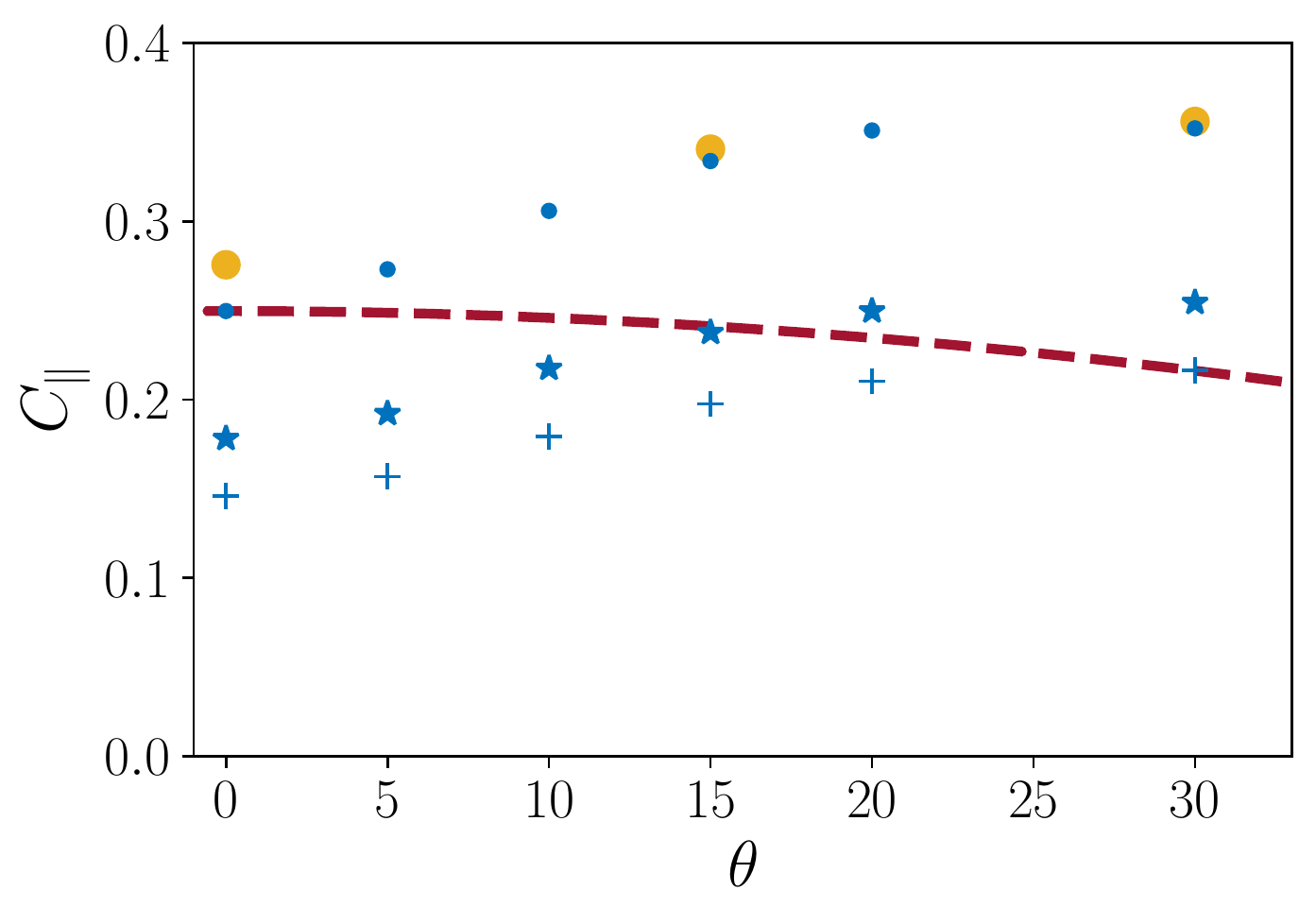}&
		\includegraphics[width=5cm]{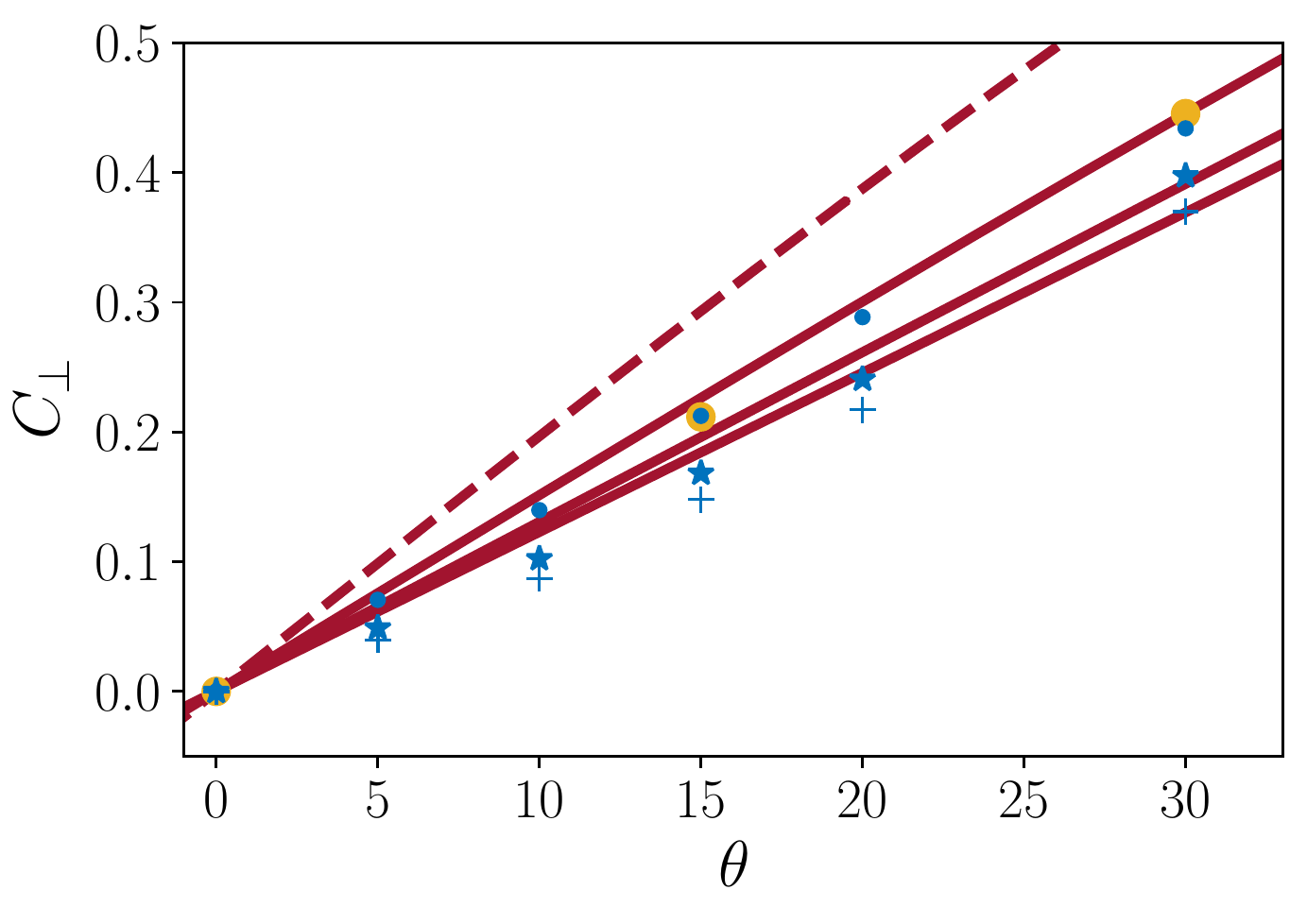}&
		&
	\vspace{-3mm}
		\\
		\hspace{-30mm}$(e)$ & \hspace{-30mm}$(f)$ \\
		\vspace{-18mm}\\
	\hspace{28mm}$\text{Re}=300$&
	\hspace{33mm}$\text{Re}=300$
	\vspace{12mm}\\
	\end{tabular}
	\caption{ Parallel and perpendicular force coefficients ($C_{\parallel},C_{\perp}$ ) vs. $\theta$ for $\bullet$: $\cchi=3$, $\star$: $\cchi=5$, +: $\cchi=7$. %$(a)-(b)$: $\text{Re}=20$; $(c)-(d)$: $\text{Re}=80$; $(e)-(f)$: $\text{Re}=300$; 
	Dashed lines: `Stokes laws' (\ref{eq:Cpara_stokes})-(\ref{eq:Cperp_stokes}) based on the value of $C_\parallel^{ \theta=0^\circ}(\cchi=3)$ and $C_\perp^{ \theta=90^\circ}(\cchi\gg1)$ at the relevant Reynolds number;  solid line: approximate fit  (\ref{corrected}). %In panel $(b)$, the `Stokes law' is virtually superimposed to the fit for $\cchi=5$. 
	The yellow bullets in panels $(c)-(d)$ and $(e)-(f)$ refer to the results of \cite{pierson2019} for $\cchi=3$ at $\text{Re}=75$ and $250$, respectively. 
		\label{fig18}}
\end{figure}

\begin{figure}%
 \centering
\includegraphics[width=0.32\textwidth]{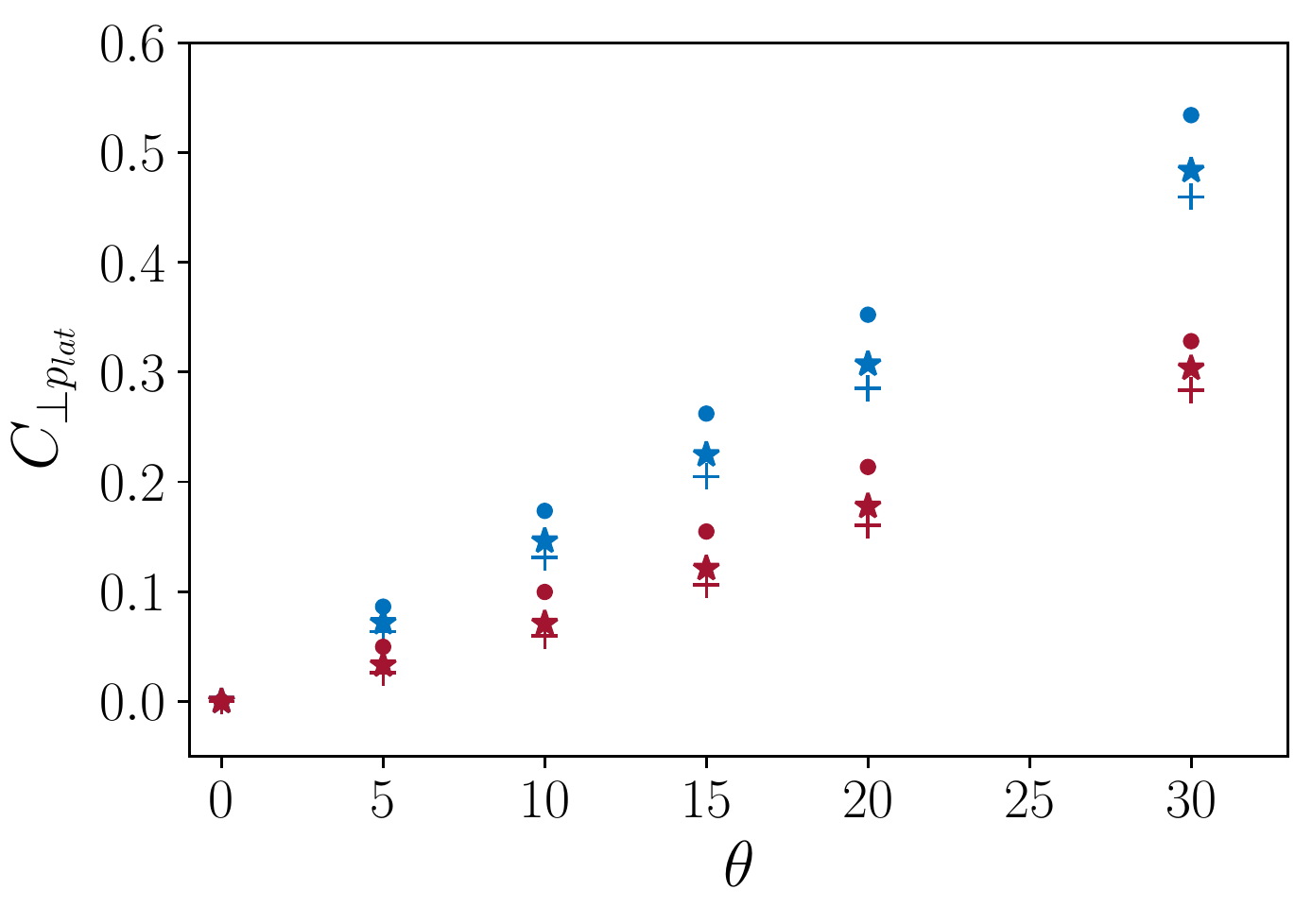}\label{fig:a}\hspace{8mm}
% \subfloat[]{\includegraphics[width=0.34\textwidth]{cnl_teta_x.eps}\label{fig:b}}
\includegraphics[width=0.32\textwidth]{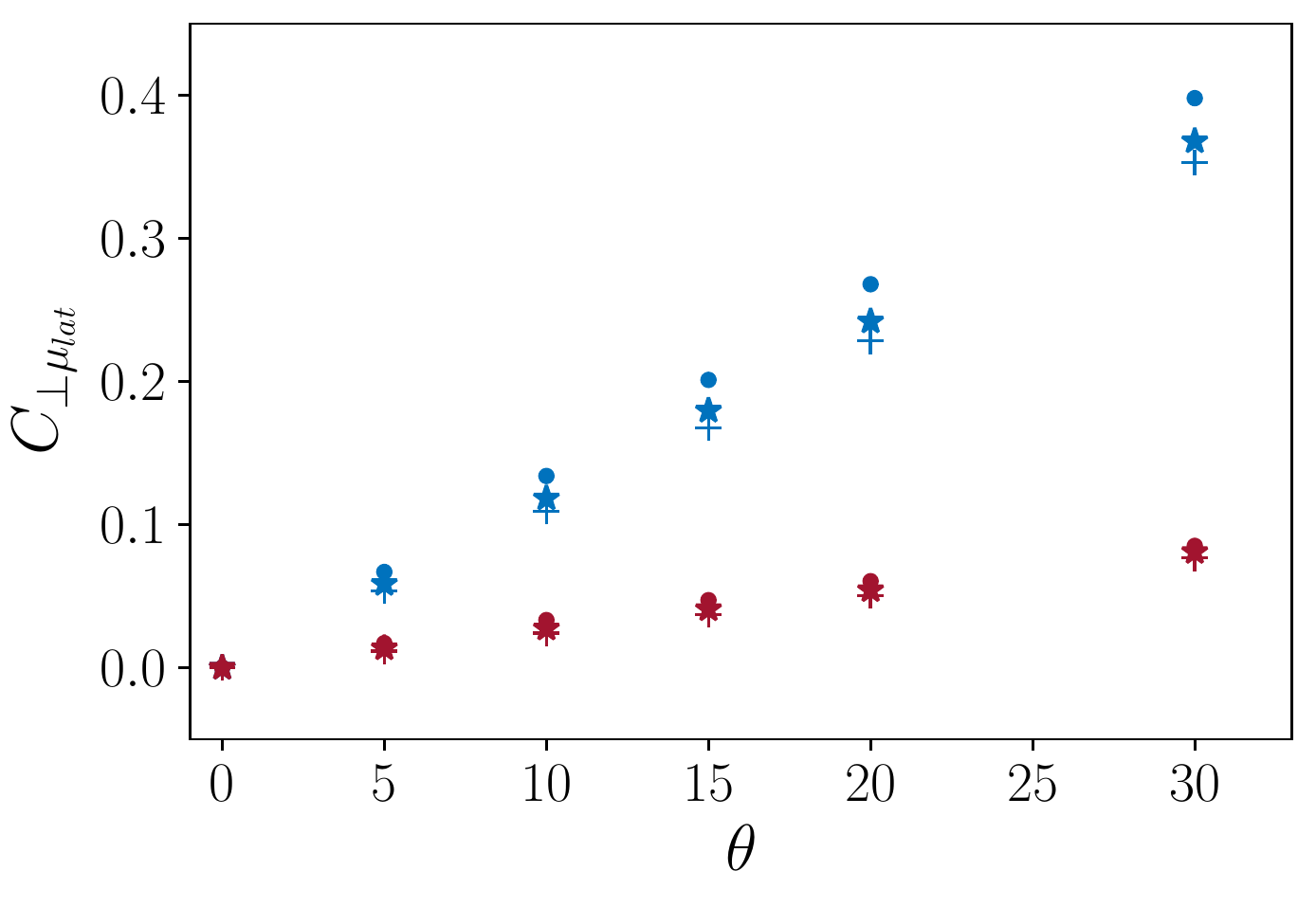}\label{fig:c}\\
\vspace{-3mm}
\hspace{-35mm} $(a)$\hspace{60mm}$(b)$\\
\vspace{3mm}
 \hspace{-3mm}\includegraphics[width=0.335\textwidth]{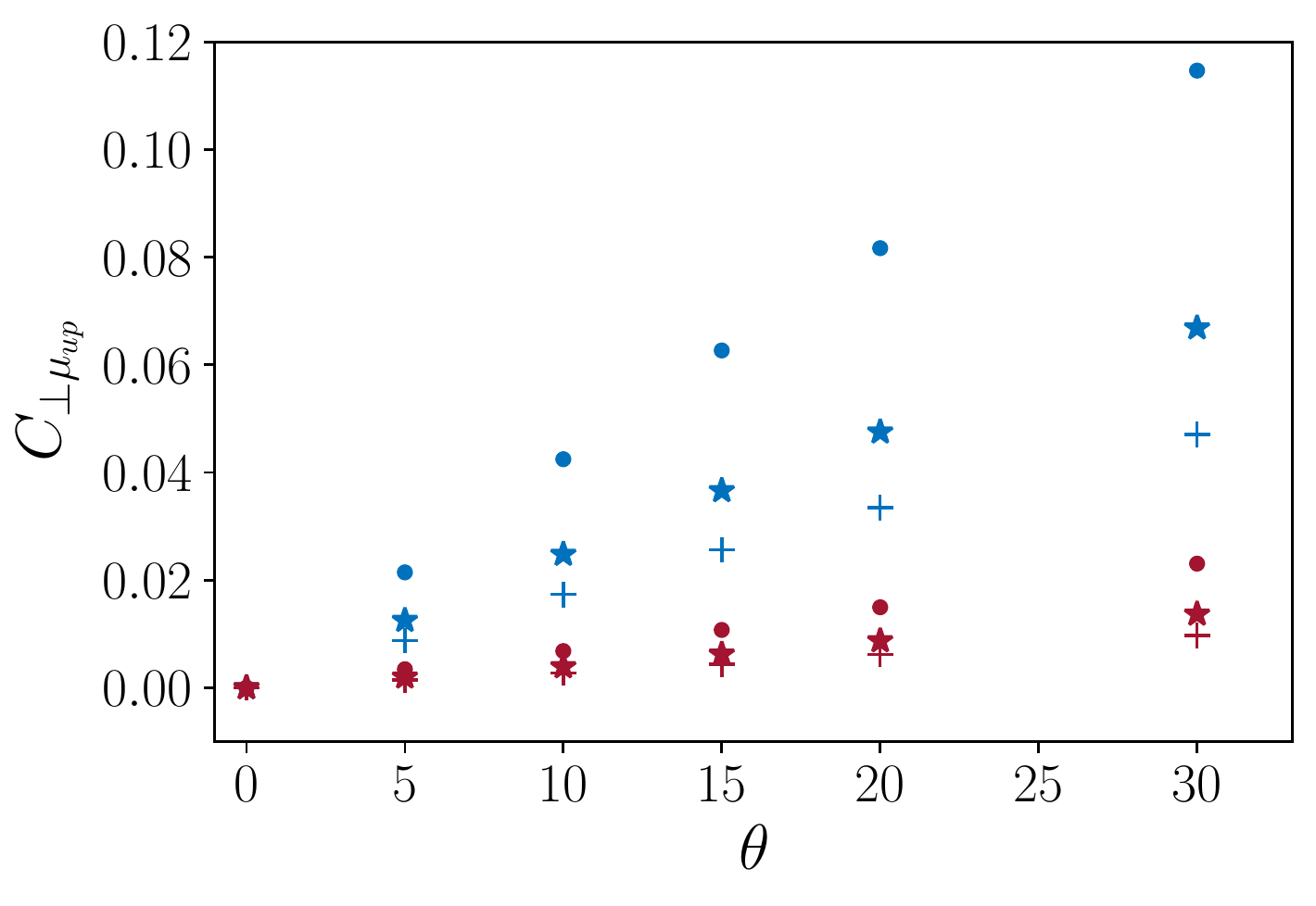}\label{fig:d}\hspace{7mm}%
\includegraphics[width=0.335\textwidth]{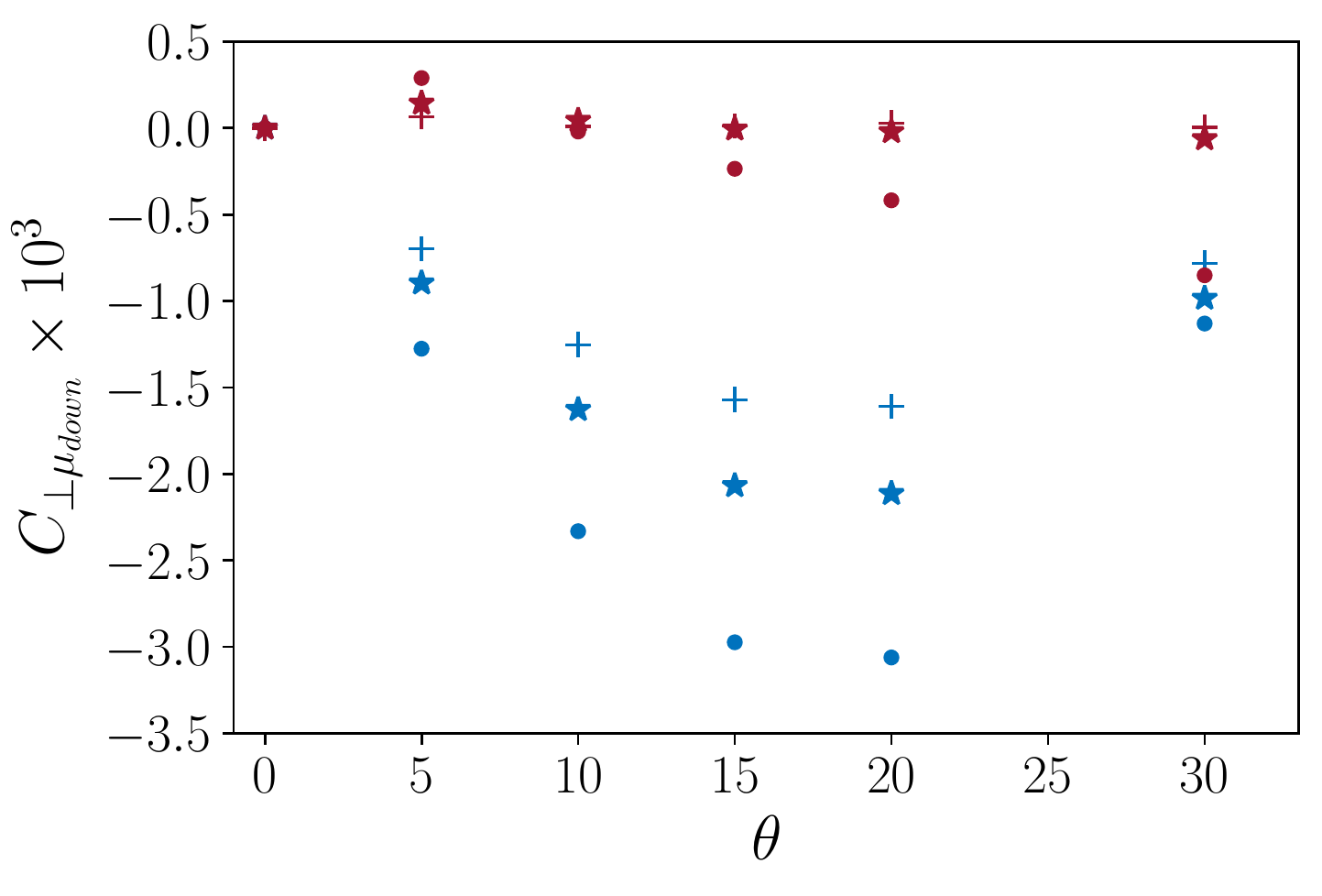}\label{fig:e}% 
\\
\vspace{-3mm}
\hspace{-35mm} $(c)$\hspace{60mm}$(d)$\\
\vspace{1mm}
 \caption{ Contributions to $C_{\perp}(\cchi,\theta,\text{Re})$ at $\text{Re}=20$ (blue) and $\text{Re}=300$ (red) for $\bullet$: $\cchi=3$, $\star$: $\cchi=5$, +: $\cchi=7$. $(a)$ pressure on the lateral surface; $(b)$ viscous stress on the lateral surface; $(c),\,(d)$ viscous stress on the upstream and  downstream ends, respectively.}%
 \label{fig:details_Cperp}
\end{figure}

%In the Stokes regime, the linearity of the loads with respect to the boundary conditions implies that the force experienced by an inclined cylinder is linearly related to the drag acting on the same cylinder in the two extreme angular configurations $\theta=0^\circ$ and $\theta=90^\circ$ through 
%\begin{eqnarray}
%\label{eq:Cpara_stokes}
%C_{\parallel}(\cchi,\theta)&=&C_\parallel^{ \theta=0^\circ}(\cchi)\cos\theta\,, \\
%\label{eq:Cperp_stokes}
%C_{\perp}(\cchi,\theta)&=&C_\perp^{ \theta=90^\circ}(\cchi)\sin\theta\,.
%\end{eqnarray}
%\vspace{0mm}\\
%Combining the purely geometric relations (\ref{eq:geom1})-(\ref{eq:geom2}) with `Stokes laws' (\ref{eq:Cpara_stokes})-(\ref{eq:Cperp_stokes}) yields under creeping flow conditions
%\begin{eqnarray}
%\label{eq:Cd_stokes}
%C_{d}(\cchi,\theta)&=&C_\parallel^{ \theta=0^\circ}(\cchi) + (C_\perp^{ \theta=90^\circ}(\cchi)-C_\parallel^{ \theta=0^\circ}(\cchi))\sin^2\theta\,,  \\
%\label{eq:Cl_stokes}
%C_l(\cchi,\theta)&=&\frac{1}{2}(C_\parallel^{ \theta=0^\circ}(\cchi) - C_\perp^{ \theta=90^\circ}(\cchi))\sin2\theta \,.
%\end{eqnarray}
To assess the validity of the `Stokes laws' in this regime, we inject the numerical results in (\ref{eq:geom1})-(\ref{eq:geom2}) and compare the resulting $C_\parallel(\cchi,\theta,\text{Re})$ and $C_\perp(\cchi,\theta,\text{Re})$ with the predictions of (\ref{eq:Cpara_stokes})-(\ref{eq:Cperp_stokes}) at the relevant Reynolds number and aspect ratio. To achieve this comparison, the two coefficients $C_\parallel^{ \theta=0^\circ}$ and $C_\perp^{ \theta=90^\circ}$ are required for every value of $\cchi$ and $\text{Re}$. $C_\parallel^{ \theta=0^\circ}(\cchi,\text{Re})$ is directly related to the drag determined in Sec. \ref{aligned} up to a normalization factor. More specifically, in the $\text{Re}$-range considered in Sec. \ref{moder}, $C_\parallel$ and the drag coefficient $C_d$ resulting from (\ref{eq:cp_theta0_up})-(\ref{eq:cd_theta0}) are linked through the relation % established for $\text{Re}\gtrsim20$ yield a drag coefficient $C_d(\cchi,\text{Re})$ based on a normalization of the force by ${\frac{\pi}{8}D^2\rho U^2}$, while the definition of $C_\parallel$ and $C_\perp$ involves a normalization by ${\frac{1}{2}LD\rho U^2}$. Therefore one merely has 
$C_\parallel^{ \theta=0^\circ}(\cchi,\text{Re})=\frac{\pi}{4}\cchi^{-1}C_d(\cchi,\text{Re})$. At lower Reynolds number, $C_\parallel^{ \theta=0^\circ}$ is readily obtained through the approximate expression (\ref{eq:slender_smallinertia2}) as $C_\parallel^{ \theta=0^\circ}(\cchi,\text{Re})=4\pi \text{Re}^{-1}\mathcal{F}(\cchi,\text{Re})$, where $\mathcal{F}(\cchi,\text{Re})$ denotes the quantity within brackets in (\ref{eq:slender_smallinertia2}). A direct graphical estimate of $\mathcal{F}(\cchi,\text{Re})$ is provided for several aspects ratios in Fig. \ref{fig:small_inertia},  noting that the relation between $\mathcal{F}(\cchi,\text{Re})$ and the normalized drag $F_d/F_{ds}$ is $\mathcal{F}(\cchi,\text{Re})=\left(\frac{81}{16}\cchi^{-2}\right)^{1/3}F_d/F_{ds}$.\\
 \indent Since we did not compute the loads on a cylinder held perpendicular to the flow, we use data from the literature to estimate $C_\perp^{ \theta=90^\circ}(\cchi,\text{Re})$. 
 %At low-but-finite Reynolds number, the drag coefficient per unit length over an infinitely long cylinder was obtained through a matched asymptotic expansion procedure in the form \citep{kaplun1957}
%\begin{equation}
%C_\perp^{ \theta=90^\circ}(\cchi\gg1,\text{Re}\lesssim1)\approx\frac{8\pi}{Re}\epsilon(1-0.87\epsilon^2)\quad \mbox{with}\quad \epsilon=\left\{\frac{1}{2}-\gamma-\log\frac{\text{Re}}{8}\right\}^{-1}\,,
%\label{kaplun}
%\end{equation}
%where $\gamma$ denotes Euler's constant. This expression is known to be accurate up to $\text{Re}=\mathcal{O}(1)$ \citep{huner1977}. For higher Reynolds numbers, 
A large number of experimental data obtained with long cylinders was compiled in \cite{pruppacher1970}. The corresponding curves for the drag per unit length were fitted in \cite{clift1978} in the form
 \begin{equation}
C_\perp^{ \theta=90^\circ}(\cchi\gg1,\text{Re})\approx 9.69\text{Re}^{-0.78}(1+a\text{Re}^{n})\,,
\label{eq:clift}
\end{equation}
with $a=0.227$ and $n=0.55$ for $5<\text{Re}\leq40$, and $a=0.084$ and $n=0.82$ for $40<\text{Re}\leq400$, respectively. 
Comparisons between predictions of (\ref{eq:clift}) and experimental results from \cite{jayaweera1965} for finite-length cylinders falling perpendicular to their axis indicate that the drag is only marginally affected by end effects as soon as $\cchi\gtrsim2$ and $\text{Re}\gtrsim10$ \citep{clift1978}. This is why we consider that (\ref{eq:clift}) provides a relevant estimate of $C_\perp^{ \theta=90^\circ}(\cchi,\text{Re})$ throughout the range of aspect ratios and Reynolds numbers of interest here. \vspace{2mm}\\
\indent The resulting comparison between numerical results and predictions of (\ref{eq:Cpara_stokes})-(\ref{eq:Cperp_stokes}) is presented in Fig. \ref{fig18}. The parallel force coefficients are seen to decrease significantly as $\cchi$ increases. In Sec. \ref{aligned}, where the drag was normalized using the frontal area $\pi D^2/4$, we found that the pressure contribution to the drag in the configuration $\theta=0^\circ$ is almost independent of $\cchi$. For this reason, the corresponding contribution to $C_\parallel^{ \theta=0^\circ}(\cchi,\text{Re})$, which involves a normalization by $LD$, behaves as $\cchi^{-1}$ and is responsible for the most part of the large variations of $C_{\parallel}(\cchi, \theta,\text{Re})$ with $\cchi$ observed in Figs. \ref{fig18}$(a), (c),(e)$. Variations of $C_{\parallel}$ with $\theta$ are remarkable in that they clearly contradict the  `Stokes law'. Indeed,  it is seen that $C_{\parallel}$ is almost independent of $\theta$ for $\text{Re}=20$ (apart from a modest decrease at $\theta=30^\circ$ for $\cchi=3$), while it increases with the cylinder inclination for larger Reynolds numbers. Still modest for $\text{Re}=80$, this increase makes $C_{\parallel}$ $40-50\%$ larger for $\theta=30^\circ$ than for $\theta=0^\circ$ at $\text{Re}=300$.  %also seen to depend strongly on the Reynolds number. \color{black} Roughly speaking, these variations are quite modest for $\text{Re}=20$ and $\text{Re}=80$ \color{red} While $C_{\parallel}$ experiences little variation up to $\theta\approx20^\circ$ for $\text{Re}=20$ before exhibiting a significant decrease for $\theta=30^\circ$, the plots corresponding to $\text{Re}=80$ and $\text{Re}=300$ reveal a gradual increase from $\theta=0^\circ$ to $\theta\approx20^\circ$ and, depending on $\cchi$ and $\text{Re}$, a decrease, a `plateau' or a further increase in the range $20^\circ\leq\theta\leq30^\circ$. \color{black}  
Clearly the `Stokes law' (\ref{eq:Cpara_stokes}) is unable to reproduce the observed trends.\\
\indent
The situation is markedly different with the perpendicular force coefficient, which is found to follow closely the $\sin\theta$-dependence predicted by (\ref{eq:Cperp_stokes}) throughout the entire range of $\theta$ and $\text{Re}$. Moreover, $C_\perp$ only mildly varies with the aspect ratio, with less than $20\%$ differences between the shortest and longest cylinders whatever $\text{Re}$ (Figs. \ref{fig18}$(b), (d),(f)$). However, it is also clear from these panels that the predictions of (\ref{eq:Cperp_stokes}) based on the expression (\ref{eq:clift}) for $C_\perp^{ \theta=90^\circ}(\cchi\gg1,\text{Re})$ over-predict $C_\perp(\cchi,\theta,\text{Re})$ for the highest two Reynolds numbers, the overestimate increasing with $\text{Re}$. This is actually no surprise, since in the $\text{Re}$-range considered here, the flow structure past an inclined cylinder with $0^\circ\leq\theta\leq30^\circ$ has little in common with that past a cylinder held perpendicular to the incoming velocity. While the wake of the latter becomes unsteady for $\text{Re}\approx47$ and three-dimensional for $\text{Re}\approx180$ \citep{williamson1996}, all inclined configurations considered here are stationary and inherently three-dimensional. Therefore, the connection between the two configurations becomes loose beyond Reynolds numbers of a few tens. Nevertheless, to keep the advantage of (\ref{eq:Cperp_stokes}) which is asymptotically correct in the creeping-flow limit, we sought an empirical $\text{Re}$-dependent correction capable of properly approaching the numerical results while vanishing for both $\text{Re}\rightarrow0$ and $\theta=90^\circ$. \color{black} We sought another correction to account for the dependence of $C_\perp$ with respect to the aspect ratio, requesting that this correction also vanishes for $\theta=90^\circ$ for the aforementioned reasons. Ideally, one would like this correction to recover the proper behavior $C_\perp\propto\text{Re}^{-1}(\ln\cchi)^{-1}$ for $\text{Re}\rightarrow0$. However, due to the singular nature of the problem in the limit $\text{Re}\rightarrow0,\,\cchi\rightarrow\infty$ (Stokes paradox), it is known that $C_\perp^{ \theta=90^\circ}(\cchi\gg1,\text{Re}\rightarrow0)\propto-(\text{Re}\log\text{Re})^{-1}$ \cite{Lamb}, which makes the ratio $C_{\perp}/C_\perp^{ \theta\approx90^\circ}$ ill-defined in this limit. Consequently, we merely built the finite-length correction on the basis of present numerical data. We found this correction to be almost $\text{Re}$-independent, and eventually obtained the approximate expression for $C_{\perp}(\cchi,\theta,\text{Re})$ in the form
%\begin{equation}
%C_{\perp}(\cchi,\theta,\text{Re})=\frac{C_\perp^{ \theta\approx90^\circ}(\cchi\gg1,\text{Re})}{1+B\text{Re}^{1/2}\cos\theta}\sin\theta\,,
%\vspace{0mm}
%\label{corrected}
%\end{equation}
\begin{equation}
%C_{\perp}(\cchi,\theta,\text{Re})=C_\perp^{ \theta\approx90^\circ}(\cchi\gg1,\text{Re})\frac{1+1.15e^{-0.45\cchi(1-\frac{3.2}{\text{Re}})}\cos\theta}{1+0.04\text{Re}^{1/2}\cos\theta}\sin\theta\,,
%C_{\perp}(\cchi,\theta,\text{Re})=C_\perp^{ \theta\approx90^\circ}(\cchi\gg1,\text{Re})\frac{1+1.15e^{-0.45\cchi}\cos\theta}{1+0.04(1-e^{-0.25\text{Re}^{1/2}})\text{Re}^{1/2}\cos\theta}\sin\theta\,,
C_{\perp}(\cchi,\theta,\text{Re})=C_\perp^{ \theta\approx90^\circ}(\cchi\gg1,\text{Re})\frac{1+1.15e^{-0.45\cchi}\cos\theta}{1+0.04\text{Re}^{1/2}\cos\theta}\sin\theta\,.
\vspace{0mm}
\label{corrected}
\end{equation}
\color{black} Figures \ref{fig18}$(b)$, $(d)$ and $(f)$ show that the main trends of the numerical data are properly captured by the above fit. \color{black} Additional improvements could easily be introduced, such as a $\sin6\theta$-correction to compensate for the slight overestimates noticed at intermediate inclinations ($\theta\approx15^\circ$) as $\cchi$ and $\text{Re}$ increase. The finite-length correction suggests that the transverse force is virtually proportional to the cylinder length beyond $\cchi\approx10$.  For shorter cylinders, the increase of $C_\perp$ as $\cchi$ decreases is qualitatively reminiscent of the $(\ln\cchi)^{-1}$ low-$\text{Re}$ behavior. However finite-length effects are much weaker in the inertial regime. For instance, Fig. \ref{fig:f_perp_stokes} indicates that $C_\perp^{ \theta=90^\circ}$ is $60\%$ larger for $\cchi=3$ than for $\cchi=10$ in the creeping-flow regime, a difference reduced to $25\%$ in the fully inertial regime according to (\ref{corrected}). \color{black} From (\ref{eq:clift}) (see also figure 3 in \cite{jayaweera1965}), it may be inferred that $C_\perp^{ \theta=90^\circ}(\cchi\gg1,\text{Re})$ only weakly decreases with the Reynolds number for $\text{Re}\gtrsim100$. Therefore, the fit (\ref{corrected}) indicates that $C_{\perp}(\cchi,\theta,\text{Re})$ is almost proportional to $\text{Re}^{-1/2}$ in this range of $\text{Re}$, suggesting that the dominant contribution to the perpendicular force arises from boundary layer effects. Although (\ref{corrected}) correctly reduces to $C_{\perp}(\cchi,\theta,\text{Re})=C_\perp^{ \theta=90^\circ}(\cchi\gg1,\text{Re})$ when $\theta=90^\circ$, it is not clear up to which maximum inclination this expression provides a reliable approximation of the actual transverse force. The numerical results of \cite{pierson2019} for $\cchi=3$ suggest that this maximum is close to $50^\circ$. Computations with higher inclinations are required to clarify this issue.\\
 \indent Interestingly, \citet{sanjeevi2017} recently concluded that `Stokes laws', especially the sine-squared \color{black} drag law $C_{d}(\cchi,\theta)=C_\parallel^{ \theta=0^\circ}(\cchi) + (C_\perp^{ \theta=90^\circ}(\cchi)-C_\parallel^{ \theta=0^\circ}(\cchi))\sin^2\theta$ which results from the combination of (\ref{eq:geom1})-(\ref{eq:geom2}) and (\ref{eq:Cpara_stokes})-(\ref{eq:Cperp_stokes}), \color{black} hold for prolate spheroids (and moderately oblate spheroids) up to Reynolds numbers (based on the diameter of the equivalent sphere) of $\mathcal{O}(10^3)$. They argued that the reason for this surprising agreement is due to a partial compensation between contributions to the pressure drag arising from the regions close to the two stagnation points, in such a way that the overall pressure drag follows a sine-squared law, while the viscous contribution to the drag is almost insensitive to the inclination. Clearly this scenario does not hold for cylinders with flat ends. In the present case, wake effects are strong, with massive separation at the back of the cylinder, even for $\theta=0^\circ$, as soon as $\text{Re}$ exceeds some tens (see Fig. \ref{fig:streamlines}). These effects are deeply influenced by the body inclination (see Fig. \ref{fig:ldc140}) and, as Fig. \ref{fig18} reveals, result in non-monotonic variations of $C_{\parallel}(\cchi,\theta,\text{Re})$ with $\cchi$ and $\text{Re}$ which cannot be reduced to a simple geometric law. Therefore, it must be concluded that the scenario suggested in \cite{sanjeevi2017} to explain the validity of the sine-squared drag law applies only to streamlined bodies for which wake effects weakly affect the surface stress distribution.\\
%We tend to explain the linearity of $C_{\perp}=f(\theta)$ observed previously by detailing the expression of $C_{\perp}$ as a sum of pressure and viscous forces applied on the two ends of the cylinder (up and down) and the curved lateral surface.\\
%\citet{pierson2019} modified the Stokes law in order to be valid on the whole range of $\theta$ and obtain : \\
%\begin{equation}
%C_{\parallel}=C_\parallel^{ \theta=0^\circ}cos\theta +0.8C_\parallel^{ \theta=0^\circ}sin\theta cos^2\theta\\
%\label{eq:Cpara_pierson}
%\end{equation}
%\begin{equation}
%C_{\perp}=C_\perp^{ \theta=90^\circ}sin\theta +0.26C_\perp^{ \theta=90^\circ}cos\theta sin^2\theta \\
%\label{eq:Cperp_pierson}
%\end{equation}
%It should be noted that \ref{eq:Cpara_pierson} and \ref{eq:Cperp_pierson} are only valid for $\chi=3$, and it would be interesting to test it for other aspect ratios.
%The match is particularly good for $\cchi=3$, confirming earlier results by \citet{pierson2019}. A slight overestimate is noticed for more slender cylinders, especially when the Reynolds number becomes large, but the agreement remains reasonable.
\indent To better understand why the perpendicular force follows the approximate law (\ref{corrected}) throughout the parameter range explored here, it is useful to isolate the contributions to $C_\perp$ provided by the various parts of the body surface, and split each of them into a pressure and a viscous stress term. Since the body ends are flat and we are focusing here on the perpendicular force, no pressure contribution arises from the ends. Figure \ref{fig:details_Cperp} displays the variations of the remaining four nonzero terms with $\theta$ for two markedly different Reynolds numbers. The viscous contribution arising from the downstream end ($C_{\perp\mu_{down}}$) is seen to be negligibly small in all cases \color{black}(note the $10^3$ magnification factor in Fig. \ref{fig:details_Cperp}$(d)$). \color{black} Hence, virtually no contribution to $C_\perp$ is provided by this part of the body surface, on which wake effects concentrate in the range of $\theta$ and $\text{Re}$ relevant here. Examining panels $(a)-(c)$ in Fig. \ref{fig:details_Cperp} makes it clear that the various contributions to $C_\perp$ exhibit little dependence with respect to $\chi$, apart from the viscous stress on the upstream end ($C_{\perp\mu_{up}}$) at $\text{Re}=20$. Nevertheless, this term is one order of magnitude smaller than the total (pressure+viscous stress) contribution from the lateral surface. Consequently, the behavior of $C_\perp$ is essentially dictated by the latter. Among the corresponding two terms, the viscous contribution ($C_{\perp\mu_{lat}}$)  is virtually independent of $\cchi$ at large $\text{Re}$, while some finite-length influence subsists in $C_{\perp p_{lat}}$. This weak dependence with respect to the aspect ratio implies that the perpendicular force  increases almost linearly with the body length, given the chosen normalization factor $\rho U^2LD/2$. %{\color{red}{We still need to explain qualitatively (or give an argument) why the contributions from the lateral surface vary linearly with $\theta$... Independence Principle? Something else?}}
 %We do note a departure of the Stokes law at low $\theta$. %Contrary to low $\text{Re}$, $C_{\parallel}$ is increasing with $\theta$ until we observe a drop at $\theta=30^\circ$. 
 The quasi-linearity of the two dominant contributions to $C_\perp$ with respect to the inclination angle and their weak $\cchi$-dependence for $\text{Re}\gg1$ imply that, for $\text{Re}\gtrsim100$, the dimensional transverse force behaves approximately as
 \begin{equation}
 F_\perp(\cchi,\theta,\text{Re})\approx\frac{C_{\perp \theta\approx90^\circ}(\cchi\gg1,\text{Re})}{1+0.04\text{Re}^{1/2}\cos\theta}\rho LD\frac{||{\bf{U}}||U_\perp}{2}\,,
 \label{perp}
 \end{equation} 
 \noindent 
where $U_\perp=||\bf{U}||\sin\theta$ is the component of the upstream velocity normal to the lateral surface and the Reynolds number $\text{Re}$ is based on the norm of the upstream velocity. %As $C_\perp^{ \theta=90^\circ}(Re)\propto \text{Re}^{-1}$ at low Reynolds number, the `Stokes law' (\ref{eq:Cperp_stokes}) is recovered in this limit. In contrast, $C_\perp^{ \theta=90^\circ}(\text{Re})$ becomes almost $\text{Re}$-independent for Reynolds number of a few hundreds, as (\ref{eq:clift}) suggests. 
In this Reynolds number range, $C_\perp^{ \theta=90^\circ}$ is almost constant according to (\ref{eq:clift}), so that (\ref{perp}) indicates that the perpendicular force is approximately proportional to the power three-half of the incoming velocity, while it still varies almost linearly with $\theta$. In contrast, the Independence Principle frequently invoked in the area of vortex-induced vibrations \citep{ramberg1983,zdravkovich1997b} suggests that the perpendicular force on a long cylinder should only depend on the normal component $U_\perp$ of the incoming flow, which would imply $F_\perp=\frac{1}{2}C_\perp^{ \theta=90^\circ}(\cchi\gg1,Re_\perp)\rho LD||{\bf{U}}_\perp||U_\perp$, with $\text{Re}_\perp=\rho||{\bf{U}}_\perp||D/\mu$. If this were true in the present situation, a $\sin^2\theta$-dependence of $C_\perp$ would be observed for $\text{Re}_\perp=\mathcal{O}(10^2)$ (\textit{i.e.} $\text{Re}=300$ here in practice), and the force would vary as the square of the incoming velocity. As Fig. \ref{fig18}$(f)$ indicates, no quadratic dependence with respect to the inclination angle is noticed, which implies that the Independence Principle does not apply to the flow configurations under consideration. This is in line with the conclusions of \cite{vakil2009} where it was observed at somewhat lower Reynolds numbers that this `principle' only holds for inclinations larger than $70^\circ$ but overestimates the force by more than $50\%$ for $\theta\leq30^\circ$, even for long cylinders with $\cchi=15$. In other words, this `principle' is approximately valid when the upstream flow is almost normal to the cylinder lateral surface but can by no means be used to approximate the transverse force when the cylinder inclination is moderate.
% but somewhat d and the gap between the two aspect ratios is too small. This linearity were also observed by \citet{pierson2019} for $\cchi=3$. 
\subsection{Approximate laws for the parallel force and spanwise torque}
\label{approxpt}
\begin{figure}%
 \centering
\hspace{-10mm}\includegraphics[width=0.33\textwidth]{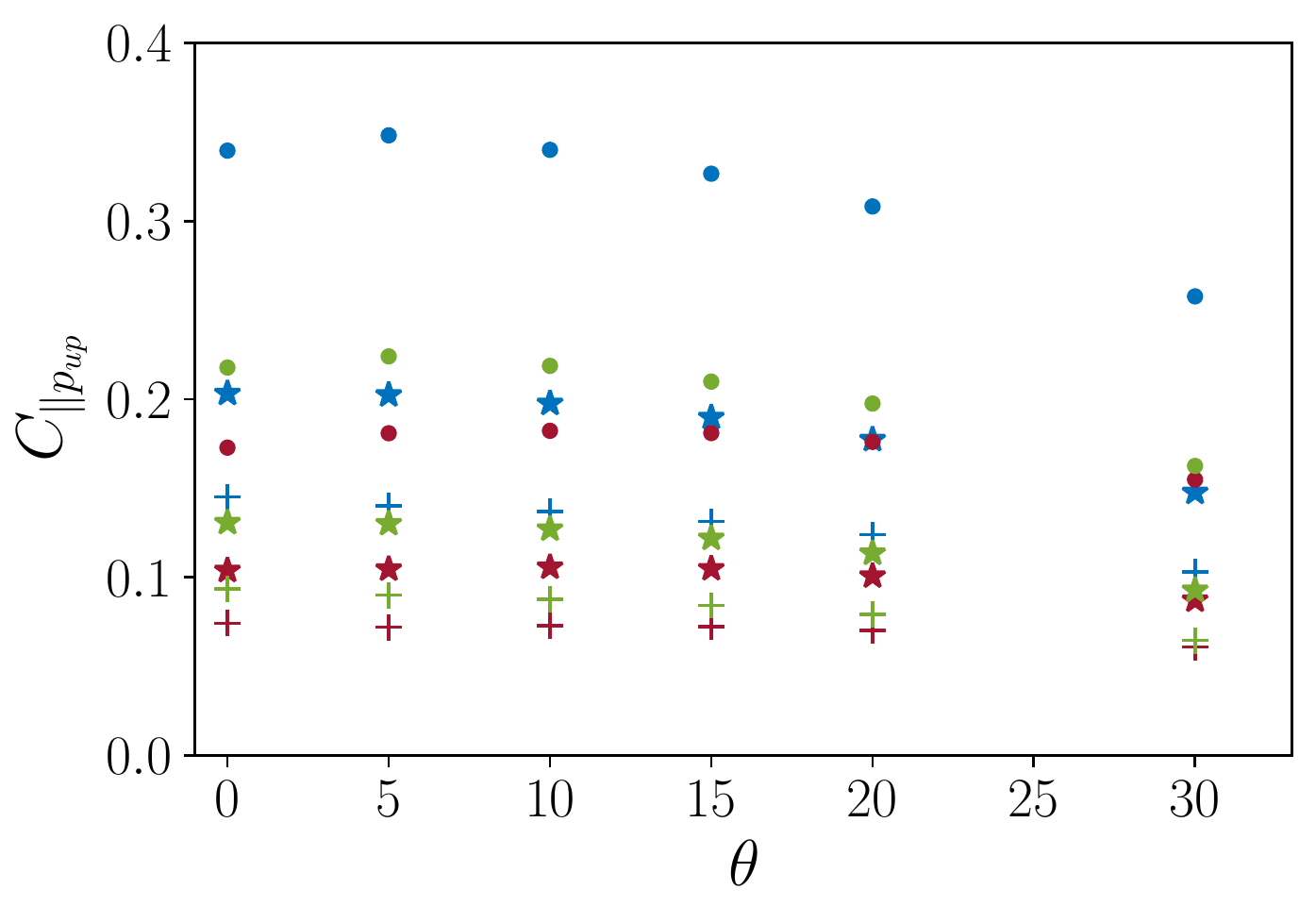}\label{fig:aa}\hspace{-1mm}
\includegraphics[width=0.335\textwidth]{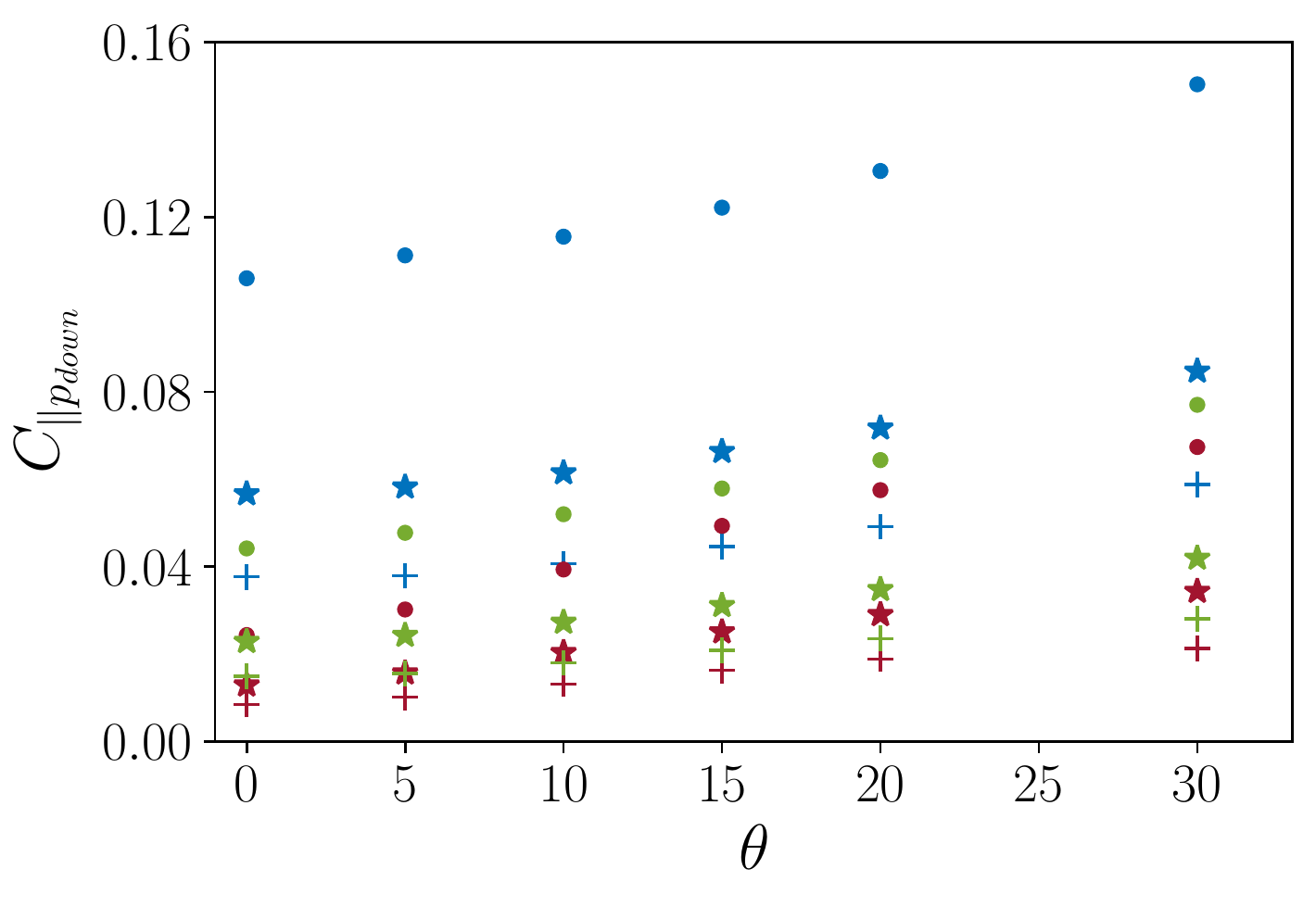}\label{fig:bb} \hspace{-2mm}
\includegraphics[width=0.33\textwidth]{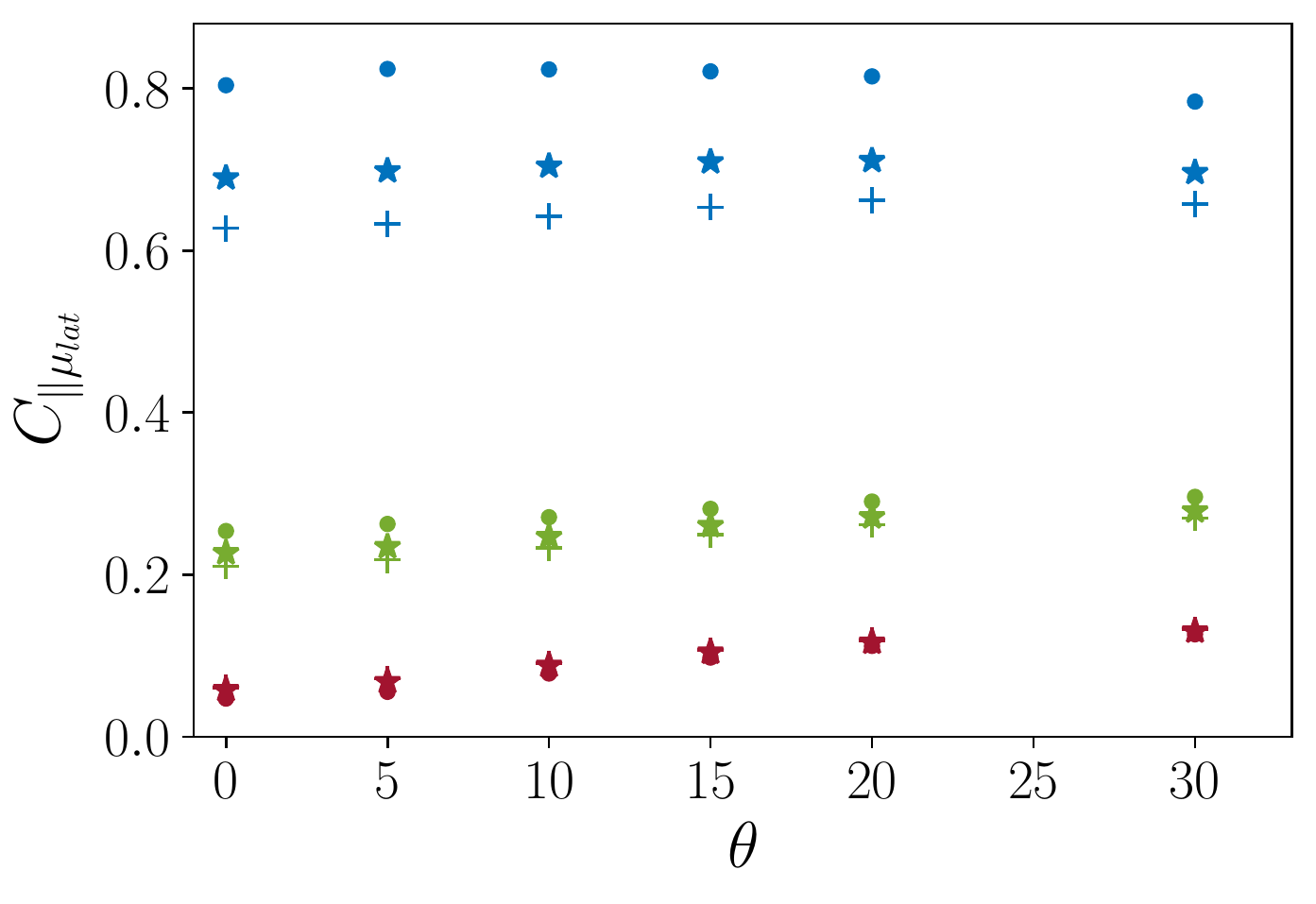}\label{fig:ee}
\vspace{-2mm}\\
%\includegraphics[width=0.35\textwidth]{cmuw_teta_x.pdf}\label{fig:cc}\hspace{6mm}
%\includegraphics[width=0.35\textwidth]{cmue_teta_x.pdf}\label{fig:dd}\\
%\vspace{5mm}
% \hspace{-3mm}\includegraphics[width=0.33\textwidth]{cmul_teta_x.pdf}\label{fig:ee}\\%
\hspace{-45mm} $(a)$\hspace{48mm}$(b)$\hspace{48mm} $(c)$\\
 \caption{ Contributions to $C_{\parallel}(\cchi,\theta, \text{Re})$ at $\text{Re}=20$ (blue), $\text{Re}=80$ (green) and $\text{Re}=300$ (red) for $\bullet$: $\cchi=3$, $\star$: $\cchi=5$, +: $\cchi=7$. $(a)-(b)$ pressure on the upstream and downstream ends, respectively; $(c)$ viscous stress on the lateral surface.}%
 \label{fig:details_Cparal}
\end{figure}

Figure \ref{fig:details_Cparal} shows the main contributions to $C_\parallel(\cchi,\theta,\text{Re})$ arising from pressure and viscous stress distributions over the various parts of the cylinder surface. \color{black} On both ends, the latter (not shown) are found to be more than one order of magnitude smaller than the former. Therefore, $C_\parallel$ is essentially controlled by the viscous stress acting on the lateral surface and the pressure distribution on both ends. \color{black}  All contributions are seen to decrease for increasing aspect ratios, similar to what happens when the body is aligned with the incoming flow. This influence weakens significantly as $\text{Re}$ increases, again in line with the observations reported for $\theta=0^\circ$. In particular, the viscous contribution arising from the lateral surface (Fig. \ref{fig:details_Cparal}$(c)$) is found to be virtually independent of $\cchi$ for $\text{Re}=300$. That these features subsist for all inclinations considered here suggests that seeking an empirical expression relating $C_\parallel(\cchi,\theta,\text{Re})$ to $C_{\parallel\theta=0^\circ}(\cchi,\text{Re})$ is reasonable. Variations with $\theta$ follow different and sometimes opposite trends on the various surfaces. For instance, Fig. \ref{fig:details_Cparal}$(a)$ indicates that $C_{\parallel p_{up}}$ decreases as the inclination increases, an effect weakening at large Reynolds number. 
\noindent %The dominant variation of $C_{\parallel p_{up}}$ is close to a $\cos^2\theta$-dependency, which may be interpreted as the fact that the force on the upstream end follows an inertial scaling with respect to $U\cos\theta$, the velocity component normal to that surface. 
In contrast, $C_{\parallel p_{down}}$ (Fig. \ref{fig:details_Cparal}$(b)$) increases gradually with $\theta$. \color{black} This is because the recirculating region at the back of the cylinder tends to shrink when $\theta$ increases, as noticed in Sec. \ref{separ}. \color{black} A significant increase of $C_{\parallel\mu_{lat}}$ is also noticed at $\text{Re}=300$ (Fig. \ref{fig:details_Cparal}$(c)$), but this trend weakens as $\text{Re}$ decreases and is no longer present at $\text{Re}=20$. \\%\color{black} Some of these trends directly stem from the influence of the inclination on the size of the separated regions described in Sec. \ref{separ}; \textit{e.g.} the increase of $C_{\parallel p_{down}}$ with $\theta$, which results from the reduction of the recirculating region at the back of the cylinder as $\theta$ increases. \color{black}
%These diverse and sometimes opposite variations with $\cchi$ of the three dominant contributions to $C_\parallel$ provide evidence that the `Stokes law' (\ref{eq:Cpara_stokes}) cannot be extended to inertia-dominated regimes. Consequently, it is necessary to consider separately the contributions of each part of the surface to obtain a sound empirical expression for $C_\parallel(\theta,\cchi,\text{Re})$. \\
\indent To approximate the variations of $C_{\parallel}$ with $\cchi,\,\theta$ and $\text{Re}$, we started from the fits (\ref{eq:cp_theta0_up})-(\ref{eq:cmu_theta0}) established for $\theta=0^\circ$ in Sec. \ref{moder}. \color{black} We took into account the constraint that $C_\parallel$ cannot depend on the sign of $\theta$ and must change sign for $\theta=90^\circ$, although this configuration is well beyond the maximum inclination considered in the simulations. This led us to assume that the leading-order angular dependence of the correction to the `Stokes law' is proportional to $\sin^2\theta\cos\theta$. Then, we first considered the case $\cchi=7$ for which finite-length effects are the weakest, and started to fit the dependence with respect to $\text{Re}$ for $\theta=30^\circ$, the maximum inclination. The behavior of $C_{\parallel}$ at median inclinations suggests that the angular dependence also involves a secondary contribution that may be approached by a term proportional to  $\sin^26\theta\cos\theta$. %We had to introduce additional angular corrections proportional to $\sin^26\theta\cos\theta$ to account for  the proper behavior of some quantities, especially $C_{{\parallel\mu}_{lat}}$, close to the intermediate inclination $\theta=15^\circ$. 
Last, we considered finite-length effects, starting with $\cchi=3$ for which they are most severe. All empirical pre-factors were constrained to vanish for $\text{Re}\rightarrow0$, so that $C_\parallel$ reduces to its low-$\text{Re}$ form in this limit. %This required the introduction of additional terms dependent on $\theta$ and $\text{Re}$ which vanish in the limit of large aspect ratios; \textit{e.g.} terms proportional to $\cchi^{-2}$ in the fits for $C_{{\parallel p}_{up}}$ and $C_{{\parallel\mu}_{lat}}$. 
The whole process was carried out iteratively, to optimize the pre-factors and exponents for the three aspect ratios over the whole range of $\text{Re}$ and $\theta$. \color{black}
 \color{black} Keeping in mind that the drag coefficient $C_d(\cchi,Re)$ determined in Sec. \ref{moder} by summing (\ref{eq:cp_theta0_up})-(\ref{eq:cmu_theta0}) has to be multiplied by a factor $\frac{\pi}{4}\cchi^{-1}$ to be used in the prediction of $C_\parallel$, the final fit takes the form
\begin{figure}
	\centering
\hspace{-9mm}\includegraphics[width=0.32\textwidth]{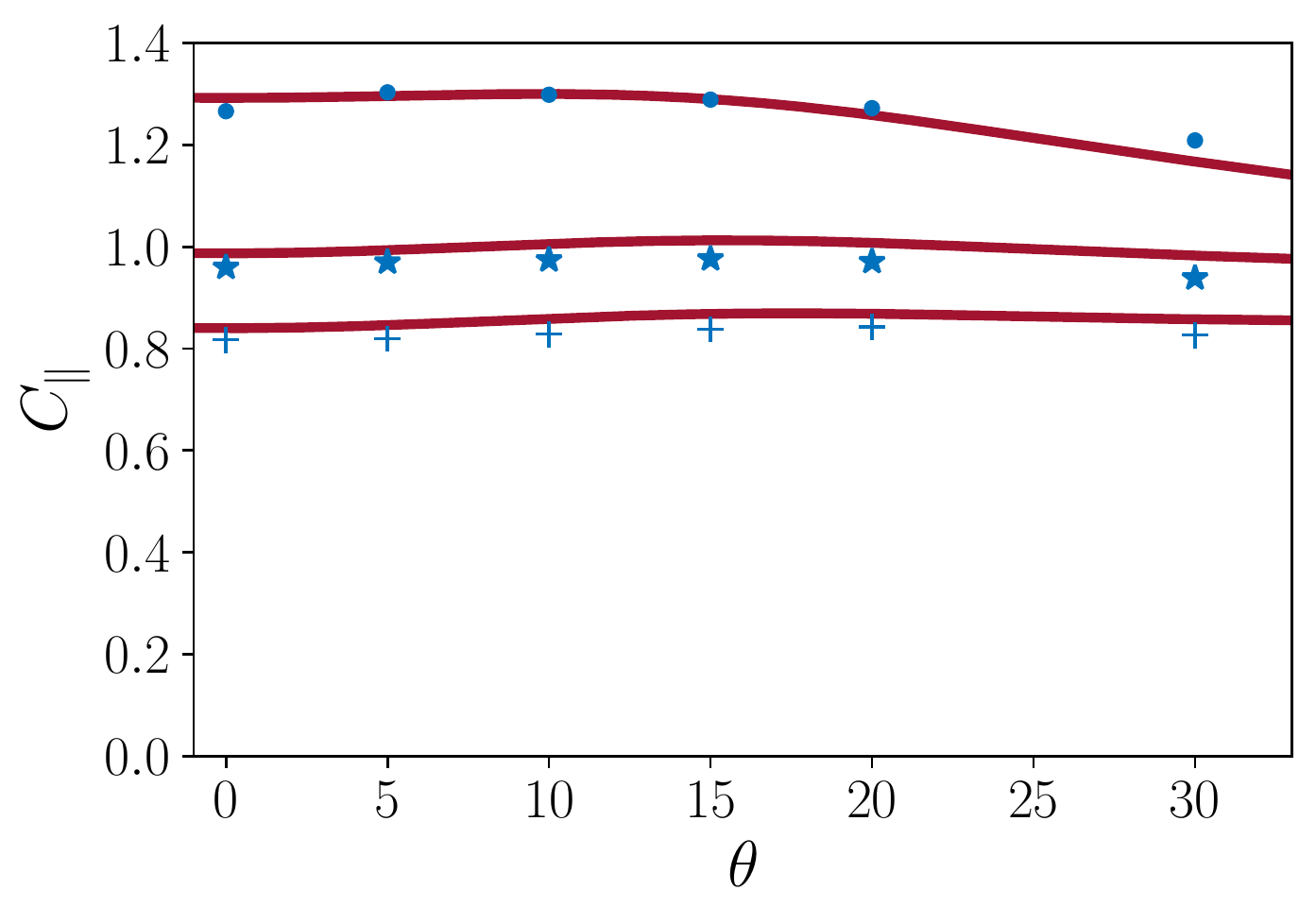}\quad
\hspace{-3.5mm}\includegraphics[width=0.32\textwidth]{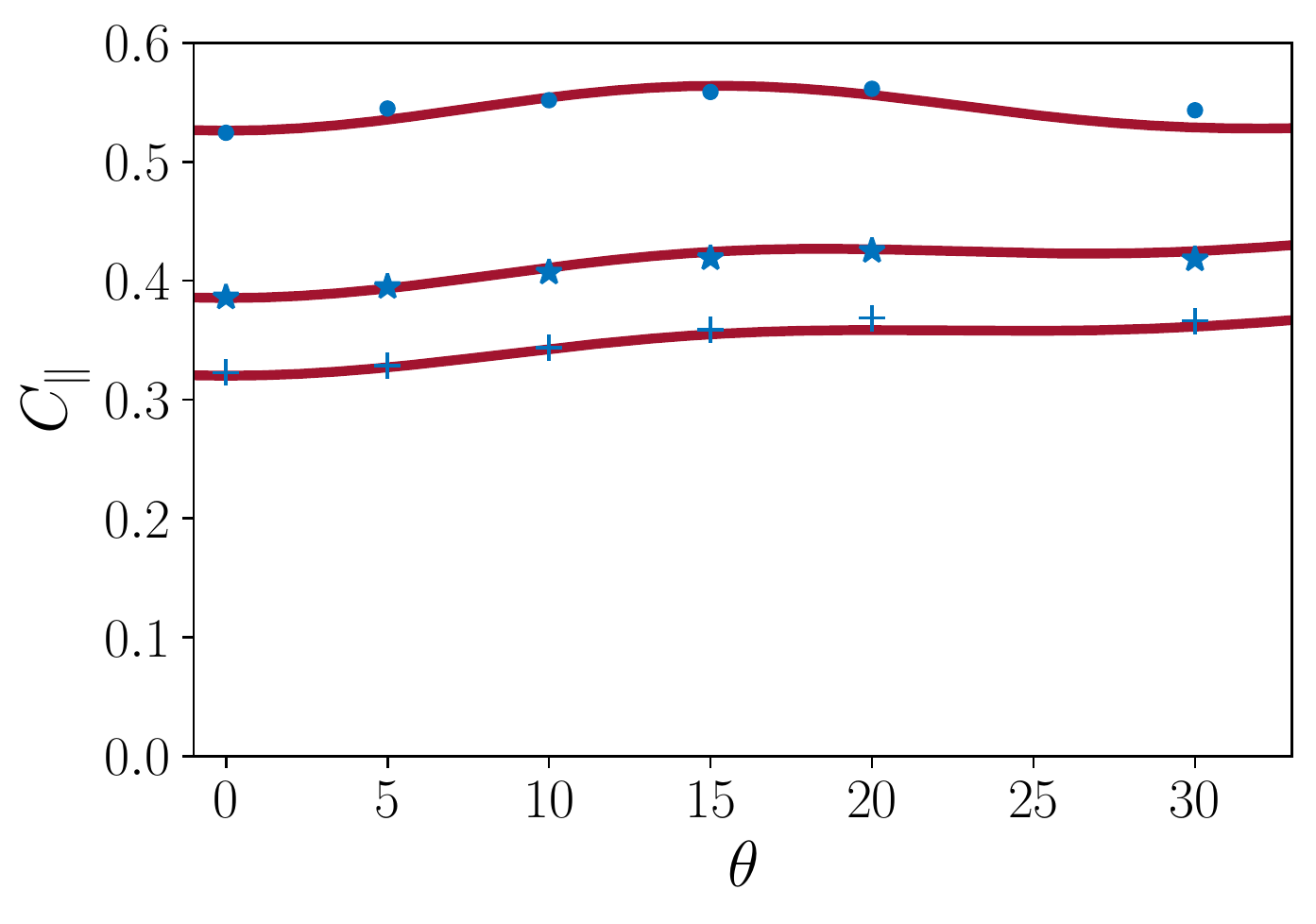}\quad
\hspace{-3.5mm}\includegraphics[width=0.32\textwidth]{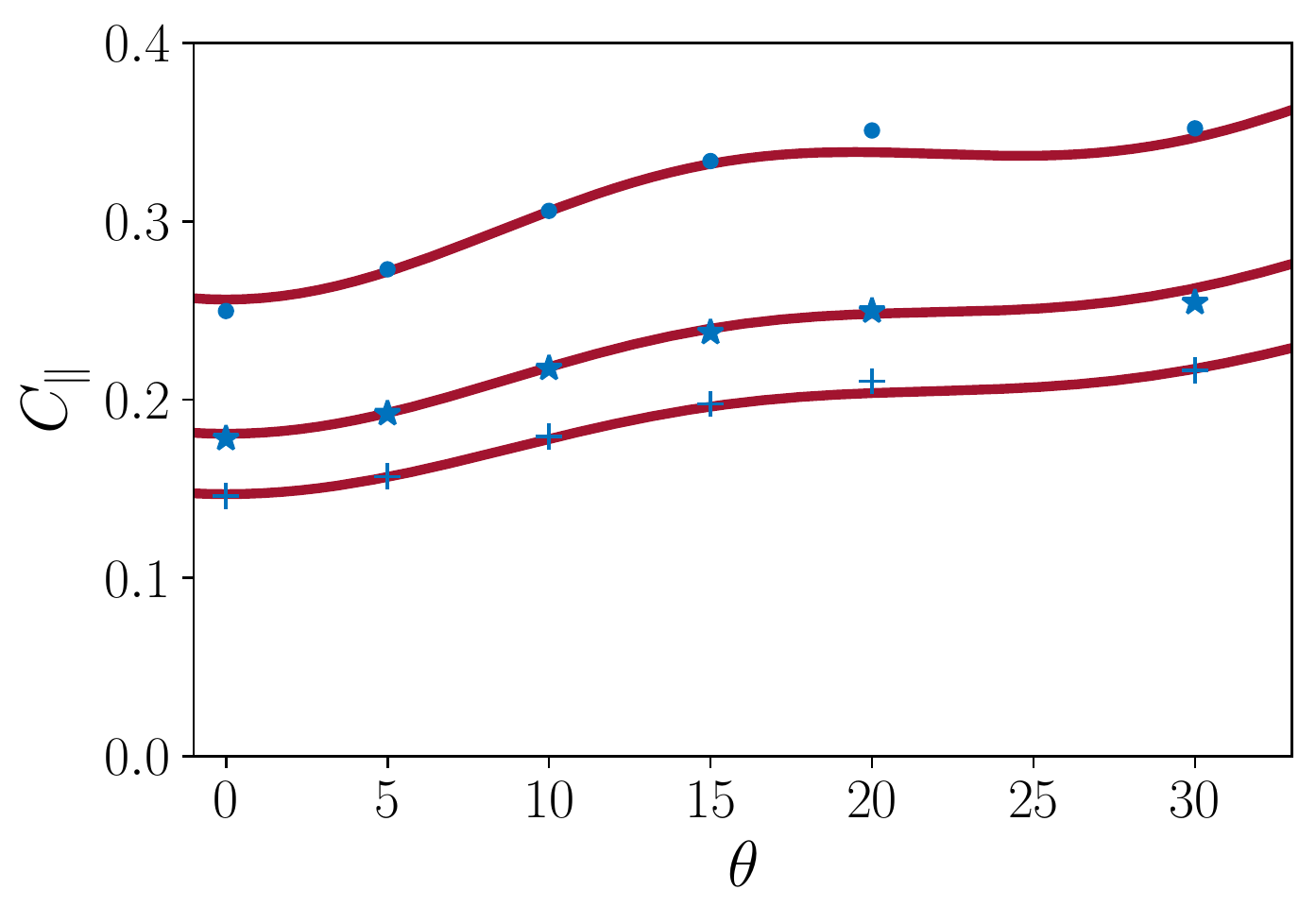}\\
\vspace{-4mm}
		\hspace{-20mm}$(a)$\hspace{50mm}$(b)$\hspace{50mm}$(c)$  \\
		\vspace{-13mm}
	\hspace{-18mm}$\text{Re}=20$\hspace{43mm}$\text{Re}=80$\hspace{43mm}$\text{Re}=300$
	\vspace{10mm}\\
\caption{Parallel force coefficient $C_\parallel(\cchi,\theta,\text{Re})$ vs. $\theta$ for $\bullet$: $\cchi=3$, $\star$: $\cchi=5$, +: $\cchi=7$. %$(a)$: $\text{Re}=20$; $(b)$: $\text{Re}=80$; $(c)$: $\text{Re}=300$;
 Solid line: empirical fit (\ref{eq:cparfin}).}
\label{fig:cparfit}
\end{figure}
%Making use of (\ref{eq:cpp_theta0_up})-(\ref{eq:cpmu_theta0_lat}), the parallel force coefficient is finally approximated as
\begin{equation}
\label{eq:cparfin}
C_{\parallel}(\cchi,\theta,  \text{Re}) \approx \frac{\pi}{4}\cchi^{-1}C_d(\cchi,Re)\biggl\{1+\left[(0.7-6.3\cchi^{-2})(1-e^{-0.15\text{Re}})+0.01 \text{Re}^{0.95}\right]\sin^2\theta+2\times10^{-3} \text{Re}^{0.8}\sin^26\theta\biggr\}\cos\theta\,.
%C_{\parallel}(\cchi,\theta,  \text{Re}) \approx \frac{\pi}{4}\cchi^{-1}C_d(\cchi,Re)\biggl\{1+\left[0.7+0.01 \text{Re}^{0.95}-6.4\cchi^{-2}\right]\sin^2\theta+2\times10^{-3} \text{Re}^{0.8}\sin^26\theta\biggr\}\cos\theta\,.
\end{equation} \color{black} 
%where $C_d(\cchi,Re)$ corresponds to the approximate expression (\ref{eq:cd_theta0}) for the drag coefficient at $\theta=0^\circ$ obtained by adding (\ref{eq:cp_theta0_up}), (\ref{eq:cp_theta0_down}) and (\ref{eq:cmu_theta0}). 
As Fig. \ref{fig:cparfit} shows, this fit provides a correct estimate of $C_\parallel$ throughout the range of parameters explored in the present investigation. \color{black} Expression (\ref{eq:cparfin}) highlights the fact that inertial effects act to increase $C_\parallel$ and counteract the $\cos\theta$-decrease associated with viscous effects, and even overtake them for high enough $\text{Re}$ (for large $\cchi$ and $\text{Re}\gtrsim10$, the dominant contribution to the term within curly brackets is $1+\left(0.7+0.01 \text{Re}^{0.95}\right)\sin^2\theta$). \color{black}
%Nevertheless, these fits are only valid within the inertia-dominated regime, since  $C_{{\parallel p}_{up}}$ and $C_{{\parallel p}_{down}}$ are assumed to follow different trends due to the presence of the wake. Therefore, considering the limit $\text{Re}\rightarrow0$ in (\ref{eq:cpp_theta0_up})-(\ref{eq:cpmu_theta0_lat}) would be irrelevant. 
Obviously, the above fit is not expected to be valid for Reynolds numbers significantly larger than the upper bound considered in the simulations, as it predicts a diverging drag in the limit $\text{Re}\rightarrow\infty$. \color{black} Similarly, (\ref{eq:cparfin}) is not expected to hold for larger inclinations: in \cite{pierson2019} it was observed that, for $\cchi=3$ and $\text{Re}=250$, $C_\parallel$ sharply decreases in the range $30^\circ<\theta<45^\circ$, a trend that the above fit is clearly unable to reproduce. \color{black}
%\color{red} Check the matching at $\mathcal{O}(1)-\text{Re}$ with the low-but-finite-$\text{Re}$ expression. \color{black}
\vspace{2mm}\\
%$C_d$ decreases with $\text{Re}$ and $\cchi$, but increases with $\theta$ because of the increase in the projected area of the cylinder on the plane perpendicular to the incoming flow. For $\cchi=7$, it is not necessary to modify the Stokes law to have a good accuracy as shown in figure \ref{fig:cd_teta_Stvsnum}(c). This a good agreement is partly due to $C_{d\theta=90^\circ}$ taken from \citet{clift1978} which is valid for long cylinder. However, the Stokes law is not accurate at small aspect ratios, then we modify it to obtain the empirical law \ref{eq:Cd_empir_law} which seems to have a good agreement for $\cchi=3 ; 5$ but it has a poor accuracy at $\theta=30^\circ$.% \color{red} Ce premier modèle de force est très approximatif, en se basant uniquement sur la theorie de Stokes le modele n'est pas robuste. Faut que j'utilise les resultats sur les forces de pressions et visqueuses. \color {black}
\begin{figure}%
 \centering
\hspace{-8mm}\includegraphics[width=0.35\textwidth]{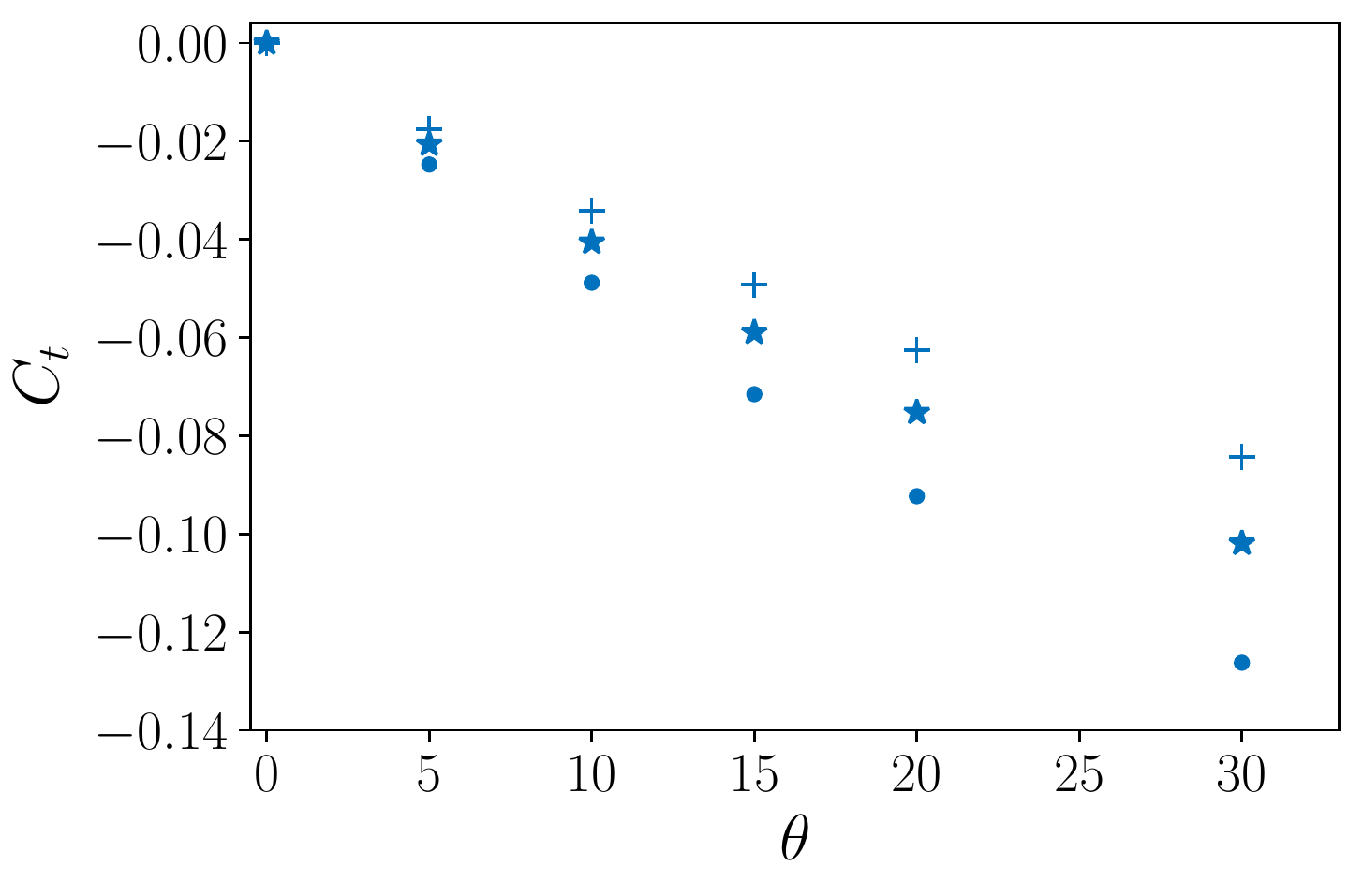}\label{fig:a}%
\hspace{-1mm}\includegraphics[width=0.35\textwidth]{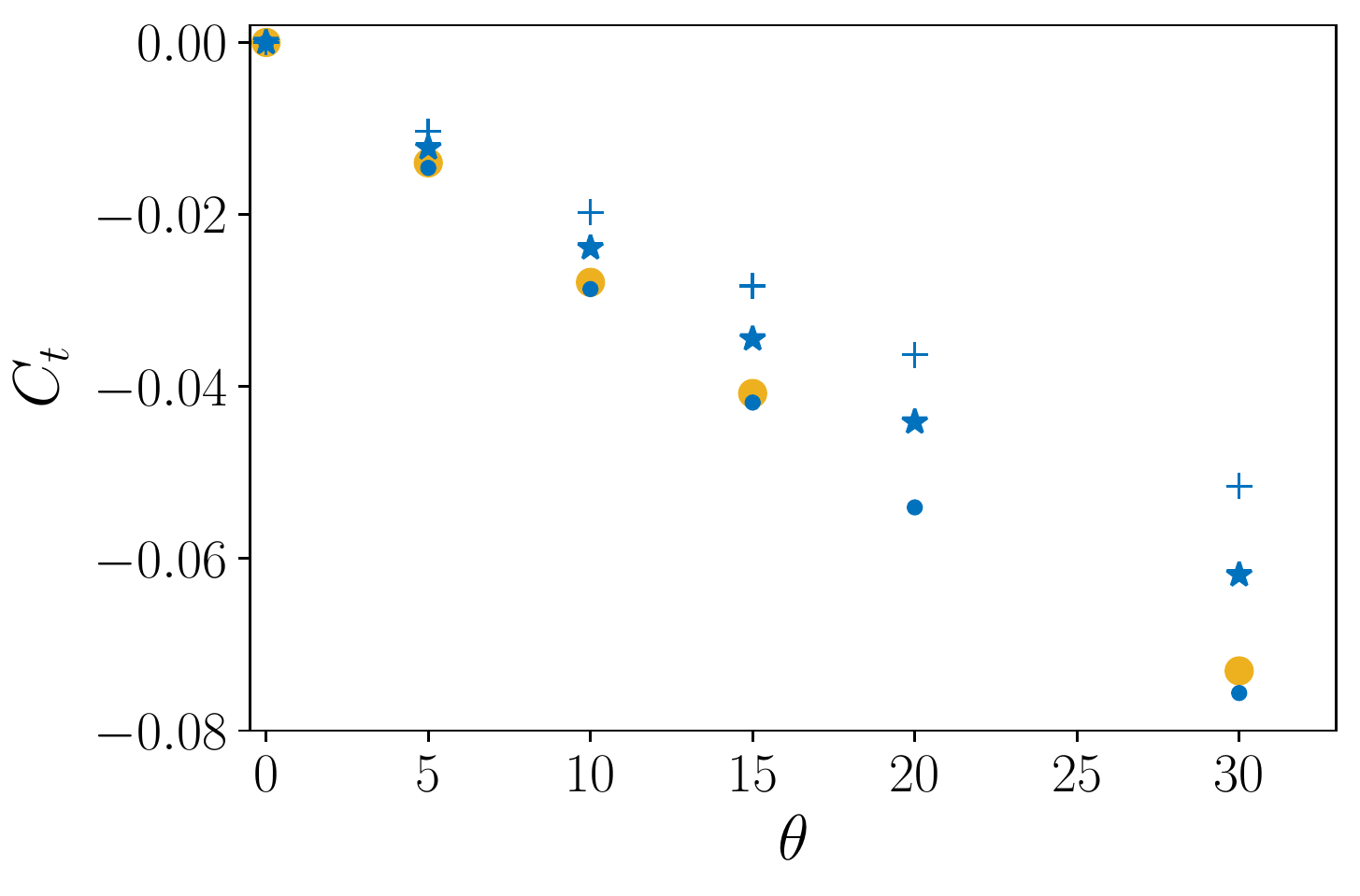}\label{fig:b}
\hspace{-2mm}\includegraphics[width=0.35\textwidth]{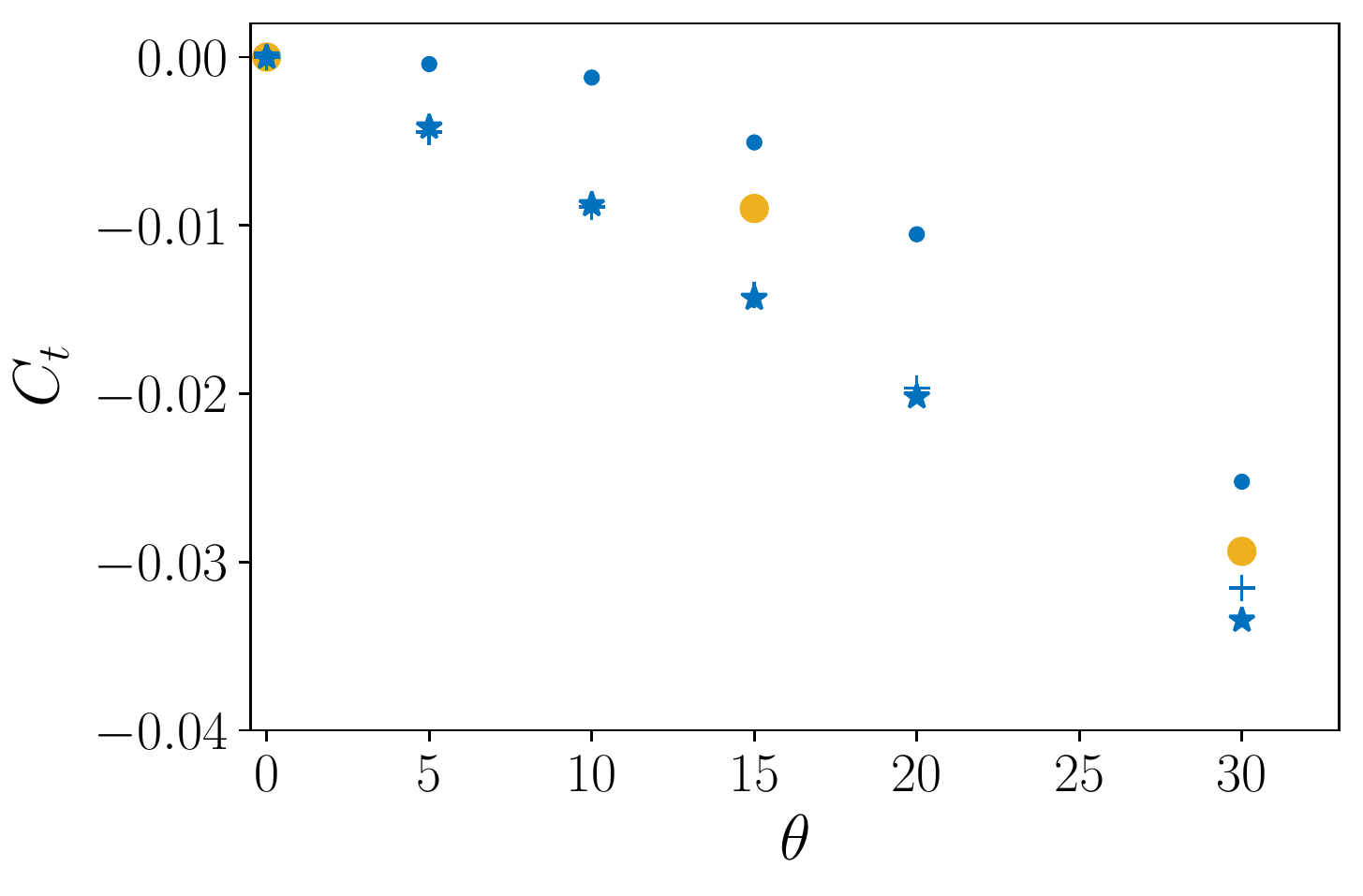}\label{fig:c}\\%\\
	\vspace{-4mm}
		\hspace{-40mm}$(a)$\hspace{54mm}$(b)$\hspace{54mm}$(c)$  \\
		\vspace{-13mm}
	\hspace{-16mm}$\text{Re}=20$\hspace{47mm}$\text{Re}=80$\hspace{47mm}$\text{Re}=300$
	\vspace{11mm}\\
 \caption{Torque coefficient vs. $\theta$ for $\bullet$: $\cchi=3$,  $\star$: $\cchi=5$, +  $\cchi=7$. The yellow bullets in panels $(b)$ and $(c)$ refer to the results of \cite{pierson2019} for $\cchi=3$ at $\text{Re}=75$ and $250$, respectively.%$(a)$ $\text{Re}=20$, $(b)$ $\text{Re}=80$, $(c)$ $\text{Re}=300$.
 }
 \label{fig:torque_theta}
\end{figure}\indent 
Figure \ref{fig:torque_theta} shows the variations of the torque coefficient as a function of $\theta$. \color{black} Similar to the low-to-moderate $\text{Re}$ regime, the torque is always negative, tending to orient the cylinder axis perpendicular to the upstream flow. \color{black} %As (\ref{KCt0}) indicates, torque variations with $\theta$ are proportional to $\sin2\theta$ in the low-but-finite Reynolds number limit. %More precisely, the leading-order prediction obtained in the limit $\cchi\gg1$ for $\cchi\text{Re}\ll1$ in \cite{khayat1989} reads in present notations
%\begin{equation}
%C_t(\cchi,\theta,Re\ll1)\approx-\frac{5\pi}{48}\frac{\cchi}{(\ln\cchi)^2}\sin2\theta\,.
%\label{KCt}
%\end{equation}
The torque coefficient exhibits a quasilinear increase with the inclination angle for $\cchi=5$ and $7$. \color{black} This may be seen as a natural extension of the $\sin2\theta$-variation characterizing the $C_t$-variations at low $\text{Re}$. %For this reason the magnitude of $C_{t}$ has an increasing evolution with $\theta$ as show in figure \ref{fig:torque_teta}. It is also found that 
$|C_t|$ is also found to decrease for increasing $\cchi$ at $\text{Re}=20$ and $\text{Re}=80$. The behavior observed at $\text{Re}=300$ is more complex, especially in the case of the shortest cylinder for which the angular dependence is strongly nonlinear for $\theta\lesssim15^\circ$. Moreover, while the corresponding $|C_t|$ is larger than those of the other two cylinders at $\text{Re}=20$ and $\text{Re}=80$, the situation is reversed at $\text{Re}=300$. \color{black} 
%This makes the torque barely vary from $\theta=20^\circ$ to $30^\circ$ at $\text{Re}=20$. As a result, for $\theta=30^\circ$, the torque coefficient for $\cchi=3$ is only $25\%$ (resp. $65\%$) of that for $\cchi=5$ at $\text{Re}=20$ (resp. $80$). \\ %For $\text{Re}=300$, the torque on the shortest cylinder is even found to change sign in the range $0^\circ<\theta\lesssim25^\circ$ (Fig. \ref{fig:torque_teta}$(c)$). Actually, this change of sign is first detected at $\theta\approx15^\circ$ for a threshold Reynolds number $\text{Re}\approx240$. These features suggest that cylinders with aspect ratios $\cchi\gtrsim5$ may almost be considered as infinitely long in inertia-dominated regimes, while finite-length effects deeply affect the torque on shorter cylinders whatever the Reynolds number.\\
Some additional insight into these variations may again be obtained by splitting the torque into pressure and viscous contributions provided by each part of the cylinder surface. \color{black} The main contributions resulting from this decomposition are displayed in Fig. \ref{fig:details_Ct}. The two dominant terms are seen to result from the pressure distribution on the lateral surface ($C_{{tp}_{lat}}$) and the viscous stress on the upstream end  ($C_{{t\mu}_{up}}$). Both terms decrease in magnitude with increasing $\cchi$ and $\text{Re}$, keeping a negative sign in all cases. %However, the former is negative while the latter is positive. Their (negative) sum remains one order of magnitude larger than any of the other contributions for $\cchi=5$ and $7$. In contrast, for $\cchi=3$, the small positive contributions provided by the pressure distributions on the upstream and downstream ends ($C_{{tp}_{up}}$ and  $C_{{tp}_{down}}$, respectively) and
 The third and fourth contributors, $C_{{tp}_{up}}$ and $C_{{t\mu}_{lat}}$, result from the pressure and viscous stress distributions on the same surfaces.  The contributions of the downstream end (not shown) are one order of  magnitude smaller than the dominant terms in all cases. %come into play, as their sum is of the same order of magnitude as $C_{{tp}_{lat}}+ C_{{t\mu}_{up}}$. All but one of the above terms exhibit a variation with the inclination angle close to the $\sin2\theta$ low-but-finite-$\text{Re}$ behavior, and they keep a constant sign whatever $\cchi, \text{Re}$ and $\theta$. 
In contrast to its viscous counterpart ($C_{{t\mu}_{up}}$), $C_{{tp}_{up}}$ is seen to keep a positive sign in all cases. The three contributions $C_{{tp}_{lat}}$, $C_{{tp}_{up}}$ and $C_{{t\mu}_{up}}$ vary almost linearly with $\theta$ whatever $\cchi$ and $\text{Re}$. %At a given $\text{Re}$, both contributions reduce as $\cchi$ increases, which is the hallmark of finite-length effects. 
The viscous contribution associated with the lateral surface, $C_{{t\mu}_{lat}}$ (Fig. \ref{fig:details_Ct}$(d)$), reveals a more complex behavior. First, its sign changes with $\cchi$ and $\text{Re}$. It stays positive whatever $\theta$ for the shortest cylinder, increasing in magnitude as $\text{Re}$ increases. Conversely, it is negative for the longest two cylinders at $\text{Re}=20$, gradually decreasing until changing sign at all inclinations for $\text{Re}=300$. Second, variations of $C_{{t\mu}_{lat}}$ with $\theta$  become increasingly nonlinear as $\cchi$ decreases and/or $\text{Re}$ increases. \color{black} 
%change exhibits a clear nonlinear variation with $\theta$ whatever $\text{Re}$ for $\cchi=3$. The same feature is observed, to a lesser extent, for the other two aspect ratios for $\text{Re}\geq80$. Moreover $C_{{t\mu}_{lat}}$ changes sign, depending on the aspect ratio and Reynolds number. 
This complex behavior is responsible for the markedly nonlinear variations of $C_t$ with $\theta$ noticed above for the shortest cylinder at $\text{Re}=300$. %, and for its change of sign within a finite range of inclinations at high enough Reynolds number (Fig. \ref{fig:torque_teta}$(c)$). 
The open separation process discussed in Sec. \ref{separ} is responsible for these features. Indeed, the dominant local viscous contribution to the spanwise torque provided by the the lateral surface is, in dimensional form, $-\mu\frac{D}{2}\frac{{\bf{r}}\cdot{\bf{e_y}}}{||{\bf{r}}||}({\bf{n}}\cdot\nabla) u_\parallel$, where $u_\parallel$ is the fluid velocity component parallel to this surface (along the $x$-direction), ${\bf{n}}$ is the unit normal directed into the fluid and ${\bf{r}}$ is the local position with respect to the cylinder geometrical center. As far as the fluid does not recirculate, this term is positive on the lower part of the surface (${\bf{r}}\cdot{\bf{e_y}}<0$) and negative on the upper part (${\bf{r}}\cdot{\bf{e_y}}>0$). However, when separation takes place, $u_\parallel$ is negative in the corresponding region of the upper part, which then provides a positive viscous contribution, making $C_{{t\mu}_{lat}}$ positive if the recirculation is strong enough. The larger the area percentage of the lateral surface corresponding to the separated region, the larger the positive value of $C_{{t\mu}_{lat}}$. \\
\indent %To approach empirically the variations of $C_t$ with the control parameters, the above observations suggest to seek separate fits for the two contributions emanating from the lateral surface ($C_{{tp}_{lat}}$ and $C_{{t\mu}_{lat}}$), plus another fit for the total contribution provided by each end, say $C_{{t}_{up}}=C_{{tp}_{up}}+C_{{t\mu}_{up}}$ and $C_{{t}_{down}}=C_{{tp}_{down}}+C_{{t\mu}_{down}}$ for the upstream and downstream ends, respectively.
 \begin{figure}%
 \centering
\hspace{-10mm}\includegraphics[width=0.33\textwidth]{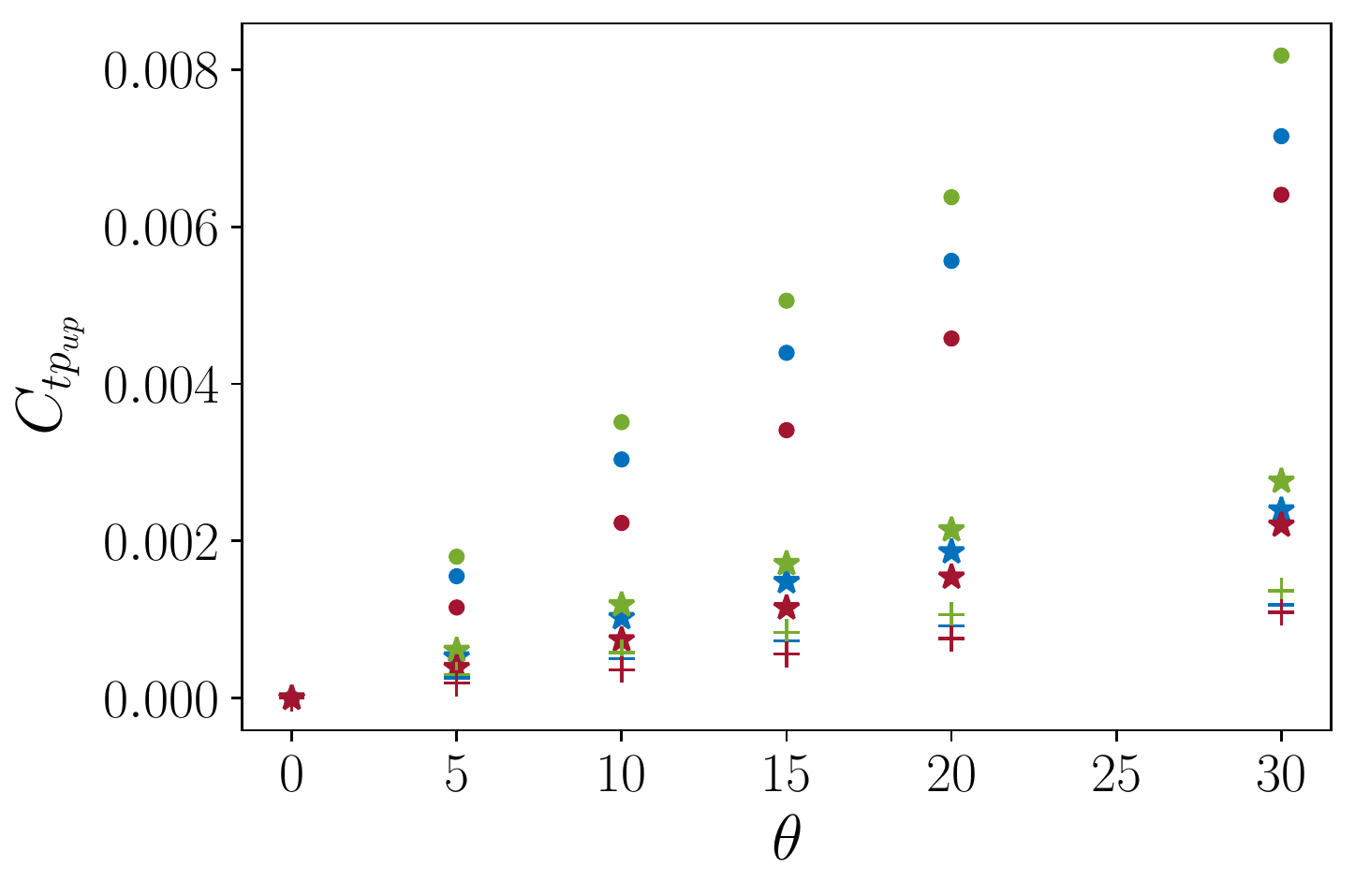}\label{fig:aat}\hspace{0mm}
\includegraphics[width=0.33\textwidth]{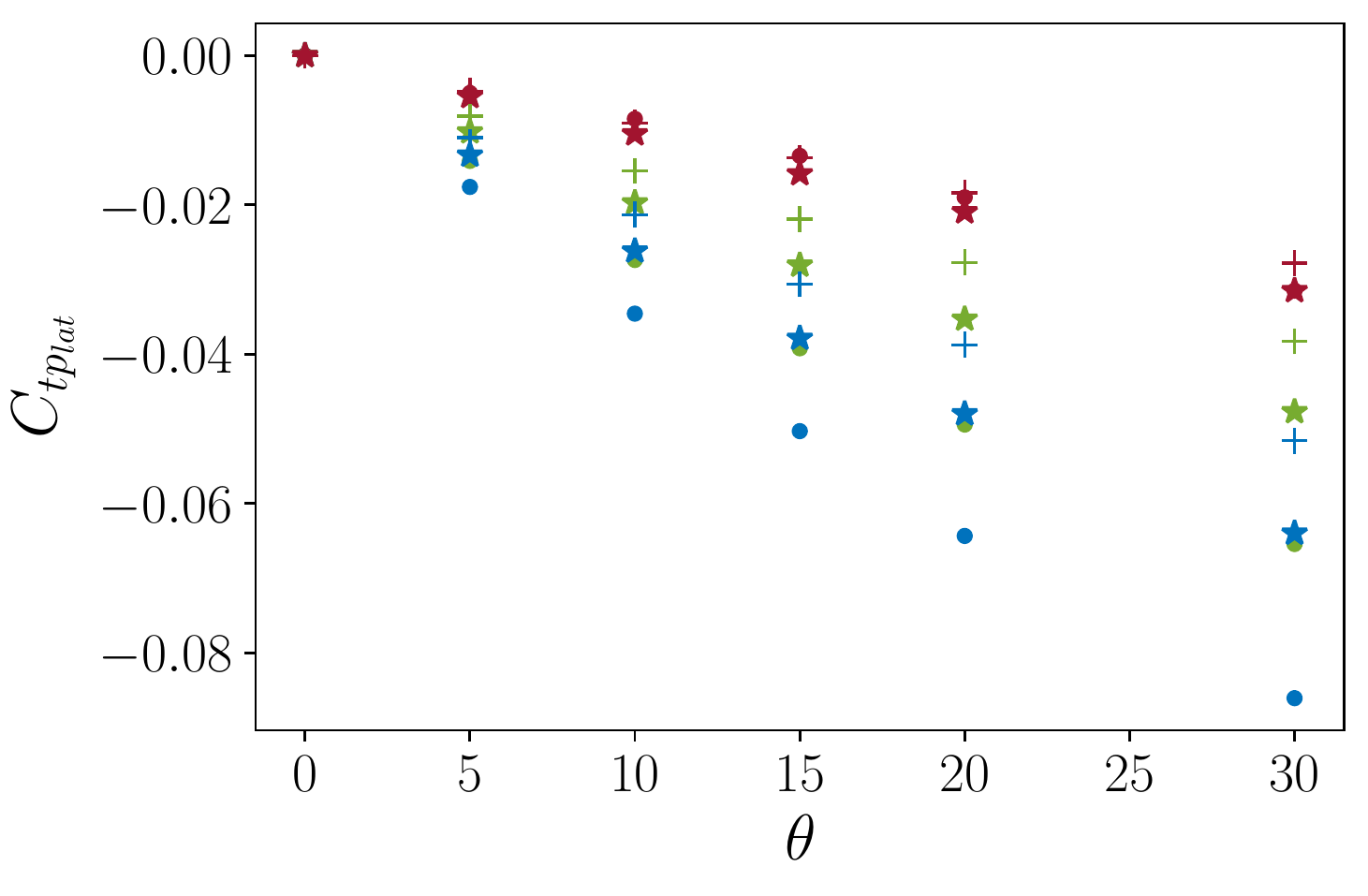}\label{fig:cct}\\
\vspace{5mm}
\hspace{-6.5mm}\includegraphics[width=0.33\textwidth]{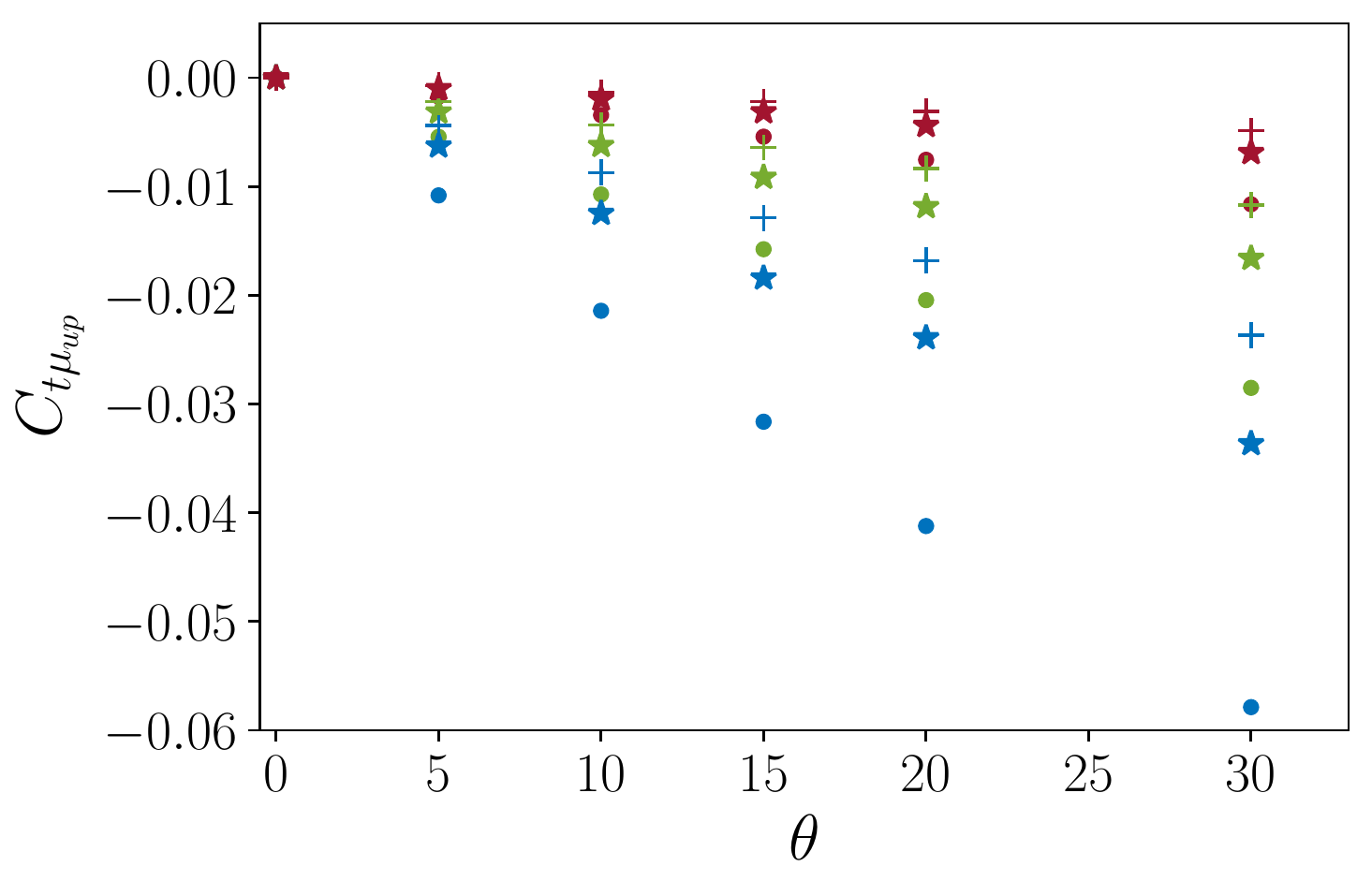}\label{fig:ddt}\hspace{-2mm}
\includegraphics[width=0.33\textwidth]{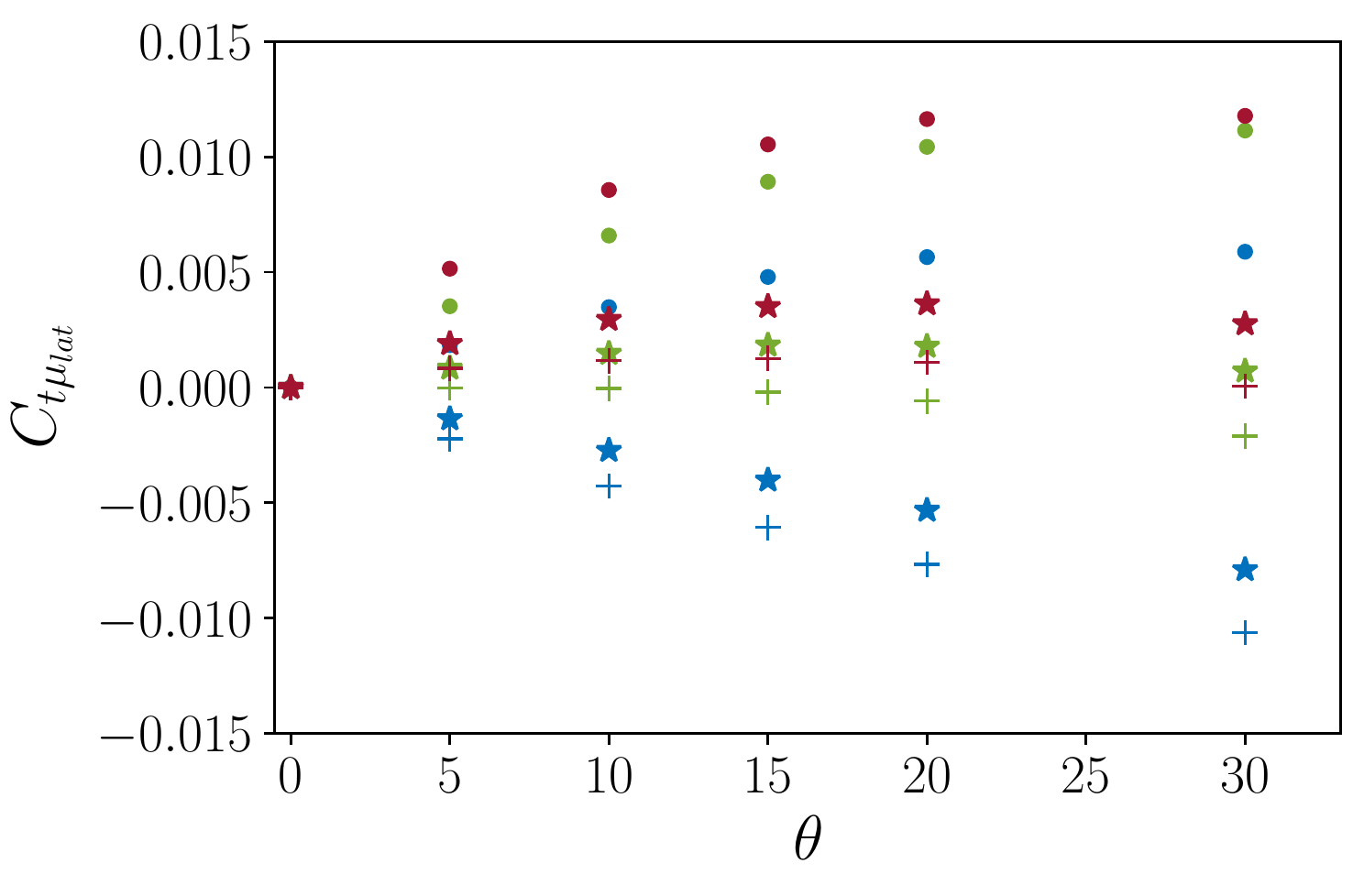}\label{fig:fft}
\vspace{-46mm}\\
\hspace{-45mm} $(a)$\hspace{55mm}$(b)$
\vspace{38.3mm}\\
\hspace{-45mm} $(c)$\hspace{55mm}$(d)$\\
\vspace{1mm}
 \caption{ Contributions to $C_t(\cchi,\theta,\text{Re})$ at $\text{Re}=20$ (blue), $\text{Re}=80$ (green) and $\text{Re}=300$ (red) for $\bullet$: $\cchi=3$, $\star$: $\cchi=5$, +: $\cchi=7$. $(a)-(b)$ pressure on the upstream end and the lateral surface, respectively; $(c)-(d)$ viscous stress on the upstream end and the lateral surface, respectively.}%
 \label{fig:details_Ct}
\end{figure}
\color{black} Following a fitting procedure similar to that described for $C_\parallel$, we approached the behaviors observed in Fig. \ref{fig:torque_theta} with the empirical expression
\begin{equation}
%C_{t}(\cchi,\theta, \text{Re})\approx C_{{tp}_{lat}}(\cchi,\theta, \text{Re})+C_{{t\mu}_{lat}}(\cchi,\theta, \text{Re})+C_{{t}_{up}}(\cchi,\theta, \text{Re})+C_{{t}_{down}}(\cchi,\theta, \text{Re})\,.
C_{t}(\cchi,\theta, \text{Re})\approx\cchi^{-0.47}\left\{-0.69\text{Re}^{-0.35-b_1(\text{Re}/\cchi)^{3.1}}\sin2\theta+1\times10^{-4}\text{Re}^{0.8+b_2(\text{Re}/\cchi)^{3.1}}\sin6\theta\right\}\,,
\label{Ctt}
\end{equation}
with $b_1=7\times10^{-8}$ and $b_2=5\times10^{-8}$.\color{black} 
\begin{figure}
	\centering
\hspace{-9mm}\includegraphics[width=0.33\textwidth]{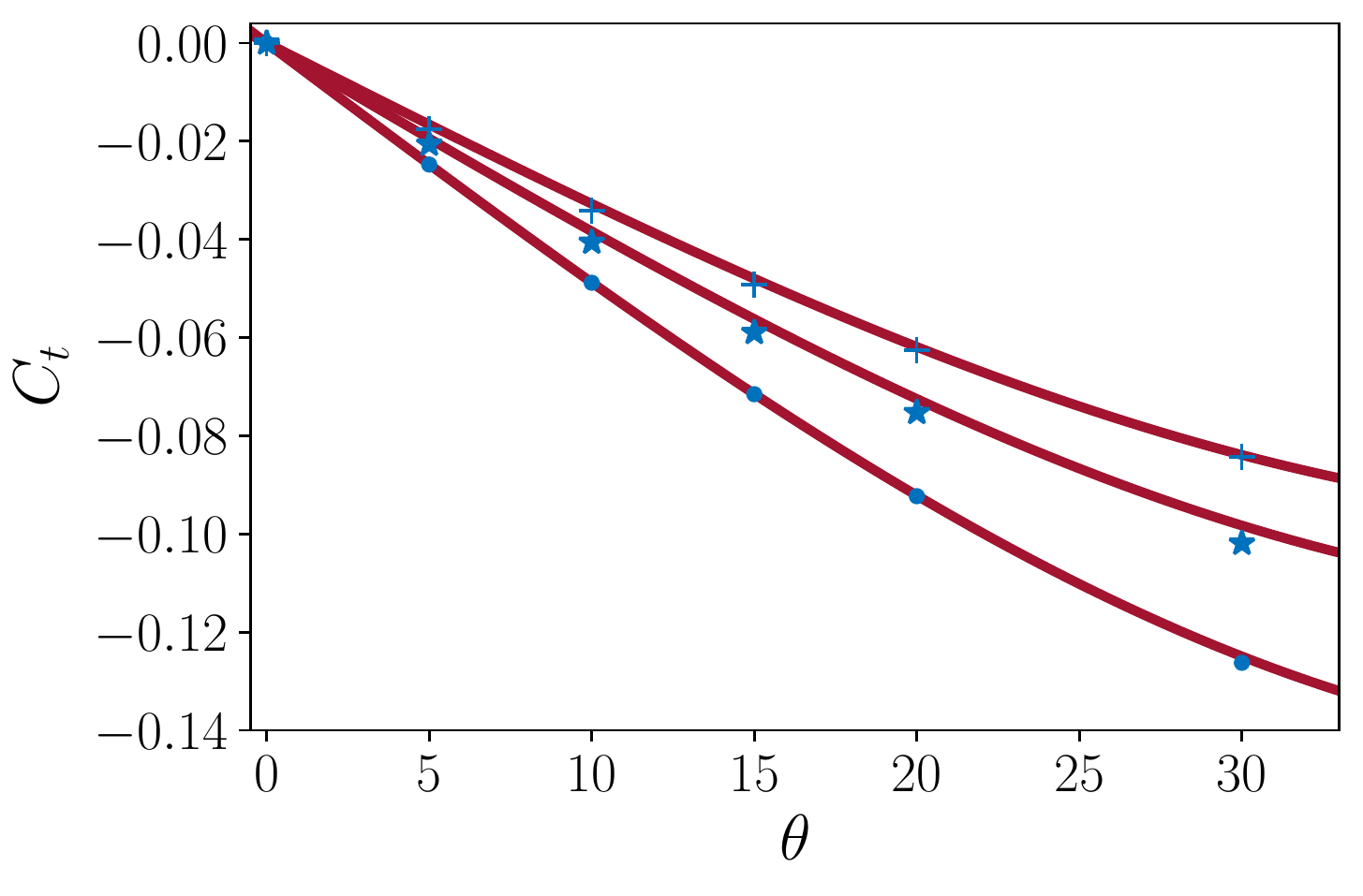}\quad
\hspace{-3.5mm}\includegraphics[width=0.33\textwidth]{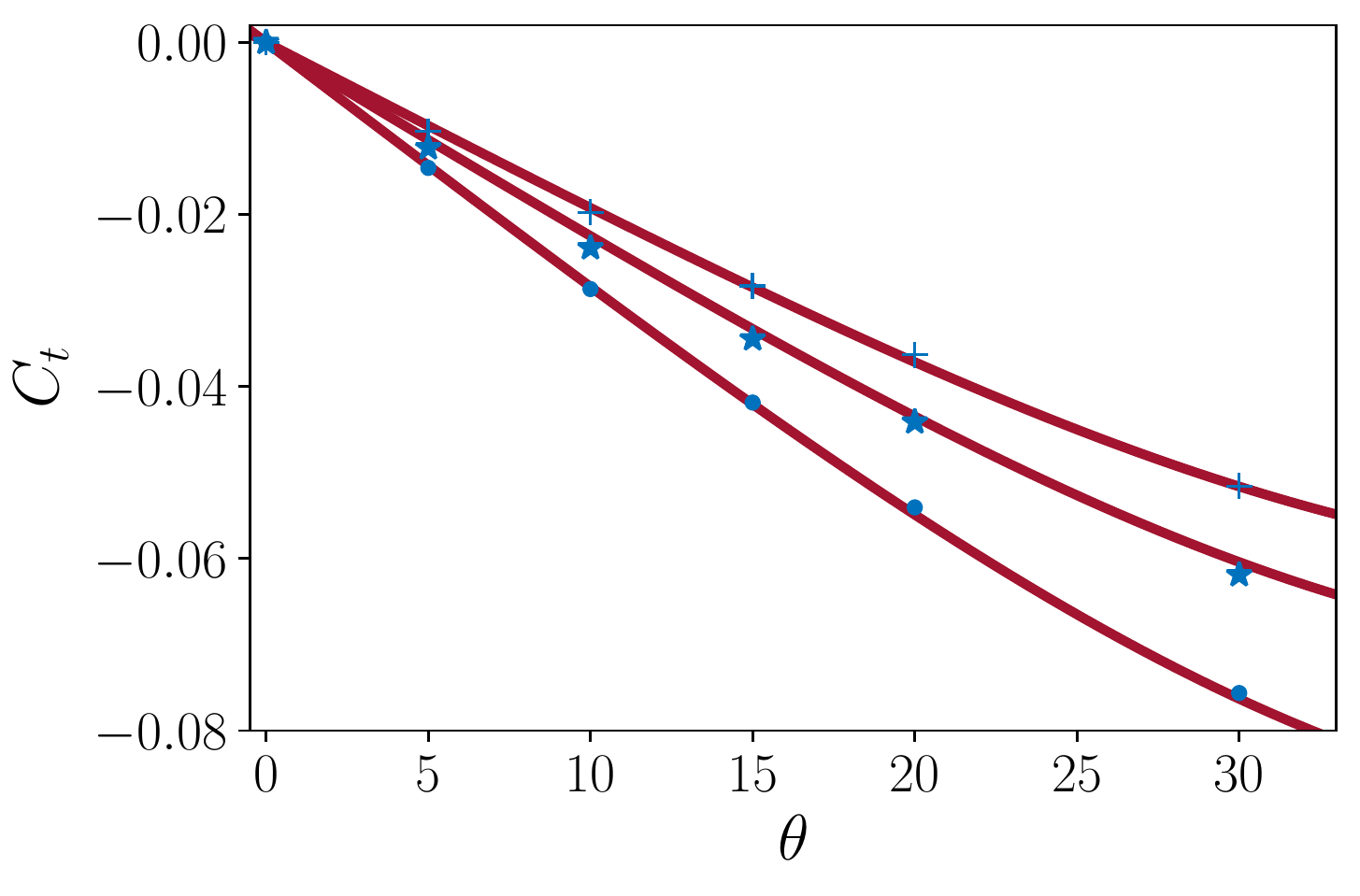}\quad
\hspace{-3.5mm}\includegraphics[width=0.33\textwidth]{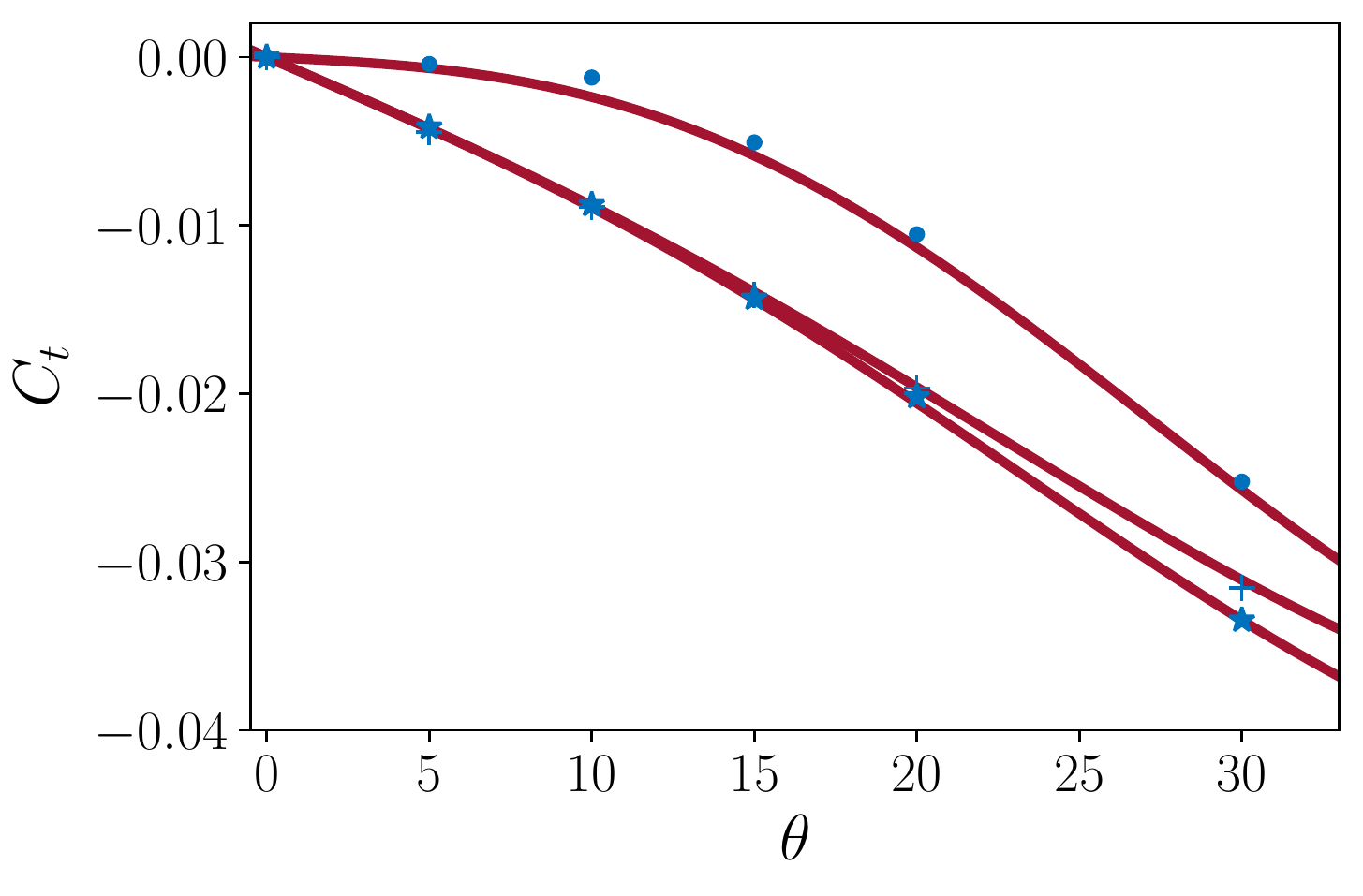}\\
\vspace{-4mm}
		\hspace{-40mm}$(a)$\hspace{52mm}$(b)$\hspace{52mm}$(c)$  \\
		\vspace{-13mm}
	\hspace{-16mm}$\text{Re}=20$\hspace{45mm}$\text{Re}=80$\hspace{44mm}$\text{Re}=300$
	\vspace{10mm}\\
\caption{Torque coefficient $C_t(\cchi,\theta,\text{Re})$ vs. $\theta$ for $\bullet$: $\cchi=3$, $\star$: $\cchi=5$, +: $\cchi=7$. % $(a)$: $\text{Re}=20$; $(b)$: $\text{Re}=80$; $(c)$: $\text{Re}=300$; 
Solid line: empirical fit (\ref{Ctt}).}
\label{fig:ctfit}
\end{figure}
As Fig. \ref{fig:ctfit} shows, the above fit provides a correct estimate of $C_t$ throughout the range of parameters explored in the present investigation. \color{black} For large enough aspect ratios, the dominant contribution to $C_t$ is still proportional to $\sin2\theta$, similar to the low-$\text{Re}$-regime, and its magnitude varies as $\cchi^{-0.47}Re^{-0.35}$. %Note that, in contrast with the case of $C_\parallel$, the main dependence of $C_t$ with respect to the aspect ratio is not a finite-length effect vanishing at large $\cchi$ for each Reynolds number. Instead, t
The torque coefficient is seen to be approximately proportional to $\cchi^{-1/2}$ whatever $\text{Re}$, which suggests that the dimensional torque behaves roughly like $L^{3/2}$ for long enough cylinders. Finite-length high-$\text{Re}$ effects associated with the open separation are translated into the above fit through the slight but sharp $(\text{Re}/\cchi)^{3.1}$-increase of the Reynolds number exponent. \color{black}  % In particular, the positive contributions associated with finite-length effects in (\ref{eq:torquelat1})-(\ref{eq:torquedown}) are seen to make the torque on the shortest cylinder positive in the range $0<\theta\lesssim30^\circ$ at $\text{Re}=300$, as observed in the computations. 
Similar to the case of $C_\parallel$, the above fit is not expected to be valid for Reynolds numbers significantly larger than the upper bound considered in the simulations. \color{black} Moreover, although (\ref{Ctt}) respects the constraint that the torque changes sign for $\theta=90^\circ$, the results of \cite{pierson2019} for a cylinder with $\cchi=3$ indicate that the behavior of $C_t$ changes significantly for $\theta\gtrsim45^\circ$, suggesting that  the present fit is appropriate only below this critical inclination. \color{black} %Check the matching at $\mathcal{O}(1)-\text{Re}$ with the low-but-finite-$\text{Re}$ expression. \color{black}
%This means that there is a critical Reynolds beyond which the flow tends to rotate the cylinder in the sens of its inclination while for lower $\text{Re}$ (figure \ref{fig:torque_teta}(a,b)) the flow resists this rotation. \color{red} Cette partie sur le torque est incomplete, dans le sens où il manquerait la loi de moment ainsi que l'explication du torque qui change de signe à $\text{Re}=300$. En regardant l'écoulement, je n'observe rien de particulier sachant que c'est à partir de $\text{Re}=240$ que le changement de signe a lieu au alentour de . Ceci laisse penser qu'il existe une zone critique qui apparait assez facilement pour les petits rapports de forme où le moment exercé tend à écarter davantage le cylindre de sa position d'equilibre. Pour l'instant je n'ai pas tout les éléments pour justifier tout cela. \color{black}
\section{Summary and concluding remarks}
\label{Summ}
With the practical objective of providing approximate laws for predicting the translation-induced drag, lift and torque acting on long cylindrical rods and fibers, we employed fully-resolved simulations to investigate the flow around a finite-length circular cylinder held fixed in a uniform stream making some angle with the body axis. \\
\indent We first focused on the specific case where the cylinder is aligned with the incoming flow. Considering the Stokes regime and the weakly inertial regime corresponding to $\text{Re}\lesssim1$, we combined numerical results with available predictions from the slender-body theory (which we slightly improved by computing the next-order term in the expansion with respect to the small parameter $1/\ln(2\cchi)$) to build the approximate drag laws (\ref{eq:slender_modified}) and (\ref{eq:slender_smallinertia2}). The former is valid down to $\cchi\approx1.5$, while the inertial corrections contained in the latter allow an accurate estimate of the drag up to $\text{Re}=\mathcal{O}(1)$, and even up to $\text{Re}=\mathcal{O}(10)$ for $\cchi\gtrsim20$. 
For larger Reynolds numbers (up to $\text{Re}=400$), the flow structure becomes more complex, although it remains stationary and axisymmetric. %The length of the standing eddy attached at the back of the body grows as $\text{Re}^{1/2}$. Then, 
Beyond a $\cchi$-dependent critical Reynolds number of the order of $200$, a second recirculating region emerges along the upstream part of the lateral surface. %Emanating from the upstream edge, it looks like a flat annulus, the length of which grows with $\text{Re}-\text{Re}_{c0}$. 
Being associated with local negative shear stresses, this lateral eddy acts to reduce the friction drag, which may even become negative if $\text{Re}$ is large enough and the cylinder is short enough. We used the numerical data to build approximate fits for this friction drag and for the pressure drag contribution of the upstream and downstream ends. %Remarkably, this pressure drag becomes virtually independent of the aspect ratio for Reynolds numbers of a few hundreds. 
With this procedure, we obtained the empirical drag law (\ref{eq:cd_theta0}) which approximates the drag well for $\cchi\gtrsim2$ in the range $20\leq \text{Re}\leq400$ and properly matches  (\ref{eq:slender_smallinertia2}) for $\text{Re}=\mathcal{O}(1)$. The friction drag still represents a substantial part of the total drag at Reynolds numbers of several hundreds if the body aspect ratio is large enough ($45\%$ at $\text{Re}=400$ for a cylinder with $\cchi=10$).\\
\indent \color{black}
In the next step, we examined the case of moderately inclined cylinders ($\theta\leq30^\circ$) in the low-to-moderate Reynolds number regime ($\text{Re}\leq5$). For Reynolds numbers less than unity and aspect ratios up to $10$, we observed that the force component parallel to the cylinder axis closely follows the $\cos\theta$-variation predicted under creeping-flow conditions. The agreement deteriorates as the length-based Reynolds number $\cchi\text{Re}$ exceeds values of $\mathcal{O}(10)$. Under more inertial conditions, the $\cos\theta$-law under-predicts the actual parallel force, the difference increasing with both $\cchi\text{Re}$ and $\theta$. The force component perpendicular to the cylinder axis was found to closely follow the $\sin\theta$-variation typical of creeping-flow conditions up to $\text{Re}=5$, irrespective of $\cchi$. However, the corresponding pre-factor deviates from the creeping-flow prediction as soon as $\cchi\text{Re}\gtrsim0.5$, beyond which inertial effects become significant. Accurate predictions are obtained up to $\cchi\text{Re}\approx10$ by estimating the pre-factor of the $\sin\theta$-law through the semi-empirical formula (\ref{eq:slender_perp_smallinertia_m}) which provides a finite-Reynolds-number approximation of $F_\perp^{\theta=90^\circ}$. Throughout the low-to-moderate Reynolds number range, the inertial torque follows the $\sin2\theta$-variation predicted by the asymptotic theory of \citet{khayat1989} in the limit $\cchi\text{Re}\ll1$. However the magnitude of the torque is correctly predicted by this theory only up to $\text{Re}=\mathcal{O}(1)$ and provided $\text{Re}/\cchi^2\lesssim0.01$. To obtain a correct estimate of the torque over a broader range of conditions, we derived the semiempirical law (\ref{KCJM}) which correctly reduces to the theoretical prediction in the limit $\cchi\text{Re}\ll1, \cchi\gg1$ and closely approaches numerical data for cylinders with $\cchi\geq5$ up to $\text{Re}=5$.\color{black}\\
\indent Last we considered the three-dimensional flow past moderately inclined cylinders with aspect ratios in the range $3-7$ for $20\leq\text{Re}\leq300$. %In this range, we found that the wake, be it steady or not, always preserves a symmetry with respect to the plane containing the body axis and the incoming velocity. 
For $\theta\leq30^\circ$, the flow remains stationary irrespective of the inclination and preserves a symmetry with respect to the plane containing the body axis and the incoming velocity. %The wake is then made of a pair of counter-rotating streamwise vortices, and a nonzero lift force lying in the symmetry plane takes place. 
For sufficiently low inclinations and Reynolds numbers, the flow separates only at the back of the body, the recirculating region looking like an open toroid. % from which fluid particles eventually escape downstream in the wake, close to a free streamline emanating from the `trailing edge' of the cylinder located at the intersection of its downstream end and its lower generatrix. 
In contrast, beyond a critical inclination decreasing as $\text{Re}$ increases, an open separated region emerges on the `extrados' of the lateral surface, near its `leading edge'.
%located at the intersection of the upstream end and the upper generatrix %Increasing $\theta$ or $\text{Re}$ makes the position of the open separation surface move further downstream, so that the fluid recirculates over a significant fraction of the upstream part of the lateral surface. Then, depending on their position with respect to the `eye' of the separated region, fluid particles escape from this region and join either the downstream part of the lateral surface or directly the downstream flow, being in this case advected close to a second open streamline emanating from the `leading edge'. 
In such configurations, the flow past the cylinder looks massively separated in between the two free streamlines emanating from its `trailing' and `leading' edges. \\%, \textit{i.e.} the intersection of its downstream end and its lower generatrix, and that of its  upstream end and its upper generatrix, respectively. \\
\indent We used numerical data collected in this fully inertial regime to obtain approximate laws for the loads acting on the cylinder. Similar to the the low-$\text{Re}$ behavior, the perpendicular force obeys essentially a $\sin\theta$-variation with a mild dependence with respect to the aspect ratio. We found that the corresponding force coefficient $C_\perp$ may be related to the drag coefficient of a cylinder held perpendicular to the incoming flow through the emprical law (\ref{corrected}) \color{black} which involves two simple independent corrections, one proportional to $\text{Re}^{-1/2}$ accounting for inertial effects, the other for finite-length effects.\color{black}%This suggests that, for $\text{Re}\gtrsim10$, the perpendicular force is controlled by boundary layer effects. 
%The scaling laws revealed by the corresponding expression (\ref{corrected}) for $C_\perp$ contradict those predicted by the so-called Independence Principle. In line with previous findings, this contradiction confirms that this `principle' may only apply to configurations with $\theta$ close to $90^\circ$ and cannot be used to predict the perpendicular force on moderately inclined cylinders. 
\begin{figure}
	\centering
\hspace{-9mm}\includegraphics[width=0.33\textwidth]{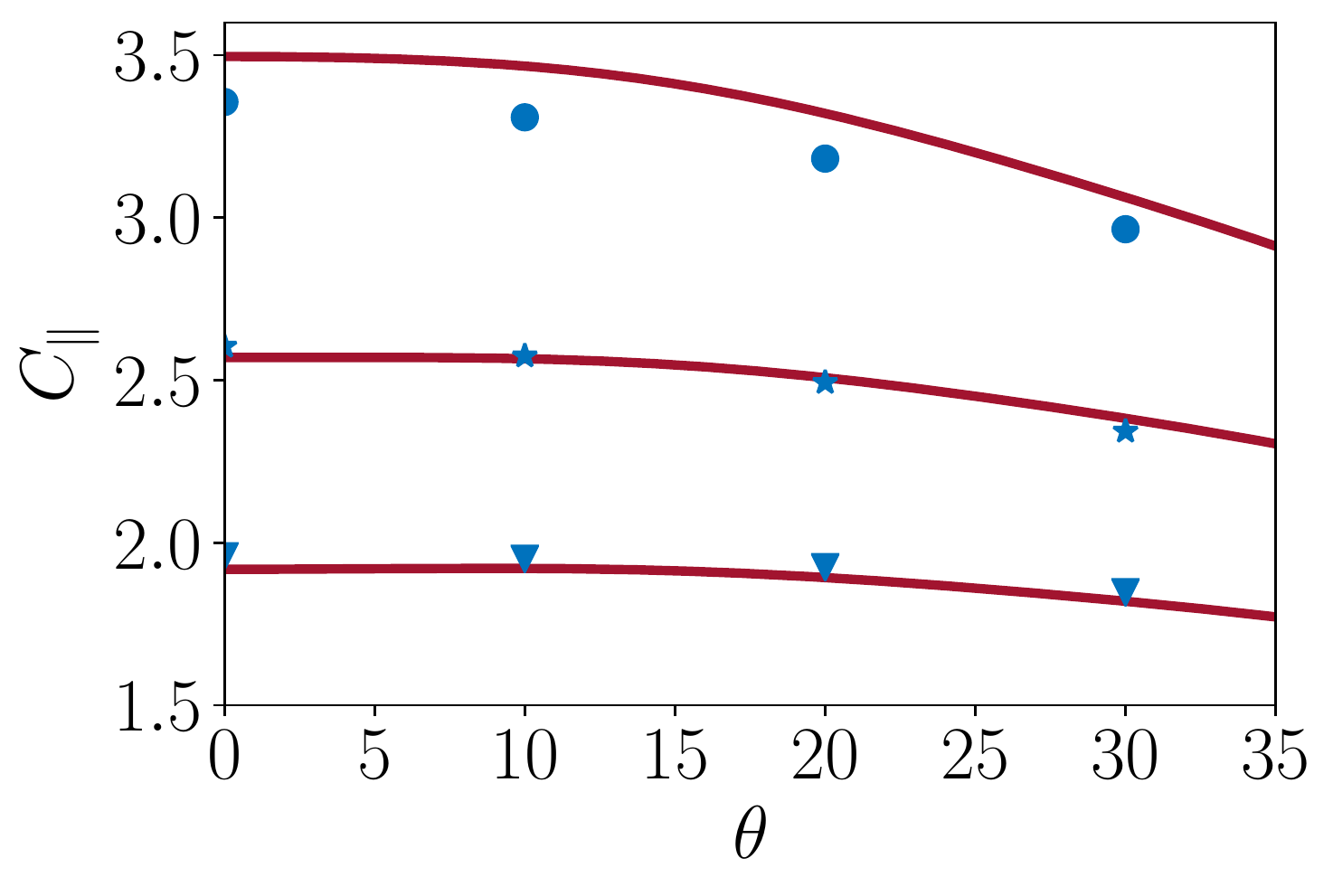}\quad
\hspace{-3.5mm}\includegraphics[width=0.33\textwidth]{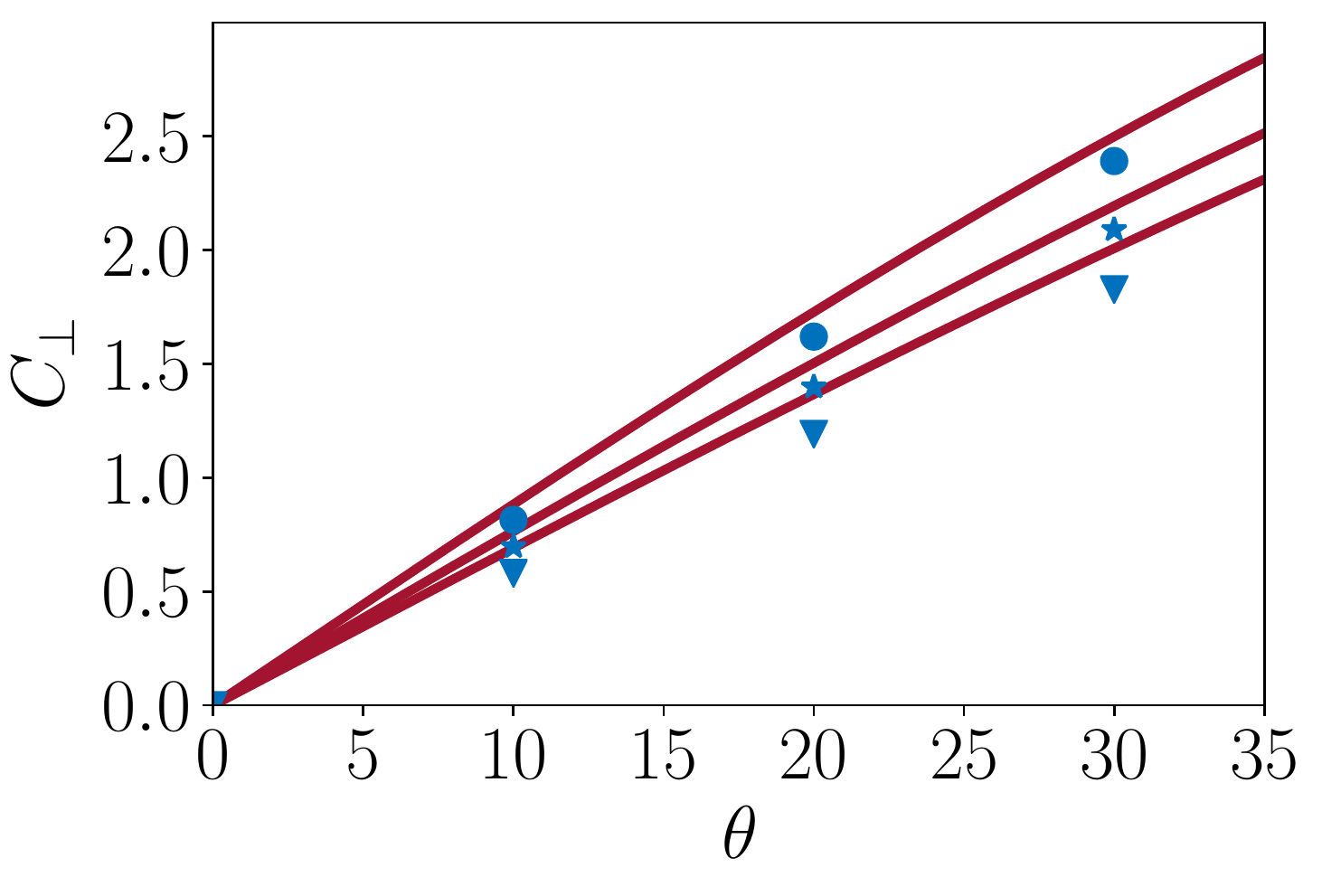}\quad
\hspace{-3.5mm}\includegraphics[width=0.33\textwidth]{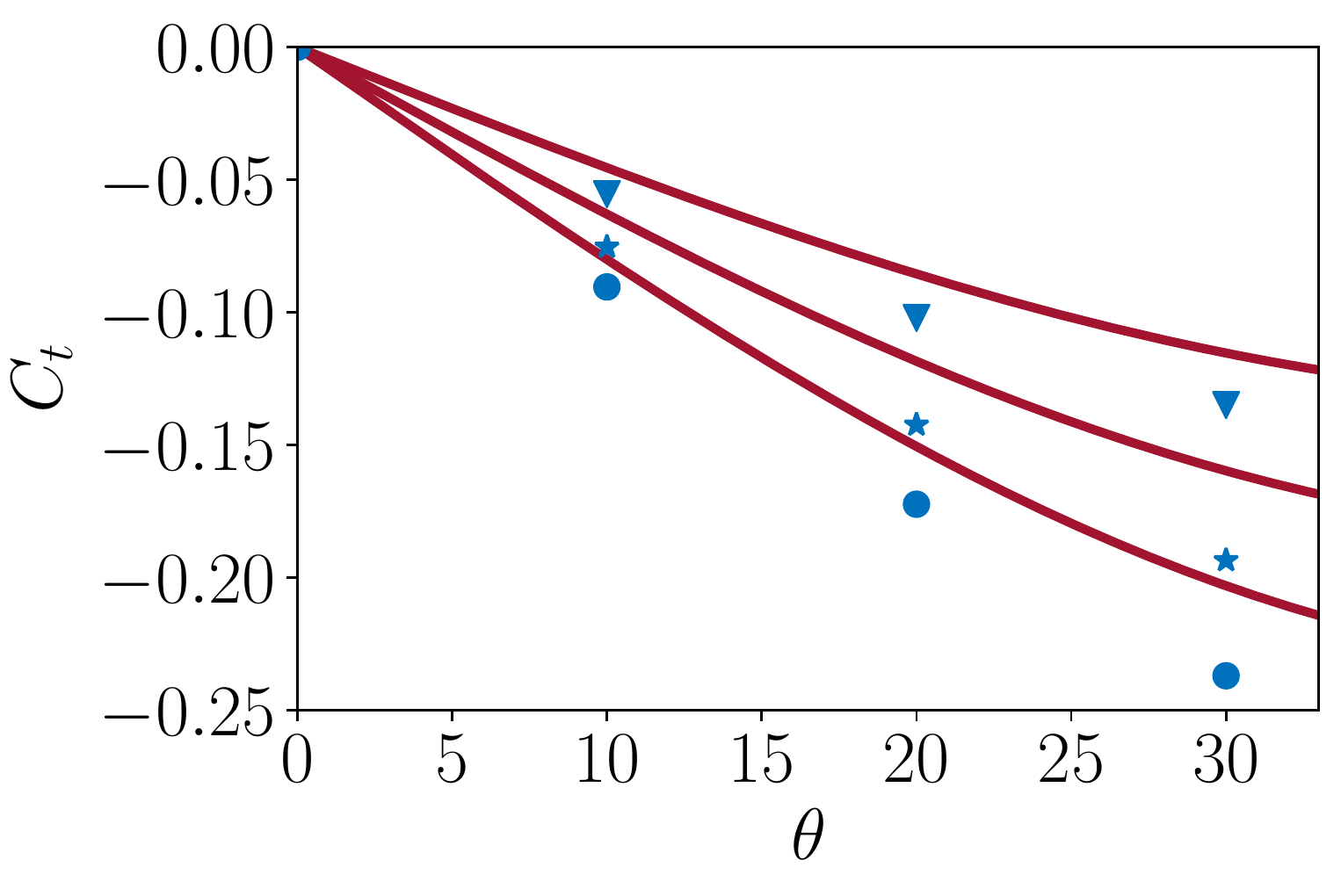}\\
\vspace{-4mm}
		\hspace{-40mm}$(a)$\hspace{52mm}$(b)$\hspace{54mm}$(c)$  \\
\caption{Extrapolated predictions at $\text{Re}=5$ of fits derived in the fully inertial regime for $\bullet$: $\cchi=3$, $\star$: $\cchi=5$, +: $\cchi=7$. $(a)$ $C_\parallel$ and fit (\ref{eq:cparfin}); $(b)$ $C_\perp$ and fit (\ref{corrected}); $(c)$ $C_t$ and fit (\ref{Ctt}).} % $(a)$: $\text{Re}=20$; $(b)$: $\text{Re}=80$; $(c)$: $\text{Re}=300$; 
\label{fig:fitsRe5}
\end{figure}
In contrast, variations of the parallel force do not follow the $\cos\theta$-law prevailing in the low-$\text{Re}$ regime. Instead, the force coefficient $C_\parallel$ barely varies with $\theta$ in the moderate-$\text{Re}$ regime and even increases with the inclination at large Reynolds number. A fitting procedure with respect to the three control parameters allowed us to mimic the influence of inertial and finite-length effects on $C_\parallel$ through the empirical law (\ref{eq:cparfin}) which reproduces all observed trends well. Variations of the spanwise torque at moderate Reynolds number are qualitatively similar to those observed for $\text{Re}\leq5$. However strong finite-length effects manifest themselves at high Reynolds number, in connection with the separation process affecting the upstream part of the lateral surface in this regime. A fitting approach similar to that employed for $C_\parallel$ yielded the empirical law (\ref{Ctt}) which correctly approximates $C_t$ throughout the explored range of $\cchi$ and $\theta$ in the fully inertial regime. \color{black}  For application purposes, it is of course of interest to know how robust these fits determined from data in the range $20\leq\text{Re}\leq300$ are when the Reynolds number is decreased to the upper limit of the low-to-moderate $\text{Re}$ regime considered in Sec. \ref{incl1}. The results or this test are summarized in Fig. \ref{fig:fitsRe5}. It turns out that the empirical formula established in the fully inertial regime still perform quite well for $\text{Re}=5$. Predictions depart from numerical data by less than $5\%$ for $C_\parallel$, $10\%$ for $C_\perp$ and $15\%$ for $C_t$. Therefore, the semi-empirical predictions derived in Sec. \ref{incl1} combined with the fits established in Sec. \ref{incl2} offer a complete and almost smooth description of load variations from the creeping-flow regime up to $\text{Re}=300$. \color{black}\\
%and spanwise torque with the three control parameters were found to be much more complex, especially due to strong finite-length effects. %On the one hand, $C_\parallel$ strongly depends on the aspect ratio whatever $\text{Re}$, while its variations with $\theta$ do not change monotonically with $\text{Re}$. 
%On the other hand, the torque becomes almost independent of the aspect ratio for $\cchi\gtrsim5$ but is dominated by finite-length effects for shorter cylinders. %The most spectacular consequence of these effects is the change of sign of the torque within a finite range of $\theta$ for $\cchi=3$ when the Reynolds number exceeds a critical value slightly larger than $200$. In such a case, the torque tends to realign the cylinder with the incoming flow instead of orienting it broadside on. 
%Most of these finite-lengths effects are directly related to the separation process that affects the upstream part of the lateral surface. 
%To obtain reasonable empirical fits for $C_\parallel$ and $C_t$, we had to decompose the total force or torque into elementary contributions provided by each part of the cylinder surface, and to split some of these contributions into pressure and viscous components. Thanks to this decomposition, we obtained 
\indent Most computations only considered aspect ratios below $10$ or even $7$ for $\text{Re}\geq20$, owing to the rapid increase of computational costs with $\cchi$. However, finite-length effects were found to decrease monotonically and sharply as $\cchi$ increases, making us confident that the various empirical laws derived in the course of this study remain valid for cylindrical particles with larger aspect ratios, and therefore apply to long fibers. Obviously this does not mean that the load coefficients become independent of $\cchi$ for $\cchi\gg1$, but simply that their asymptotic dependence with respect to $\cchi$ in the limit of large aspect ratios is already captured by considering $\mathcal{O}(10)$-aspect ratios as we did. The situation is less clear regarding their range of validity with respect to the inclination angle. All of them were calibrated in the range $|\theta|\leq30^\circ$ and satisfy the required geometrical constraints for $|\theta|=90^\circ$ and the associated symmetry conditions. Nevertheless, as soon as the Reynolds number exceeds a few tens, the dynamics of the flow past a cylinder in the configuration $\theta=90^\circ$ drastically differs from that past the same cylinder for $\theta=0^\circ$. Hence, physical features that are not present in the low-to-moderate inclination range considered here take place in the near-body flow when the inclination exceeds $45^\circ$ or so, which is likely to make the empirical laws proposed here invalid for such large inclinations.\vspace{2mm}\\
\indent The present investigation leaves several important configurations and parameter ranges unexplored. First, for Reynolds numbers similar to those considered in Secs. \ref{incl1} and \ref{incl2}, the above discussion calls for a specific study focused on large inclinations, say $45^\circ\lesssim|\theta|\leq90 ^\circ$, for which the flow past the cylinder is expected to be massively separated and most of the time unsteady. 
Another series of questions arises when the cylinder is allowed to rotate about an axis perpendicular to its symmetry axis and passing through its geometrical center, as rodlike particles and fibers customarily do. The torque on a slender rotating cylinder was predicted in the creeping flow limit in \cite{batchelor1970,cox1970} but no theoretical attempt to derive inertial corrections in this configuration has been reported so far. This is even more true for the general situation in which the cylinder undergoes both a translation and a rotation. In such a case, inertial effects couple the two types of motion, yielding specific contributions to the loads, which are for instance responsible for the well-known Magnus effect on a spinning sphere. To the best of our knowledge, such couplings have not been considered for slender cylinders. We are currently investigating numerically the configuration in which the cylinder undergoes an imposed rotation, and plan to apply the same methodology to the combined translation+rotation case in the near future.
\section*{Acknowledgments}
M. K.'s fellowship was provided by IFP Energies Nouvelles whose financial support is greatly appreciated. The authors thank Annaïg Pedrono for her continuous support with the use of the JADIM code. Part of the computations were carried out on the national supercomputers operated by the GENCI organization under allocation A0072B10978.

\appendix
\section{Specific numerical validations}
\label{app:num}
As mentioned in Sec. \ref{num}, the JADIM code was extensively used in the past to compute flows past axisymmetric bodies. In particular, transitional flows past disks and short cylinders were considered in \cite{fabre2008,auguste2010bif,auguste2013}. Nevertheless, we performed additional validations relevant to the present physical configuration by considering the flow past an inclined cylinder of aspect ratio $\cchi=3$ for different Reynolds numbers and inclination angles. 

% our reference case is the first one (10 cells per boundary layer thickness $\delta$). We choose the mean value of the drag force coefficient $<C_d>$ as a criterion to compare the three cases.  
%  \begin{table}[H]
%	\centering
%\begin{tabular}{l|lll|ll}
%	\hline
%	& \multicolumn{3}{c}{Number of cells along the boundary layer} & \multicolumn{2}{c|}{$E(C_d) \%$} \\
%	\cline{2-6} \cline{3-6}
%	   & 10  & 8  &5 & $ $  \\
% \hline
%	$(Re=100,\theta=5^\circ)$   & 0.481 &  0.479 & 0.478 & 0.4& 0.62 \\
%	$(Re=100,\theta=15^\circ ) $  & 0.532 &   0.534 & 0.546 & 0.37&2.2       \\
%	$(Re=200,\theta=15^\circ)$    & 0.41 &  0.411  &  0.416 & 0.2& 1.4 \\
%	\hline
%\end{tabular}
%		\caption{Comparison of the mean drag coefficient $<C_d>$ for three configurations : 10 ; 8 ; and 5 grid points in the thickness of the boundary layer $\delta$. $E(C_d)$ represent the relative error on the drag coefficient compared to the case of $10$ cells per $\delta$  \\}
%\label{table1}
%\end{table}

\begin{table}[H]
\centering
\begin{tabular}{cccc}
\hline 
                    & Number of cells across the boundary layer & $C_d$ & Error \% \\
$\text{Re}=100,\theta=5^\circ$ & 5 & 0.478 & 0.62\\
                        & 8 & 0.479 & 0.4\\ 
                        & 10& 0.481 & - \\ 

\hline 
$\text{Re}=100,\theta=15^\circ$ & 5 & 0.546 & 2.2\\
                         & 8 & 0.534 & 0.37\\ 
                         & 10& 0.532 & - \\ 
\hline 
$\text{Re}=200,\theta=15^\circ$ & 5 & 0.416 & 1.4\\  
                         & 8 & 0.411 & 0.2\\ 
                         & 10& 0.410 & - \\
\hline
\end{tabular} 
		\caption{Drag coefficient obtained with three different grids (with 5, 8 and 10 cells across the boundary layer, respectively), and three different flow configurations. The relative error is based on the most refined grid.}
\label{table1}
\end{table}
\noindent We first performed runs with an increasing number of cells across the boundary layer, the thickness of which is estimated as $D\text{Re}^{-1/2}$. Table \ref{table1} shows the effect of the grid refinement on the drag coefficient (here defined as the drag force normalized by${LD\rho U^2}/2$) for three different configurations. Clearly, eight cells across the boundary layer suffice to properly capture viscous effects, since the relative difference with the drag obtained on the most refined grid is less than 1\% in each configuration. 
%We could choose the third case (5 cells) which is not too wrong since the gap is around $3\%$ but if we take into account wake pattern, 8 cells seem to be more adapted to have a good resolution of vortices in the cylinder's wake. 
\begin{table}[H]
	\centering
\begin{tabular}{clc l clc l  }
	\hline
	Re     & $\theta$& \multicolumn{2}{c}{$C_d$} & \multicolumn{1}{c}{$E(C_d) \%$} \\
	%\cline{3-4} \cline{4-5}
	     &   & \cite{pierson2019}  & Present results & $ $  \\
 \hline
	100         & $5^\circ$ &  {0.472} & 0.479 & 1.4  \\
	            & $10^\circ$ &   0.499 & 0.501 & 0.4       \\
	            & $15^\circ$ &  0.536  &  0.534 & 0.3  \\
	            & $30^\circ$ &   0.693  &  0.680 & 1.4  \\
	\hline
	200         & $10^\circ$ &  0.350 & 0.345 &$ 1.4$  \\
	            & $15^\circ$ &   0.405 & 0.411 & 1.5       \\
\hline
\end{tabular}
		\caption{Comparison of the drag coefficient obtained with the present numerical methodology with results of \cite{pierson2019} for $5 ^\circ \leq \theta \leq 30^\circ$ and two moderate Reynolds numbers. $E(C_d)$ is the relative difference between the drag coefficients provided by the two sources. }
\label{table2}
\end{table}
\noindent Then we checked present results obtained with eight cells across the boundary layer against those of \cite{pierson2019} based on the PELIGRIFF code \citep{Wachs2009}. Table \ref{table2} shows how the two sets of the results compare for six different flow configurations. The drag coefficients are seen to differ by less than $1.5\%$ in all cases. %The results give strong confidence in the present numerical methodology.

% indicating that our numerical domain and grid points in the boundary layer are good enough  

\section{Higher-order zero-Reynolds-number slender-body prediction for the drag on a finite-length cylinder aligned with the flow or perpendicular to it}
\label{app:slender}
The hydrodynamic force experienced by a slender body immersed in a nonuniform flow was derived independently by \citet{batchelor1970}, \citet{cox1970} and \citet{keller1976} in the form of an expansion with respect to the small parameter $ 1/\ln(2\cchi)$ for $\cchi \gg 1$. References \cite{batchelor1970} and \cite{keller1976} provide an expansion up to order 3 in this small parameter. However, the logarithmic dependence of the force with respect to $\cchi$ makes higher-order contributions significant as soon as $\cchi$ becomes of $\mathcal{O}(10)$ or less. This is why a higher-order prediction is desirable to obtain a more accurate evaluation of the force on moderately-long cylinders. In this appendix, restricting ourselves to the case where the cylinder is aligned with the incoming flow, we provide the expression for the drag valid up to order 4, based on the expansion carried out in \cite{keller1976}. \\
\indent The total force experienced by a slender fiber of length $L$ immersed in a viscous flow may be expressed in the form 
\begin{equation}
\mathbf{F} = -8 \pi \mu L \int _0 ^1 \mathbf{f}(s) \mathrm{d} s\,,
\label{tot}
\end{equation}
where $\mathbf{f}(s)$ is the density of the Stokeslet distribution along the body centerline and $s$ denotes the arc length. The density $\mathbf{f}$ is obtained through a matched asymptotic procedure, the details of which may be found in \cite{keller1976}. If the body is a circular cylinder aligned with the flow direction, one has $\mathbf{f}(s) = f _x(x)  \mathbf{e_x}$ with 
%the  needed to match the boundary boundary conditions on the fiber
\begin{equation}
\displaystyle f_x(x)=-\frac{1}{2\ln\left(2\cchi\right)}\left(\frac{U_x}{2}+f_x(x)\left(\ln(4x(1-x))-1\right)+\int_{-x}^{1-x}\frac{f_x(x+t)-f_x(x)}{\vert t \vert}dt\right)\,,
\label{eq:slender_f_x}
\end{equation}
where $U_x=\bf{U}\cdot\bf{e_x}$. 
An approximate solution of (\ref{eq:slender_f_x}) may be obtained by successive approximations. Setting $f_x=0$ in the right-hand side, the first-order approximation is found to be $f _x^{(1)} = -U_x / (4 \ln\left(2\cchi)\right) $. The iterative solution was obtained in \cite{keller1976} up to order 3 in the form

%An iterative solution can be obatined by succisively injecting the las itration on the RHS of ...
\begin{align}
f_x^{(3)}(x)=-\frac{U_x}{4\ln\left(2\cchi\right)}&\left( 1-\frac{1}{2\ln\left(2\cchi\right)}\left(\ln(4x(1-x))-1\right)+\frac{1}{(2\ln\left(2\cchi\right))^2}\left(\ln(4x(1-x))-1\right)^2\right. \nonumber\\
&\left. +\frac{1}{(2\ln\left(2\cchi\right))^2}\int_{-x}^{1-x}\frac{\ln((x+t)(1-x -t))-\ln(x(1-x))}{\vert t \vert}dt\right).
\label{eq:slender_f3}
\end{align}
%+\frac{1}{2\ln\left(\frac{R}{L}\right)}\int_{-z}^{1-z}\frac{f_z^{(1)}(z+t)-f_z^{(1)}(z)}{\vert t \vert}dt
%&-\frac{U_x}{4\ln\left(2\cchi\right)}\left(\frac{1}{(2\ln\left(2\cchi\right))^2}\int_{-x}^{1-x}\frac{f_x^{(2)}(x+t)-f_x^{(2)}(x)}{\vert t \vert}dt\right).
%We do not explicit the last term in the integral since it does not contribute to the force \citep{keller1976}. 

\noindent Integrating the penultimate term in (\ref{eq:slender_f3}), we obtain

%\begin{align}
%F_x^{(3)}=&-8 \pi \mu L \int _0 ^1 f ^{(3)}(x) \mathrm{d} x
%=2 \pi \mu L U_x \left( \frac{1}{\ln(2\cchi)} + \frac{3/2 - \ln2}{(\ln(2\cchi))^2} + \frac{13/4-\pi^2/12+\ln2 (\ln2 -3)}{(\ln(2\cchi))^3} \right)&
%\end{align}

\begin{align}
F_x^{(3)}=&-8 \pi \mu L \int _0 ^1 f_x^{(3)}(x) \mathrm{d} x
=2 \pi \mu L U_x \left( \frac{a_x^{(1)}}{\ln(2\cchi)} + \frac{a_x^{(2)}}{(\ln(2\cchi))^2} + \frac{a_x^{(3)}}{(\ln(2\cchi))^3} \right)\,,
\end{align}

\noindent with $a_x^{(1)} = 1 $, $a_x^{(2)} = 3/2 - \ln2 \approx 0.80685$, and $a_x^{(3)} = 13/4-\pi^2/12+\ln2 (\ln2 -3) \approx 0.82854$. This result agrees with those of \cite{batchelor1970} and \cite{keller1976}. At next order, the force density may be obtained by inserting (\ref{eq:slender_f3}) in the right-hand side of (\ref{eq:slender_f_x}), yielding

%Following the same procedure the fourth order solution can be obtained  :

\begin{align}
f_x^{(4)}(x)&=-\frac{U_x}{4\ln\left(2\cchi\right)}\left(1-\frac{1}{2\ln\left(2\cchi\right)}\left(\ln(4x(1-x))-1\right)+\frac{1}{(2\ln\left(2\cchi\right))^2}\left(\ln(4x(1-x))-1\right)^2\right. \nonumber\\
&\left.-\frac{1}{(2\ln\left(2\cchi\right))^3}\left[\left(\ln(4x(1-x))-1\right)^3+\left(\ln(4x(1-x))-1\right)\int_{-x}^{1-x}\frac{\ln((x+t)(1-x -t))-\ln(x(1-x))}{\vert t \vert}dt\right]\right.\nonumber\\
&\left.-\frac{1}{(2\ln\left(2\cchi\right))^3}\int_{-x}^{1-x}\frac{f_x^{(3)}(x+t)-f_x^{(3)}(x)}{\vert t \vert}dt\right).
\label{eq:slender_f4}
\end{align}

\noindent The drag acting on the body is eventually obtained by making use of the previous expression in (\ref{tot}) and integrating along the body centerline. Since the last term in (\ref{eq:slender_f4}) does not contribute to the force \citep{keller1976}, one is left with

\begin{equation}
F_x^{(4)}= F_x^{(3)}  + 2\pi\mu L U_x \left(\frac{a_x^{(4)}}{(\ln(2\cchi))^4}\right)\,,
\end{equation}

%\begin{equation}
%F_x^{(4)}= F_x^{(3)}  + 2\pi\mu L U_x \left(\frac{-10 \zeta (3)+79+\pi ^2 (\log 4-3)-\log 4 (39+((\log 4)^2-9 \log 4))}{8(\ln(2\cchi))^4}\right),
%\end{equation}

%\begin{equation}
%F_x^{(4)}= F_x^{(3)}  +2\pi\mu L U_x \left(\frac{79/8 - 5/4 \zeta (3)+\pi ^2 ((\log 2)/4-3/8)-\log 2 (39/4+((\log 2)^2-9/2 \log 2))}{8(\ln(2\cchi))^4}\right)
%\end{equation}

\noindent with $a _x^{(4)} = \left[ -10 \zeta (3)+79+\pi ^2 (\ln 4-3)-\ln 4 (39+(\ln 4)^2-9 \ln 4) \right]/8 \approx 1.45243$, $\zeta$ denoting the Riemann zeta function. The main difficulty in the integration required to obtain $a^{(4)}$ results from the first integral in the right-hand side of (\ref{eq:slender_f4}). A formal computation using Mathematica\textsuperscript{TM} indicates that this term provides a contribution of $2 \zeta(3)$.
%                 2
%Out[11]= -29 - Pi  (-1 + Log[4]) + Log[4] (15 + (-3 + Log[4]) Log[4]) + 12 Zeta[3]

%It reads 
%\begin{align}
%f_y^{(4)}(x)=&-\frac{U_y}{2\ln\left(2\cchi\right)}\left(1-\frac{1}{2\ln\left(2\cchi\right)}\left(\ln(4x(1-x))+1\right)+\frac{1}{(2\ln\left(2\cchi\right))^2}\left(\ln(4x(1-x))+1\right)^2\right. \nonumber\\
%&\left.-\frac{1}{(2\ln\left(2\cchi\right))^3}\left[\left(\ln(4x(1-x))+1\right)^3+\left(\ln(4x(1-x))+1\right)\int_{-x}^{1-x}\frac{\ln((x+t)(1-x -t))-\ln(x(1-x))}{\vert t \vert}dt\right]\right.\nonumber\\
%&\left.-\frac{1}{(2\ln\left(2\cchi\right))^3}\int_{-x}^{1-x}\frac{f_y^{(3)}(x+t)-f_y^{(3)}(x)}{\vert t \vert}dt\right).
%\label{eq:slender_fy4}
%\end{align}
\color{black}
\section{Drag force on a cylinder held perpendicular to the flow: Slender-body approximation and semiempirical laws at zero and low-but-finite Reynolds number}
\label{perps}

Due to the linearity of the Stokes equation, the force acting on an arbitrarily inclined cylinder in the low-$\text{Re}$ regime may be obtained by suitably combining linearly the drag forces corresponding to the aligned ($\theta=0^\circ$) and perpendicular ($\theta=90^\circ$) configurations.
This is why an accurate estimate of the zero-$\text{Re}$ drag force on a finite-length cylinder held perpendicular to the incoming flow is desirable. To our surprise, such an estimate does not seem to be available in the literature. \citet{clift1978} proposed an empirical relationship accurate for moderate aspect ratios but did not match it with the prediction of the the slender-body theory in the limit of large aspect ratios. In this appendix, we first use the methodology employed in appendix \ref {app:slender} to establish the $4^{th}$-order slender-body approximation of the corresponding drag force at $\text{Re}=0$. Then we modify the corresponding expression in an \textit{ad hoc} manner to extend its validity to short cylinders, before incorporating the finite-$Re$ correction derived in \cite{khayat1989}.\\
% to be accurate at moderate aspect ratio.
\indent Duplicating the technique used in appendix \ref {app:slender}, the density of the Stokeslet distribution $f_y$ required to obtain the force $F_y$ on the cylinder is obtained by replacing $\ln(4x(1-x))-1$ everywhere with $\ln(4x(1-x))+1$, and $U_x$ with $2 U_y$ in (\ref{eq:slender_f4}) \citep{keller1976}. Then the total force is found to be

\begin{align}
F_y^{(4)}=&-8 \pi \mu L \int _0 ^1 f_y ^{(4)}(x) \mathrm{d} x
=4 \pi \mu L U_y \left( \frac{a_y^{(1)}}{\ln(2\cchi)} + \frac{a_y^{(2)}}{(\ln(2\cchi))^2} + \frac{a_y^{(3)}}{(\ln(2\cchi))^3}  + \frac{a_y^{(4)}}{(\ln(2\cchi))^4}\right)\,,&
\label{eq:fperp4}
\end{align}
with $a_y^{(1)} = 1 $, $a_y^{(2)} = 1/2 - \ln2 \approx -0.19315$, $a_y^{(3)} = 5/4-\pi^2/12+\ln 2 (\ln 2 -1) \approx 0.21484$ and $a_y^{(4)} = \left[ -10 \zeta (3)+29+\pi ^2 (\ln 4-1)-\ln 4 (15+(\ln 4)^2-3 \ln 4) \right]/8 \approx 0.38735$.

\begin{figure}[h]
\centering
\includegraphics[width=6cm]{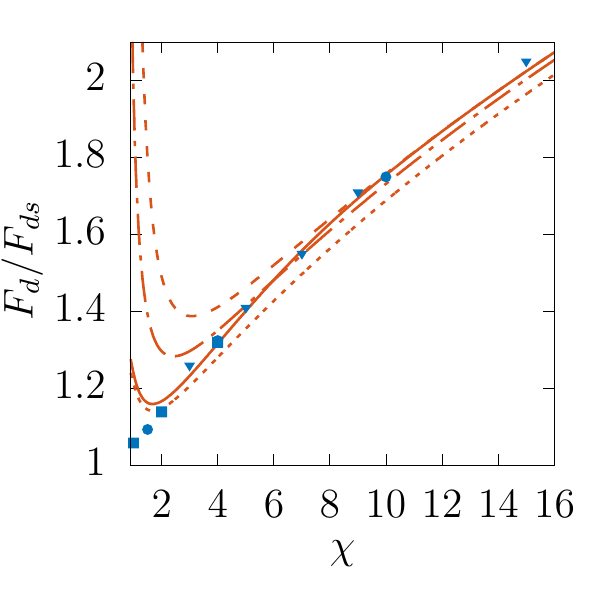}
\vspace{-2mm}
\caption{Drag on a finite-length cylinder held perpendicular to the flow direction, normalized by the drag $F_{ds}$ of a
sphere of same volume. Dotted, dash-dotted and dashed lines: slender-body approximation (\ref{eq:fperp4}) truncated at $2^{nd}$, $3^{rd}$ and $4^{th}$ order, respectively; solid line: semiempirical formula (\ref{eq:fperp4m}); $\bullet$: numerical results of \cite{ui1984}; $\blacksquare$: experimental results of \cite{heiss1952}; $\blacktriangledown$: experimental results of \cite{kasper1985}.}
\label{fig:f_perp_stokes}
\end{figure}

Figure \ref{fig:f_perp_stokes} displays the corresponding successive predictions %drag force (normalized by the drag force on a sphere of the same volume) in this configuration. Making use of coefficients $a_y^{1}-a_y^{4}$ determined in (\ref{eq:fperp4}), predictions from the slender-body approximation at successive orders are 
and compares them with experimental and numerical data. The $3^{rd}$-order approximation provides a fairly good agreement for small-to-moderate aspect ratios. However, neither the $2^{nd}$-order nor the $3^{rd}$-order approximation properly matches available data in the limit of high aspect ratios. The $4^{th}$-order approximation provides a better prediction at high $\chi$ but quickly diverges as $\chi$ becomes less than $\approx4$. Based on these observations, we empirically modify (\ref{eq:fperp4}) by weighting the third- and fourth-order terms with a pre-factor that quickly varies from $1$ for moderate-to-large $\cchi$ to $0$ for $\chi\rightarrow1/2$, in such a way that the behavior of the $3^{rd}$-order expansion is recovered for moderate aspect ratios. The corresponding modified drag law reads

\begin{equation}
F_y^{(4)}=-8 \pi \mu L \int _0 ^1 f_y ^{(4)}(x) \mathrm{d} x
=4 \pi \mu L U_y \left( \frac{a_y^{(1)}}{\ln(2\cchi)} + \frac{a_y^{(2)}}{(\ln(2\cchi))^2} + \left(1-e^{-c_1\cchi^{c_2}}\right)\left(\frac{a_y^{(3)}}{(\ln(2\cchi))^3}  + \frac{a_y^{(4)}}{(\ln(2\cchi))^4}\right)\right)\,,
\label{eq:fperp4m}
\end{equation}
with $c_1=0.01$ and $c_2=2.5$.
As Fig. \ref{fig:f_perp_stokes} shows, (\ref{eq:fperp4m}) properly approximates available experimental and numerical results down to $\chi \approx2$. \\
\indent A second step is to capture the drag increase due to finite-$\text{Re}$ effects. \citet{khayat1989} computed such effects up to second order with respect to $1/\text{ln}(2\cchi)$ and obtained (see also \cite{lopez2017} and \cite{roy2019}) 
 \begin{equation}
F_y^{\cchi\text{Re}=\mathcal{O}(1)}\approx  4 \pi \mu L U \left( \frac{a_y^{(1)}}{\ln\cchi} + \frac{a_y^{(2)}-\ln2+f_\perp}{(\ln\cchi)^2} \right)\,,
\label{eq:slender_perp_smallinertia}
\end{equation} 
 with 
  \begin{equation}
 f_\perp = E_1(\frac{\cchi \text{Re}}{2})+\ln(\frac{\cchi \text{Re}}{2})-2\left(\frac{e^{-{\chi\text{Re}}/2 }-1}{\cchi \text{Re}}\right)+\gamma -1  \,,\nonumber
 \end{equation}
so that $f_\perp\rightarrow\frac{1}{4}\cchi\text{Re}$ when $\cchi\text{Re}\rightarrow0$. As in the $\theta=0^\circ$-case, the main shortcoming of (\ref{eq:slender_perp_smallinertia}) is the second-order truncation with respect to $1/\text{ln}(2\cchi)$. To partly alleviate this limitation, we take advantage of the higher-order corrections present in (\ref{eq:fperp4m}) and merely add the second-order finite-$\text{Re}$ correction while leaving higher-order terms unchanged. This yields the empirical composite approximation
 \begin{equation}
F_y^{\cchi\text{Re}=\mathcal{O}(1)}\approx  4 \pi \mu L U_y \left( \frac{a_y^{(1)}}{\ln(2\cchi)} + \frac{a_y^{(2)}+f_\perp}{(\ln(2\cchi))^2} + \left(1-e^{c_1\cchi^{c_2}}\right)\left(\frac{a_y^{(3)}}{(\ln(2\cchi))^3}  + \frac{a_y^{(4)}}{(\ln(2\cchi))^4}\right)\right).
\label{eq:slender_perp_smallinertia_m}
\end{equation}

\begin{figure}[h]
\centering
\includegraphics[width=5cm]{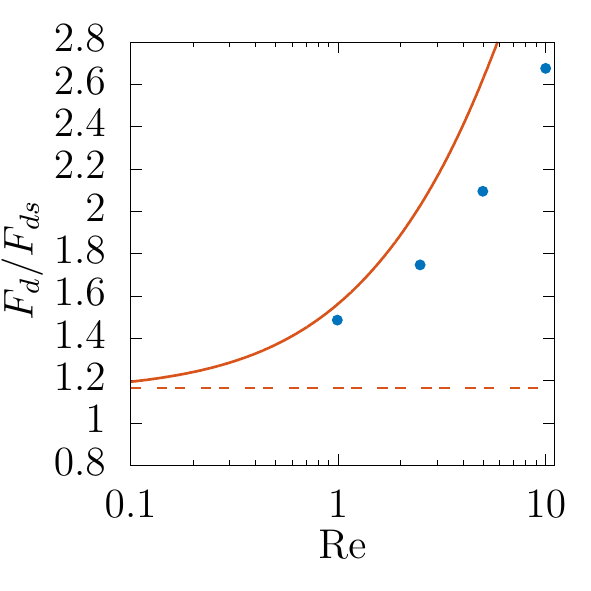}\includegraphics[width=5cm]{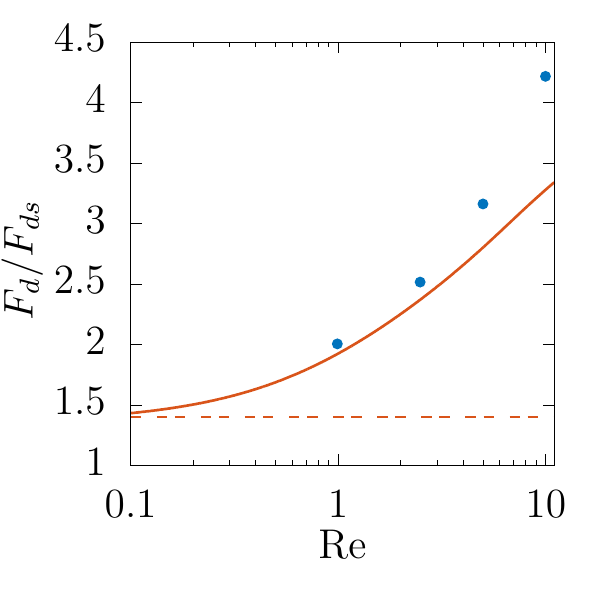}\includegraphics[width=5cm]{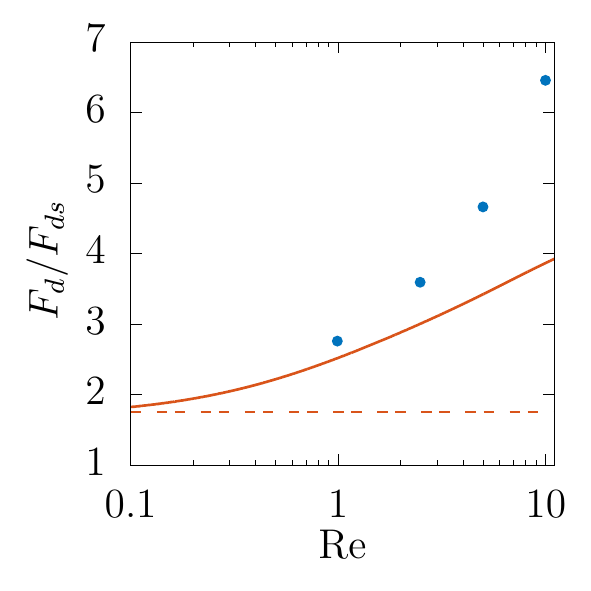}\\
\vspace{-5mm}
 \hspace{-3cm}{$(a)$}\hspace{4.6cm}{$(b)$}\hspace{4.6cm}{$(c)$}\\
 \vspace{-43mm}
 \hspace{-10mm}$\cchi=2$ \hspace{41mm}$\cchi=5$ \hspace{41mm}$\cchi=10$\\
  \vspace{40mm}
\caption{Influence of inertial effects on the drag of finite-length cylinders held perpendicular to the upstream flow. The drag is normalized by that of a sphere of same volume. %$(a)$: $\cchi=2$, $(b)$: $\cchi=5$, $(c)$: $\cchi=10$. 
Solid line: semiempirical prediction (\ref{eq:slender_perp_smallinertia_m}); dashed line: prediction (\ref{eq:fperp4m}); $\bullet$: numerical results of \cite{vakil2009}.}
\label{fig:ca_perp_moderate}
\end{figure}
\noindent As Fig. \ref{fig:ca_perp_moderate} indicates, predictions from (\ref{eq:slender_perp_smallinertia_m}) almost match the numerical results of \cite{vakil2009} for $\text{Re}=1$. Not surprisingly, they increasingly deviate from these results as $\text{Re}$ increases beyond this point,  over- (under-) predicting the actual drag for $\cchi=2$ ($\cchi=5,\,10$). 
\color{black}
%For larger Reynolds number an important deviation is observed. For the smallest aspect ratio, we observe an important scattering of the experimental vs numerical data. It is thus difficult fo make an accurate prediction. Also figures \ref{fig:c_perp_moderate} show that the increase of the drag force with the Reynolds number is particularly important for high aspect ratios. 
%\bibliographystyle{apalike}
\bibliography{biblio}
\bibliographystyle{unsrtnat}

\end{document}